\documentclass[%
 reprint,
 onecolumn,
 aps
]{revtex4-2}

\usepackage{graphicx}
\usepackage{dcolumn}
\usepackage{bm}
\usepackage{float}
\usepackage{amsmath,amssymb,amsfonts}
\usepackage{todonotes}
\def\Xint#1{\mathchoice
	{\XXint\displaystyle\textstyle{#1}}%
	{\XXint\textstyle\scriptstyle{#1}}%
	{\XXint\scriptstyle\scriptscriptstyle{#1}}%
	{\XXint\scriptscriptstyle\scriptscriptstyle{#1}}%
	\!\int}
\def\XXint#1#2#3{{\setbox0=\hbox{$#1{#2#3}{\int}$}
		\vcenter{\hbox{$#2#3$}}\kern-.5\wd0}}

\usepackage{hyperref}
\usepackage{color}  
\definecolor{darkGreen}{rgb}{0,0.45,0}
\definecolor{darkBlue}{rgb}{0,0,0.7}
\definecolor{darkRed}{rgb}{0.76, 0.13, 0.28}
\hypersetup{
	colorlinks=true, 
	linktoc=all,    
	linkcolor=darkBlue, 
	citecolor=darkGreen,
	urlcolor = darkRed,
}
\usepackage{comment}

\renewcommand{\d}{{\mathrm{d}}}

\renewcommand{\Re}{\mathrm{R}}
\renewcommand{\Im}{\mathrm{I}}
\newcommand{\real}{\mathrm{Re}}
\newcommand{\imag}{\mathrm{Im}}
\newcommand{\nonlin}{\mathrm{nonlin}}
\usepackage[margin=2.3cm]{geometry}
\begin{document}

\title{Eigenmode analysis of membrane stability in inviscid flow}
\author{Christiana Mavroyiakoumou}%
\email[Electronic address: ]{chrismav@umich.edu}
\author{Silas Alben}
\email[Electronic address: ]{alben@umich.edu}
\affiliation{\vspace{.1cm}Department of Mathematics, University of Michigan, Ann Arbor, MI 48109, USA\vspace{.1cm}}

\date{\today}

\begin{abstract}
We study the instability of a thin membrane (of zero bending rigidity)
to out-of-plane deflections, when the membrane is immersed in an inviscid fluid flow and sheds a trailing vortex-sheet wake. We solve the nonlinear eigenvalue problem iteratively with large ensembles of initial guesses, for three canonical boundary conditions---both ends fixed, one end fixed and one free, and both free. Over several orders of magnitude of membrane mass density, we find instability by divergence or flutter  (particularly at large mass density, or with one or both ends free). The most unstable eigenmodes generally become ``wavier" at smaller mass density and smaller tension, but with regions of nonmonotonic behavior.
We find good quantitative agreement with unsteady time-stepping simulations at small amplitude, but only qualitative similarities with the eventual steady-state large-amplitude motions.
\end{abstract}

\maketitle

\section{Introduction\label{sec:intro}}

When extensible membranes of zero bending rigidity are placed in a fluid flow, the interaction between membrane inertia, resistance to stretching, and external fluid forces can result in complex time-dependent deformations and dynamics. This holds both for large-amplitude motions and the initial small-amplitude motions that determine the stability of undeflected membranes. Predicting the onset of membrane instability across parameter space, either by flutter, divergence, or a combination of the two~\cite{sygulski2007stability,tiomkin2017stability,mavroyiakoumou2020large}, is fundamental to a wide range of applications. As lightweight, deployable structures that are stable in a variety of configurations, membranes are used, for example, in sails~\cite{nielsen1963theory,newman1984two,newman1987aerodynamic,colgate1996fundamentals,kimball2009physics}, parachutes~\cite{pepper1971aerodynamic,stein2000parachute}, micro-air vehicles~\cite{shyy1999flapping,lian2003membrane,albertani2007aerodynamic,hu2008flexible,stanford2008fixed}, ballutes for space exploration~\cite{scott2007aeroelastic,rohrschneider2007survey}, roofs in civil engineering~\cite{haruo1975flutter,knudson1991recent,sygulski1996dynamic,sygulski2007stability}, and the wings of flying animals~\cite{swartz1996mechanical,song2008aeromechanics,cheney2015wrinkle}. 

In an early work, Nielsen~\cite{nielsen1963theory} studied a membrane with both edges fixed in a two-dimensional flow, and determined the critical value of the pretension parameter that gives rise to a fully convex membrane shape. An overview of early models based on potential-flow aerodynamics can be found in~\cite{newman1987aerodynamic}. Since then, theoretical~\cite{newman1991stability,sygulski2007stability}, computational~\cite{jaworski2012high,tiomkin2017stability,nardini2018reduced,mavroyiakoumou2020large}, and experimental~\cite{le1999unsteady} studies of membrane stability have revealed various
aspects of membrane stability and dynamics with various boundary conditions. 

In~\cite{mavroyiakoumou2020large}, we used
a non-linear time-stepping algorithm to compute the stability thresholds for membranes with three sets of boundary conditions: ``fixed-fixed," ``fixed-free," and ``free-free" leading and trailing edges. Membrane tension has a stabilizing effect in all cases. The ratio of membrane-to-fluid inertia has a less obvious effect---heavier membranes may be unstable when a lighter membrane was not, but the instability grows more slowly as membrane mass increases, to the point where it is difficult to determine whether the membrane is stable or not. Nonlinear time-stepping simulations with evolving vortex sheet wakes are expensive when large simulation times are required (i.e.\ to assess the stability of heavy membranes), and when the membrane develops fine deformations (as occurs for lighter membranes and smaller pretension values). 
In the latter case a fine grid on the membrane is required, increasing the size of the coupled system of equations that is solved implicitly, and
making it more ill-conditioned, slowing 
convergence at each time step.
 
Therefore, in this work, we develop a less expensive alternative to study the stability problem---a nonlinear eigenmode solver. We solve for an ensemble of eigenmodes and corresponding eigenvalues (growth rates and frequencies) corresponding to small-amplitude deformations. By comparing with unsteady simulations, we find that the modes accurately capture the early stages of the unsteady motion starting from the undeflected state. By comparing at later times, we find that the mode shapes qualitatively resemble those of the steady-state large amplitude motions to varying degrees.

Due to the vortex wake, simple exact eigenmode solutions are difficult to obtain, and the physical mechanisms that underlie the membrane instability are somewhat elusive, but in the present study we are able to present a comprehensive characterization of the modes and growth rates in the vicinity of the stability boundary.
 The eigenmode approach has been used previously to study membrane stability with fixed-fixed~\cite{nielsen1963theory,sygulski2007stability,tiomkin2017stability,nardini2018reduced} and periodic~\cite{newman1991stability} boundary conditions. We use our method on the fixed-fixed case, as well as the fixed-free and free-free cases introduced in~\cite{mavroyiakoumou2020large}, where a wider range of dynamics can occur. In each case, we study a much wider range of membrane mass density and pretension values than previous studies.
A version of the present method was previously used to study the flutter instability of bending beams in inviscid flows~\cite{alben2008ffi}. There solutions were obtained by continuation, starting from the known oscillation modes of a beam in a vacuum. Here we study membranes (with zero bending rigidity), and find that the continuation approach is now more susceptible to jumping between different eigenmode branches as we vary parameters. Therefore, we solve the nonlinear eigenvalue problem using dense meshes of initial eigenvalue guesses that cover the range of lower-mode states. As a result, we obtain a larger ensemble of eigenmodes at each parameter value set. We obtain good agreement with the stability results in~\cite{mavroyiakoumou2020large}, but are able extend the results to much larger and smaller values of the membrane-to-fluid density ratio, and resolve shapes with finer structures. 


The structure of the paper is as follows. In \S~\ref{sec:model} we present the membrane and vortex sheet model and in~\S~\ref{sec:linearized} its linearized, small-amplitude version, along with a summary of the numerical method for determining the eigenvalues and eigenmodes (\S~\ref{sec:numerical}). In \S\S~\ref{sec:linearizedfifi}--\ref{sec:linearizedfrfr} we present our results for an extensive range of parameters for each of the three boundary conditions. We then turn to simulations of the initial value problem and examine how the unsteady motions compare to the eigenmode shapes from the linearized model (\S~\ref{sec:comparisonIVP}). 
\S~\ref{sec:conclusions} presents conclusions.


\section{Membrane-vortex-sheet model}\label{sec:model}

\begin{figure}[H]
    \centering
     \includegraphics[width=.49\textwidth]{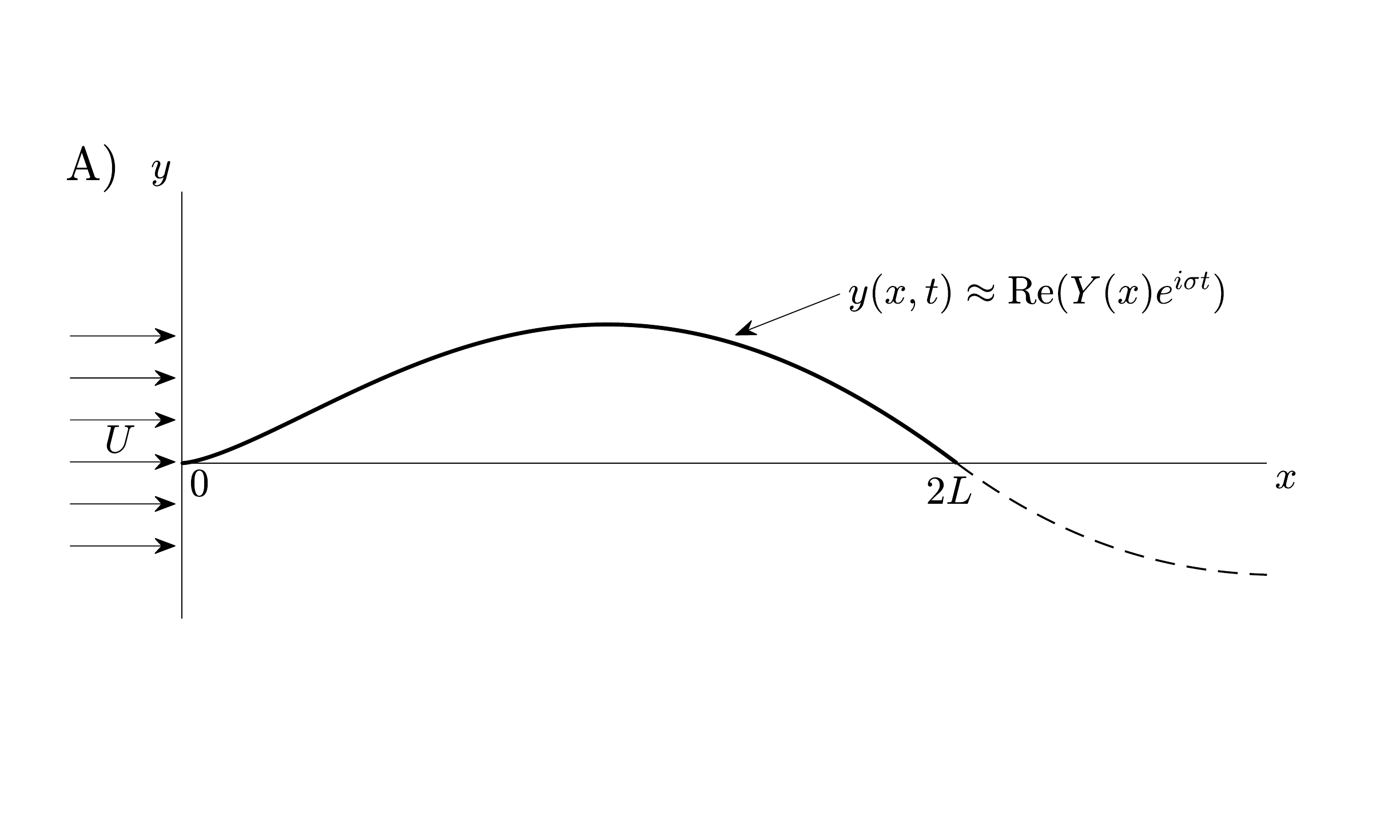}
    \includegraphics[width=.49\textwidth]{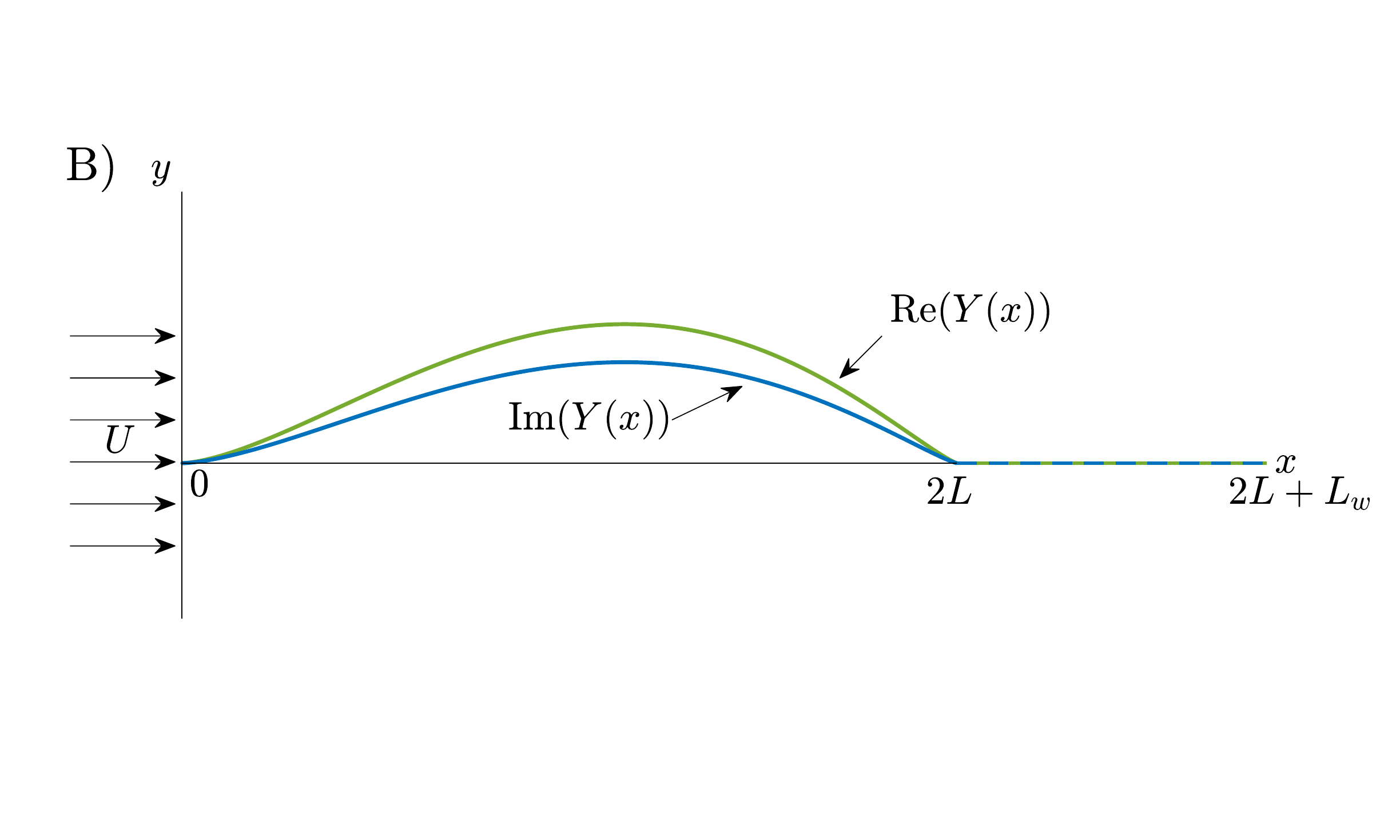}
     \includegraphics[width=.49\textwidth]{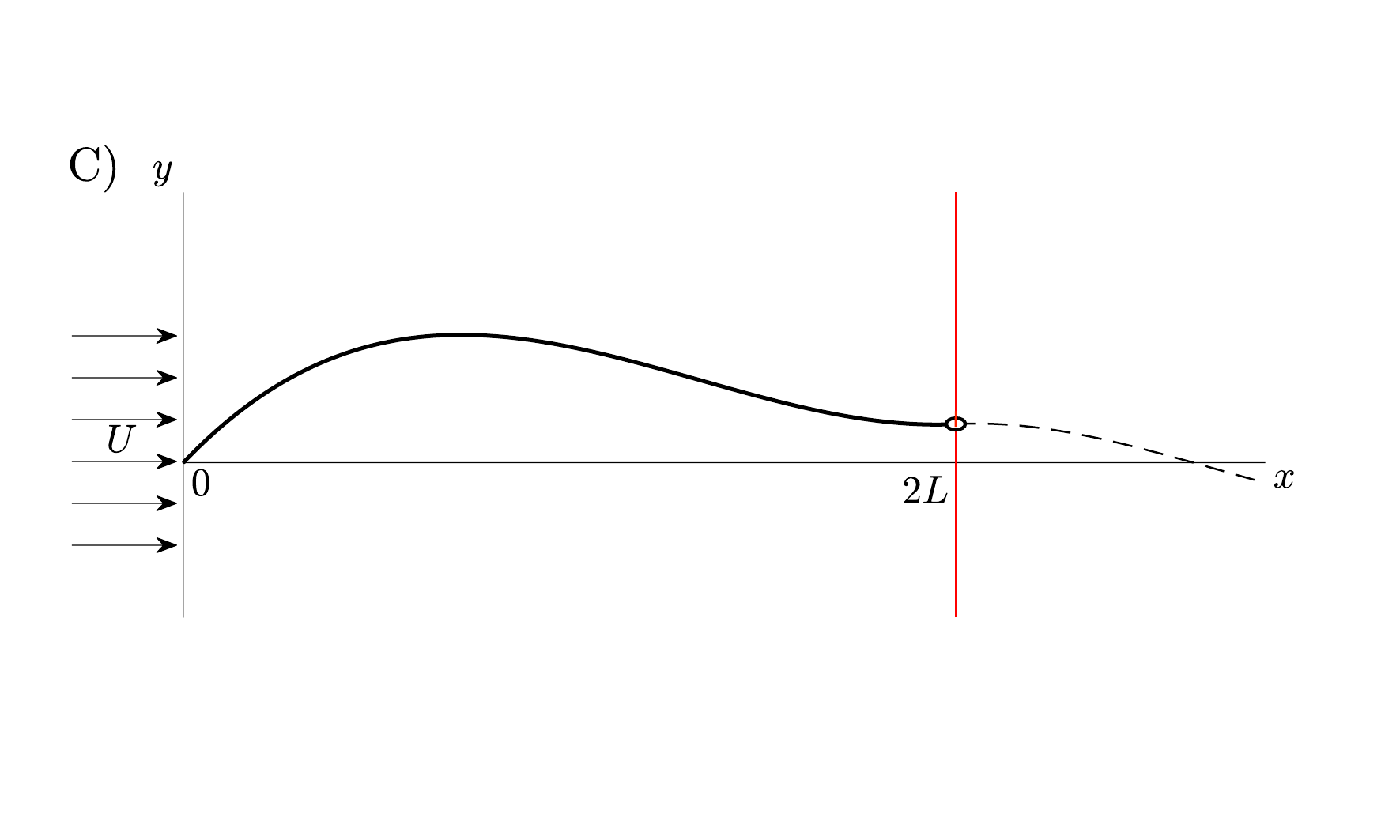}
     \includegraphics[width=.49\textwidth]{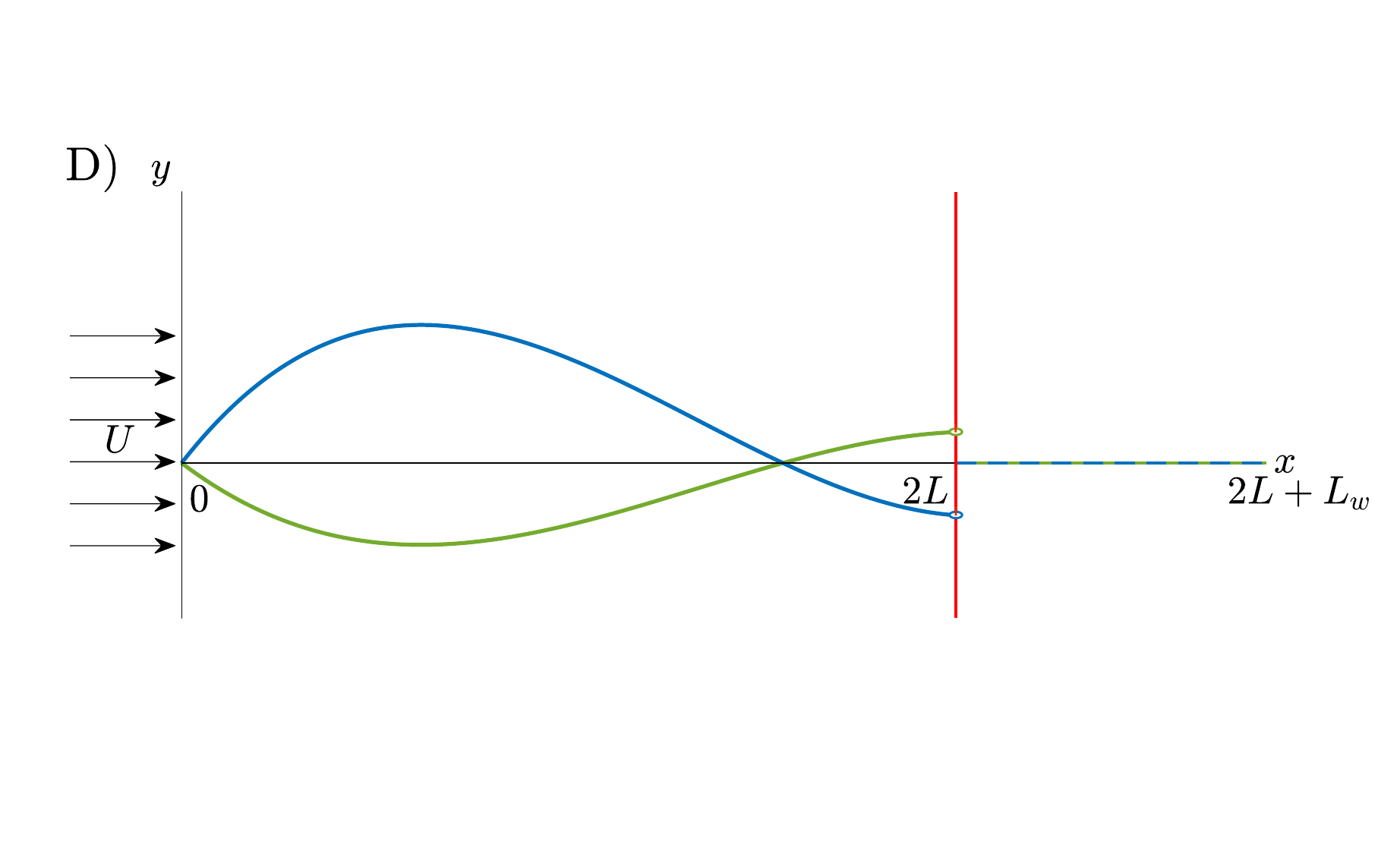}
     \includegraphics[width=.49\textwidth]{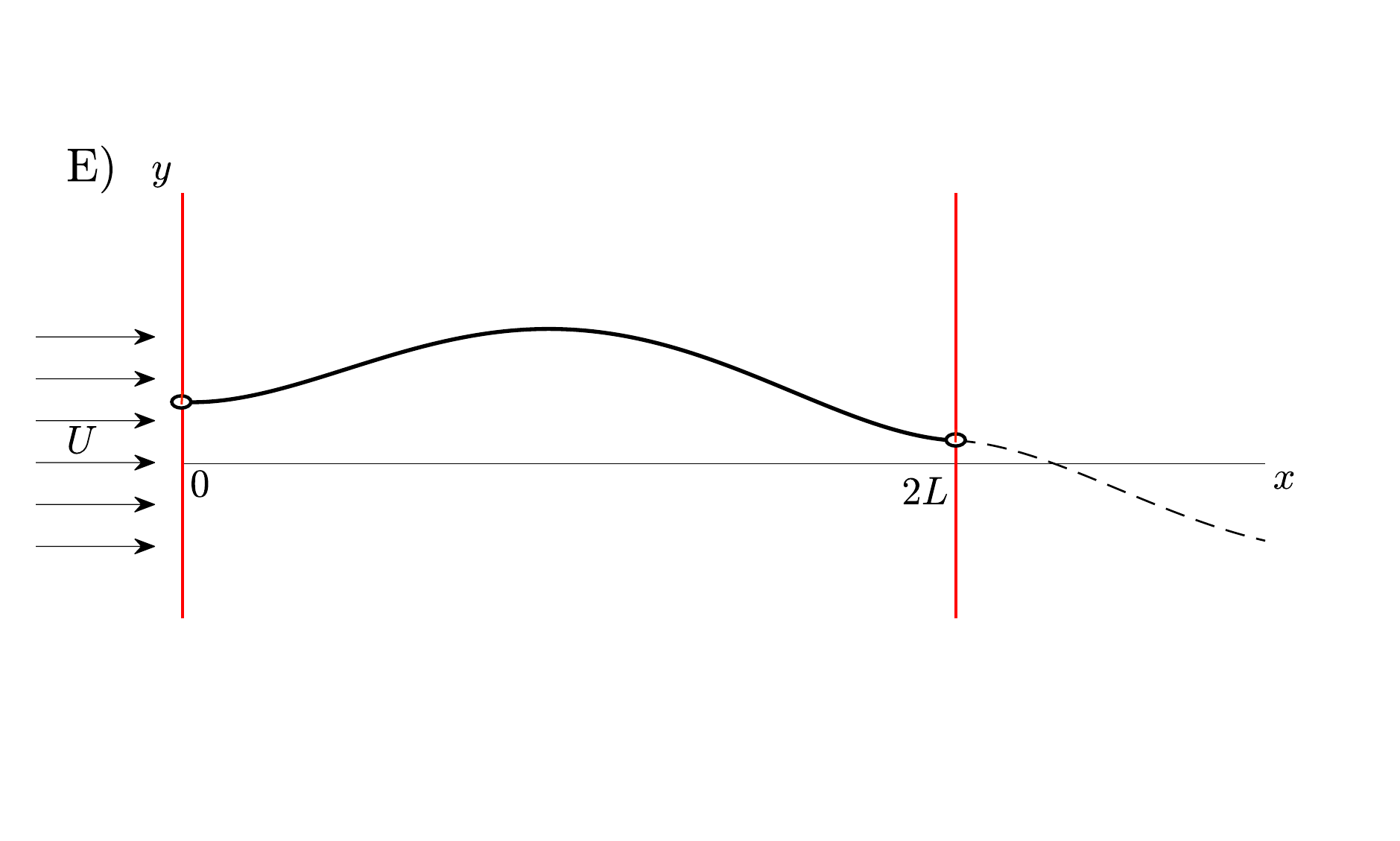}
    \includegraphics[width=.49\textwidth]{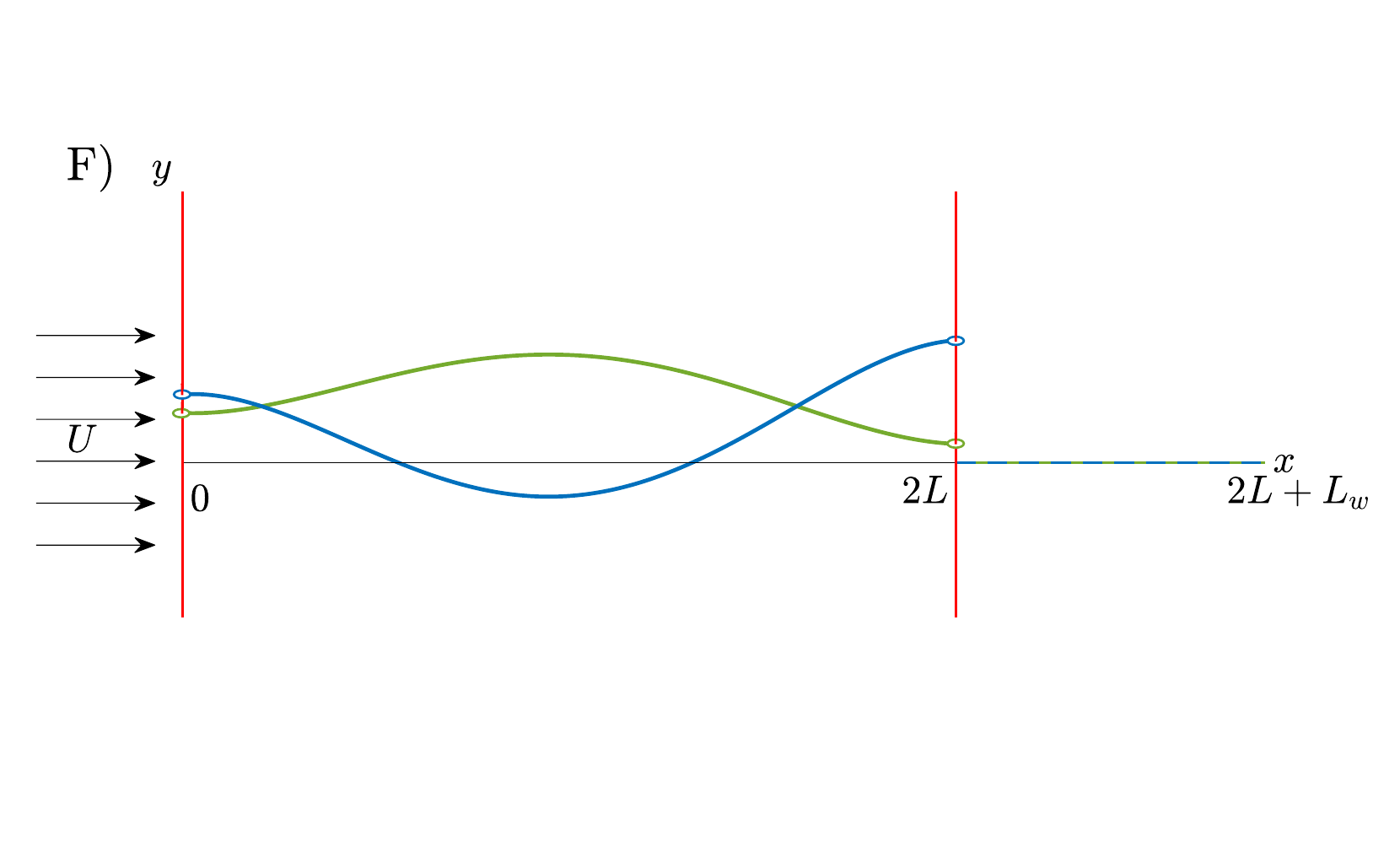}
    \caption{Schematic diagram of flexible membranes (solid curved black lines) at an instant in time. Here $2L$ is the chord length (the distance between the endpoints) and $U$ is the oncoming flow velocity. For the nonlinear, large-amplitude model (panels A, C, and E), 
    $y(x,t)$ is the membrane deflection and the dashed line is the free vortex wake. The right columns show the corresponding linearized, small-amplitude eigenvalue problems, where the motions are represented by the real and imaginary parts of the eigenmodes $Y(x)$, shown as green and blue lines, respectively, and flat vortex wakes of fixed length $L_w$ shown as dashed lines at $y=0$ (panels B, D, and F). The boundary conditions shown are: fixed-fixed membranes (panels A and B), fixed-free membranes (panels C and D), and free-free membranes (panels E and F).}
    \label{fig:schematicMembrane}
\end{figure}

We model the dynamics of an extensible membrane that is nearly aligned with a two-dimensional background fluid flow with speed $U$ in the far field (see figure~\ref{fig:schematicMembrane}).

In figure~\ref{fig:schematicMembrane}, we illustrate schematically the three cases of boundary conditions at the two ends of the membrane that we investigated in~\cite{mavroyiakoumou2020large}: fixed-fixed (panels~A and~B), fixed-free (panels~C and~D), and free-free (panels~E and~F). In all three cases, the $x$-coordinates of the ends are fixed at $0$ and $2L$. In the ``fixed-fixed'' case, we set the deflection to zero at both ends of the membrane; most previous studies of membrane flutter considered this boundary condition~\cite{le1999unsteady,sygulski2007stability,rojratsirikul2010unsteady,tiomkin2017stability,nardini2018reduced}. In the ``fixed-free'' case, the leading edge deflection is again set to zero but the trailing edge is allowed to deflect freely in the vertical direction. This is the classical free-end boundary condition for a membrane~\cite{graff1975wave,farlow1993partial}. The membrane end is fixed to a massless ring that slides along a vertical frictionless pole (represented by the red lines in figure~\ref{fig:schematicMembrane}). Since the pole is frictionless and the ring is massless, the membrane can exert no vertical force on the free end by tension, and hence the membrane slope must be zero.

Free-end boundary conditions have been implemented in various problems in classical mechanics such as beam flutter~\cite{kornecki1976aeroelastic,allen2001energy,ZP2002,tang2003flutter,shelley2005heavy,AM2005,de2006flutter,huang2007simulation,connell2007flapping,jia2007coupling,michelin2008vortex,alben2008flapping,eloy2008aeroelastic,manela2009stability,kim2013flapping,mougel2016synchronized,orrego2017harvesting,goza2018global,jin2019flow,tosi2019modeling},
but they have not been used to a great extent in membrane flutter problems.
Recently, an experimental study determined that membrane wing flutter can be enhanced by the vibrations of flexible leading and trailing edge supports~\cite{arbos2013leading}. For membrane wings with partially free trailing edges, trailing edge fluttering may occur at relatively low angles of attack~\cite{hu2008flexible}. Partially free edges occur also in sails. In~\cite{kimball2009physics}, it is shown that by altering the tension in cables running along its free edges one can control the shape of a sail membrane and when the tension in these edges is sufficiently low,  flutter can occur~\cite{colgate1996fundamentals}. A related application is to energy harvesting by membranes mounted on tensegrity structures 
and placed in fluid flows~\cite{sunny2014optimal,yang2016modeling}. 

The authors in~\cite{triantafyllou1994dynamic,manela2017hanging} consider the dynamics and flutter of membranes and cables under gravity with free ends. In~\cite{mavroyiakoumou2020large} and in the current work, to focus on the basic flutter problem~\cite{shelley2011flapping}, we do not include gravity in the model. However, we still need to ensure that the problem remains well-posed by requiring some restriction on the motion of the free membrane ends to eliminate the possibility of membrane compression~\cite{triantafyllou1994dynamic}. This restriction is provided by the vertical frictionless poles. This has been carried out experimentally by representing a membrane as an extensional spring tethered by steel wires to vertical supports~\cite{kashy1997transverse}, for example.

The model here is the same as in~\cite{mavroyiakoumou2020large} and we summarize it briefly for completeness. The nonlinear, extensible membrane equation is
\begin{equation}\label{eq:membrane}
R_1\partial_{tt}\zeta -\partial_\alpha ((T_0+R_3(\partial_\alpha s-1))\mathbf{\hat{s}})=-[p]_-^+\partial_\alpha s\mathbf{\hat{n}}.
\end{equation}
Here $\zeta(\alpha,t) = x(\alpha,t)+iy(\alpha,t)$ is the membrane position in the complex plane, parameterized by the material coordinate $\alpha$, $-L\leq \alpha\leq L$ ($L$ is half the chord length) and time $t$. The pressure jump across the membrane is $[p]_-^+$, the unit vectors tangent and normal to the membrane are $\mathbf{\hat{s}}$ and $\mathbf{\hat{n}}$, respectively, and $\partial_\alpha s$ is the local stretching factor. We use $+$ to denote the side towards which the membrane normal~$\mathbf{\hat{n}}$ is directed, and~$-$ for the other side. However, for the remainder of this paper, we drop the $+$ and $-$ for ease of notation. Equation~\eqref{eq:membrane} is made dimensionless by nondimensionalizing length by the membrane's half-chord $L$, time by~$L/U$, and pressure by $\rho_f U^2$, where $\rho_f$ is the density of the fluid and $U$ is the oncoming flow velocity. The dimensionless membrane mass is $R_1=\rho_sh/(\rho_f L)$, the dimensionless stretching rigidity is $R_3=Eh/(\rho_f U^2 L)$, and finally, $T_0=\overline{T}/(\rho_f U^2L W)$ is the dimensionless pretension. Here $\overline{T}$ is the tension in the (initial) undeflected equilibrium state. The membrane has length $2L$, uniform thickness $h$, mass per unit volume $\rho_s$, and Young's modulus~$E$. The model is linearized for small-amplitude membrane deflections in \S~\ref{sec:linearized} (shown schematically in figure~\ref{fig:schematicMembrane}, right column). 

The complex conjugate of the fluid velocity at any point $z$ in the flow is a sum of the horizontal background flow with dimensionless speed unity and the flow induced by the bound and free vortex wakes:
\begin{equation}
u_x(z)-iu_y(z)= 1 +\frac{1}{2\pi i}\int_{-1}^1\frac{\gamma(\alpha,t)}{z-\zeta(\alpha,t)}\partial_\alpha s \d\alpha+\frac{1}{2\pi i}\int_0^{s_{\max}} \frac{\gamma(s,t)}{z-\zeta(s,t)}\d s,
\end{equation}
where $s$ is the arc length along the free sheet starting at 0 at the membrane's trailing edge and extending to~$s_{\max}$ at the free sheet's far end. To determine the bound vortex sheet strength $\gamma$ we require that the fluid does not penetrate the membrane, i.e.\ the kinematic boundary condition. Here $\gamma$ also represents the jump in the component of the flow velocity tangent to the membrane from the $-$ to the $+$ side, i.e.\ $\gamma=-[(u_x, u_y)\cdot\mathbf{\hat{s}}]$. The normal components of the fluid and membrane velocities are equal:
\begin{equation}\label{eq:kinematic}
    \mathrm{Re}(\mathbf{\hat{n}}\partial_t\overline{\zeta}(\alpha,t)) =\mathrm{Re}\left( \mathbf{\hat{n}}\left( 1 +\frac{1}{2\pi i}\int_{-1}^1\frac{\gamma(\alpha,t)}{z-\zeta(\alpha,t)} \partial_\alpha s\d\alpha+\frac{1}{2\pi i}\int_0^{s_{\max}} \frac{\gamma(s,t)}{z-\zeta(s,t)}\d s \right)\right),
\end{equation}
\noindent where $\mathbf{\hat{n}}$ is written as a complex scalar.
Solving~\eqref{eq:kinematic} for
$\gamma$ on the body requires an additional constraint that the total circulation is zero for a flow started from rest. At each instant the part of the circulation in the free sheet, or alternatively, the strength of $\gamma$ where the free sheet meets the trailing edge of the membrane, 
is set by the Kutta condition which makes the flow velocity finite at the trailing edge. At every other point of the free sheet, $\gamma$ is set by the criterion that circulation (the integral of $\gamma$) is conserved at material points of the free sheet.
The vortex sheet strength $\gamma(\alpha,t)$ is coupled to the pressure jump $[p](\alpha,t)$ across the membrane using a version of  the unsteady Bernoulli equation written at a fixed material point on the membrane:
\begin{equation}\label{eq:pressure}
    \partial_\alpha s\partial_t\gamma +\partial_\alpha\left(\gamma(\mu-\tau)\right)+\gamma(\partial_\alpha \tau-\nu\kappa\partial_\alpha s)=\partial_\alpha [p],
\end{equation}
where $\mu$ is the average flow velocity tangent to the membrane, $\tau$ and $\nu$ are the tangential and normal components of the membrane velocity, respectively, and $\kappa$ is the curvature. At the trailing edge, $[p]|_{\alpha=1}=0$. The derivation of~\eqref{eq:pressure} is included in~\cite[appendix A]{mavroyiakoumou2020large}.



\section{Small-amplitude linearization}\label{sec:linearized}

The large-amplitude, nonlinear system described in \S~\ref{sec:model} becomes more amenable to analysis in the small-amplitude regime. Here we focus on the computation of eigenmodes and eigenvalues for the three boundary conditions studied in~\cite{mavroyiakoumou2020large}: ``fixed-fixed," ``fixed-free," and ``free-free" membranes. We are thereby able to present the small-amplitude motions of the membranes at larger and smaller membrane density than in the previous work, and in much greater detail. 
A similar linearized model was derived in~\cite{alben2008ffi} for the dynamics of a flapping flag.
We consider small deflections $y(x,t)$ from the straight configuration, aligned with the flow. Since the membrane stretching factor is $\partial _\alpha s\approx 1+\partial_x y^2/2$, to linear order $\alpha \approx s \approx x$, 
all $\alpha$-derivatives in~\eqref{eq:membrane} are $x$-derivatives, and $\zeta(\alpha,t) \approx \zeta(x,t)=x+iy(x,t)$. At linear order, the tangent and normal vectors are:
\begin{equation}\label{eq:linearNormal}
    \mathbf{\hat{s}}\approx(1,\partial_xy)^\top,\quad \mathbf{\hat{n}}\approx(-\partial_xy,1)^\top.
\end{equation}
The linearized version of the membrane equation is
\begin{equation}\label{eq:linearmembrane}
R_1\partial_{tt}y-T_0\partial_{xx} y = -[p].
\end{equation}

The term in the tension force $T(\alpha,t)=T_0+R_3(\partial_\alpha s-1)$ involving $R_3$ (dimensionless stretching rigidity) is of quadratic order, so the linear dynamics are governed by the dimensionless membrane mass $R_1$ and the dimensionless pretension $T_0$. The boundary conditions are:
\begin{align}
    \text{fixed-fixed:} &\quad  y(\pm 1,t)=0, \label{eq:bcfifi}\\
    \text{fixed-free:}\,\,\, &\quad  y(-1,t)=0,\,\partial_x y( 1,t)=0,\label{eq:bcfifr}\\
    \text{free-free:}\quad &\quad  \partial_x y(\pm 1,t)=0.\label{eq:bcfrfr} 
\end{align}
The dynamics of the membrane are coupled to the fluid flow through the pressure jump term $[p](x,t)$. The linearized version of the pressure jump equation is
\begin{equation}\label{eq:linearpressure}
\partial_t\gamma+\partial_x \gamma = \partial_x [p].
\end{equation}
The set of equations is closed by relating the vortex sheet strength $\gamma(x,t)$ back to the membrane position $y(x,t)$, through the kinematic condition, in linearized form:
 \begin{equation}\label{eq:linearKinematic}
 \partial_t y(x,t) =-\partial_x y(x,t) + \frac{1}{2\pi}\Xint- _{-1}^1 \frac{v(x',t)}{\sqrt{1-x'^2}(x-x')}\,\d x'+\frac{1}{2\pi} \int_{1}^{\ell_w+1}\frac{\gamma(x',t)}{x-x'}\,\d x',\quad -1<x<1.
 \end{equation}
Here, we use that $\partial_t\overline{\zeta}(x,t) \approx -i\partial_t y$ and from~\eqref{eq:linearNormal}, the normal velocity component $\mathrm{Re}(\mathbf{\hat{n}} \partial_t\overline{\zeta})\approx \partial_t y$. The general solution $\gamma(x,t)$ has inverse square-root singularities at $x=\pm 1$ and so we define $v(x,t)$, the bounded part of $\gamma(x,t)$ by $\gamma = v/\sqrt{1-x^2}$. The second integral in~\eqref{eq:linearKinematic} represents the velocity induced by the vortex sheet wake, which extends downstream from the membrane on the interval $1<x<\ell_w+1$, $y=0$. 
Therefore, the eigenvalue problem assumes a free vortex
wake of a given fixed length~$\ell_w$, which we take to be large (i.e.\ we assume we start with a deflection that is sufficiently small that we remain in the small-amplitude regime for a long time).

The circulation in the wake, 
\begin{equation}
\Gamma(x,t)=-\int_x^{\ell_w+1} \gamma (x',t)\,\d x',
\end{equation}
is conserved along material points of the wake by Kelvin's circulation theorem. At linear order,
the wake moves at the constant speed (unity) of the free stream; self-interaction is negligible.

At each time $t$, the total circulation in the wake, $\Gamma(1,t)$, is set by the Kutta condition, which in linearized form is unchanged, i.e.\ \begin{equation}\label{eq:kutta}
    v(1,t)=0.
\end{equation}

Using the system of equations~\eqref{eq:linearmembrane},~\eqref{eq:linearpressure},~\eqref{eq:linearKinematic}, and~\eqref{eq:kutta} we solve for the following unknowns: the motion of the membrane and the strengths of the vortex sheets along the membrane and in the wake.

For the linearized system, we may write solutions in the following form:
\begin{align}
y(x,t)&=Y(x)e^{i\sigma t},\label{eq:yexp}\\
\gamma(x,t)&=g(x)e^{i\sigma t},\label{eq:gammaaexp}\\
v(x,t)&=V(x)e^{i\sigma t},\label{eq:vexp}\\
\Gamma(1,t)&=\Gamma_0 e^{i\sigma t},\label{eq:Gammaexp}
\end{align}
where $Y$, $g$, $V$, and $\Gamma_0$ are components of eigenmodes with complex eigenvalues $\sigma=\sigma_\Re
+ i \sigma_\Im\in \mathbb{C}$. The real parts of the eigenvalues are the angular frequencies and the imaginary parts are the temporal growth rates. 
If $\sigma_{\Im} >0$, small perturbations decay exponentially and the mode is stable, while if $\sigma_{\Im}<0$, small perturbations grow exponentially and the
mode is unstable. If $\sigma_{\Im}$ = 0 the mode is neutrally stable.

We wish to identify the region of $R_1$--$T_0$ space in which unstable eigenmodes exist, and when there are multiple unstable modes, identify the fastest growing mode. 

Since $\Gamma$ is conserved at material points of the free vortex sheet as they move downstream (at speed 1), and the material point at location $x\geq 1$ at time $t$ was at location $x=1$ at time $t-(x-1)$, we can write
\begin{align}
    \Gamma(x,t) &= \Gamma_0e^{i\sigma(t-(x-1))}=\Gamma_0e^{-i\sigma(x-1)}e^{i\sigma t},\quad 1<x<\ell_w+1,\\
    \gamma(x,t) &= \partial_x\Gamma(x,t) = -i\sigma\Gamma_0e^{-i\sigma(x-1)}e^{i\sigma t},\quad\,\,\, 1<x<\ell_w+1,
\end{align}
using~\eqref{eq:Gammaexp}.
Inserting the eigenmodes~\eqref{eq:yexp}--\eqref{eq:Gammaexp} into the governing equations,~\eqref{eq:linearmembrane} and~\eqref{eq:linearKinematic}, yields
\begin{equation}\label{eq:Ymembrane}
-\sigma^2R_1 Y=T_0\partial_{xx} Y-i\sigma \int_{-1}^1 g\,\d x-g, 
\end{equation}
and
\begin{equation}\label{eq:2intB}
i\sigma Y=-\partial_x Y + \frac{1}{2\pi}\Xint- _{-1}^1
\frac{V(x')}{\sqrt{1-x'^2}(x-x')}\,\d x'-\frac{1}{2\pi}i\sigma \Gamma_0 \int_1^{\ell_w+1} \frac{e^{-i\sigma (x'-1)}}{x-x'}\,\d x', \quad -1<x<1,
\end{equation}
respectively. Because $\sigma$ appears in the exponential in the second integral in~\eqref{eq:2intB}, this is a nonlinear eigenvalue problem. 

\subsection{Numerical method for finding the eigenvalues and eigenmodes}\label{sec:numerical}
We solve the nonlinear eigenvalue problem iteratively. At each iteration, we have an approximation $\sigma_0$ to a given eigenvalue $\sigma$. We approximate the
equations as a quadratic eigenvalue problem:
\begin{equation}\label{eq:quadevalue}
[\sigma^2 A_2+\sigma A_1+A_0(\sigma_0)]w=0,
\end{equation}
where the matrices $A_2$, $A_1$, $A_0$
are known from equations~\eqref{eq:Ymembrane}, \eqref{eq:2intB}, and $g(x) = V(x)/\sqrt{1-x^2}$.
The eigenvector~$w$ consists of: (a) values of the eigenmodes, defined as $Y(x)$ on the Chebyshev grid $\{x_j=\cos\theta_j,\,\theta_j=(j-1)\pi /m,\, j=1,\dots,m+1\}$ and (b) the scalar $\Gamma_0$. The term $A_0(\sigma)w$ includes the exponential integral involving~$\sigma$ in~\eqref{eq:2intB} as well
as terms that are constant in $\sigma$. In the exponential integral, $\sigma$ is fixed at $\sigma_0$, the value of $\sigma$ from the previous iteration, resulting in the quadratic eigenvalue problem~\eqref{eq:quadevalue}, which is solved using \texttt{polyeig} in \textsc{Matlab}.
Equation~\eqref{eq:quadevalue} has $2m+4$ eigenvalue solutions. As in~\cite{alben2008ffi}, we define an error function as the difference between $\sigma_0$ and the eigenvalue (out of the $2m+4$ possibilities) closest to it. We also compute the derivatives of the error function (i.e.\ the Jacobian matrix) with respect to $\sigma_\Re$ and $\sigma_\Im$  using finite differences at the initial iterate, and update it at
subsequent iterates using Broyden's approximate formula~\cite{ralston2001first}. The error function and Jacobian define the search direction (via Newton's formula) for the next iterate. With this approach we obtain superlinear convergence to a given eigenvalue. By using a wide range of initial guesses, we obtain convergence to various eigenvalues and corresponding eigenmodes.


The numerical solution procedure followed in the current work differs from~\cite{alben2008ffi}.
There we used a continuation method, which for the current problem would
start from the analytical solution for each eigenvalue 
in the limit $R_1,T_0\gg 1$, and use the solution at a given
$(R_1,T_0)$ as an initial guess for slightly smaller $(R_1,T_0)$ (continuing to smaller and smaller $(R_1,T_0)$). We find that this method fails to find solutions at certain
$(R_1,T_0)$ and therefore at smaller values also, so we use a more robust approach here.
We compute a large set of eigenvalues at each $(R_1,T_0)$ using a large grid of initial eigenvalue guesses in the complex plane covering in most cases $\sigma_\Re\in[-8,8]$ and $\sigma_\Im\in[-3,-0.5]$. For each initial guess we perform the eigenvalue iteration described above until it converges. This reveals the basins of attraction of the eigenvalues under Broyden's iteration, which shows that the imaginary part is not as important as the real part of the eigenvalue guess (especially for large~$R_1$ values). We note that in the system of equations~\eqref{eq:Ymembrane}--\eqref{eq:2intB} the eigenvalue $\sigma$ appears in powers of~$i\sigma$. For each solution $\{i\sigma, w\}$, 
the complex conjugate
$\{-i\bar{\sigma}, \bar{w}\}$ is also a solution, so we need only compute one member of the pair, and obtain the other by conjugation. For the eigenvalue $i\sigma=i\sigma_{\Re}-\sigma_{\Im}$, the conjugate is $-i\bar{\sigma}
=-i\sigma_{\Re}-\sigma_{\Im}$; i.e.\ the sign of $\sigma_{\Re}$ is reversed. Thus we can restrict to 
$\sigma_{\Re} \geq 0$.


We now present typical examples of our eigenmode computations. Throughout the paper we use $m$ = 120 for the Chebyshev grid, unless noted otherwise. Comparisons between $m$ = 80 and 120 (as well as 240) are presented in appendix~\ref{sec:convergencem}. Figure~\ref{fig:pcolorfifiMinus1Minus027} shows results for $(R_1,T_0)=(10^{-1},10^{-0.27})$ with both membrane edges fixed. The coloring in panels~A and~B indicates the converged values of~$\sigma_{\Re}$ (panel A) and~$\sigma_{\Im}$ (panel B) over a
grid of initial eigenvalue guesses in the complex plane spanning 1600 values in the real direction and 4 values in the imaginary direction. In panel~C we plot the 25 distinct eigenvalues found with this set of initial guesses and in~D, 
the corresponding eigenmodes from the most unstable (smallest---or most negative---$\sigma_{\Im}$) on the left to the most stable (largest~$\sigma_{\Im}$) on the right.
The vertical black line in~D separates the (only) unstable mode (on its left) from the stable modes (on its right). The unstable mode loses stability through divergence as is evident from panel C, where the associated eigenvalue has $\sigma_{\Re}\approx 10^{-9}$ 
and $\sigma_{\Im}<0$. We also illustrate with a red circle in panels A and~B an instance of an initial guess that gives rise to this mode. The converged $\sigma$ values are more sensitive to
the real than to the imaginary part of the initial guess, which motivates the wider range of $\sigma_{\Re}$ used here and subsequently.

\begin{figure}[H]
    \centering
    \includegraphics[width=\textwidth]{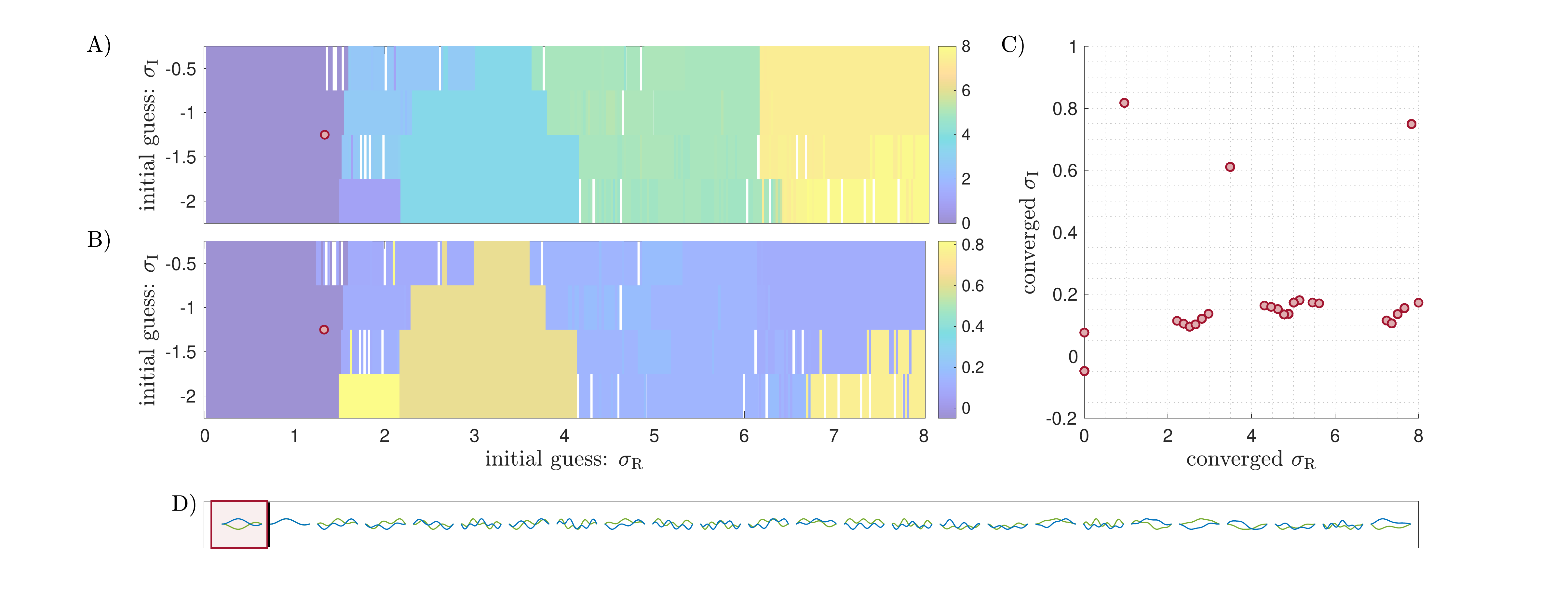}
    \caption{Fixed-fixed eigenvalues and eigenmodes with $R_1=10^{-1}$ and $T_0=10^{-0.27}$. 
    Computed~$\sigma_{\Re}$ (panel~A, values in colorbars at right) and computed~$\sigma_\Im$ (panel~B, values in colorbars at right), both plotted over the initial guess complex plane. 
    C) The distinct eigenvalues  generated by the numerical method replotted as red dots in the $(\sigma_\Re,\sigma_\Im)$ plane. D) The corresponding eigenmodes ($\mathrm{Re}(Y(x))$ in green, $\mathrm{Im}(Y(x))$ in blue) from the only unstable one (with negative~$\sigma_{\Im}$) on the left to the most stable one (largest $\sigma_{\Im}$) on the right. The vertical black line separates the unstable mode (on its left) and stable modes (on its right).}
    \label{fig:pcolorfifiMinus1Minus027}
\end{figure}
\begin{figure}[H]
    \centering
    \includegraphics[width=\textwidth]{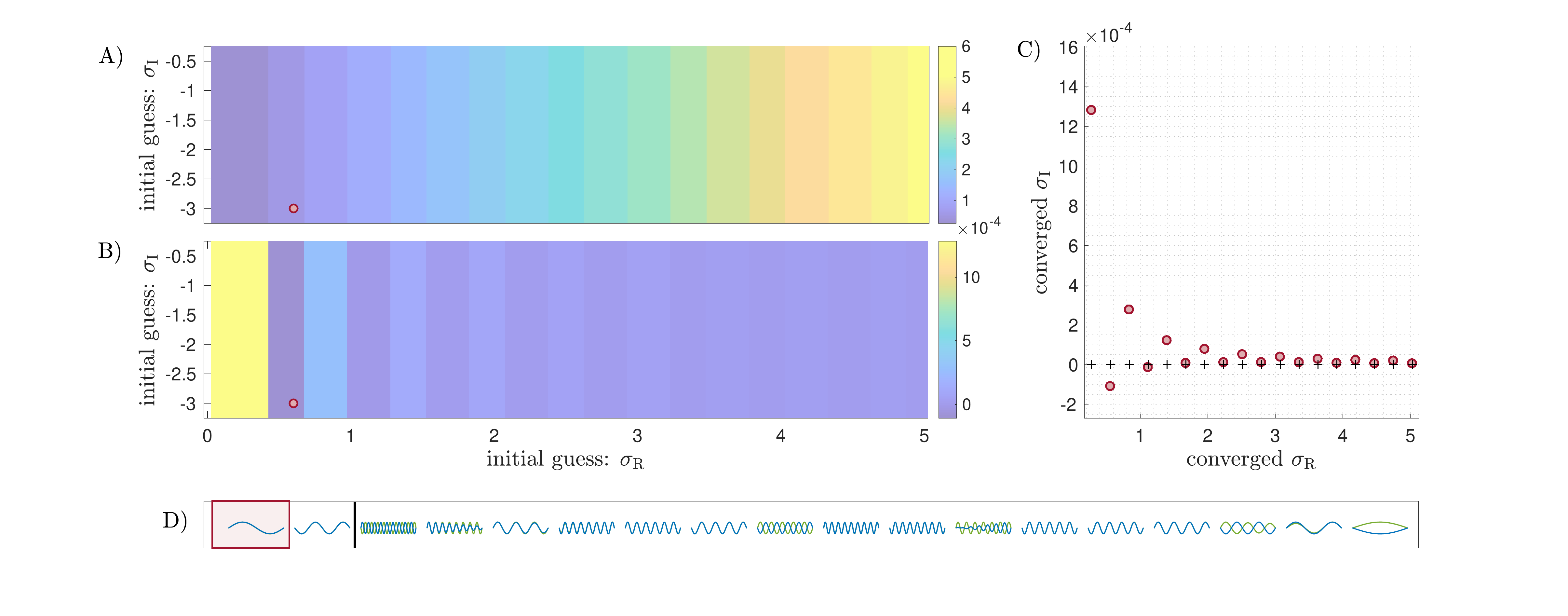}
    \caption{Fixed-fixed eigenvalues and eigenmodes with $R_1=10^{3}$ and $T_0=10^{1.5}$ and other quantities as described in figure~\ref{fig:pcolorfifiMinus1Minus027}.
    }
    \label{fig:pcolorfifi3and15}
\end{figure}

In figure~\ref{fig:pcolorfifi3and15} we show another example of the eigenvalue computation for fixed-fixed membranes, with larger membrane mass and pretension: $(R_1,T_0)=(10^{3},10^{1.5})$. In panels A and~B we use a grid of initial eigenvalue guesses spanning 640 values in the real direction and~6 values in the imaginary direction. For smaller $R_1$ (as in figure~\ref{fig:pcolorfifiMinus1Minus027}) the converged $\sigma$ vary more with the initial choice of $\sigma_{\Im}$ compared to the larger~$R_1$ here, where the converged eigenvalues are independent of the initial $\sigma_{\Im}$, and
depend only on the initial $\sigma_{\Re}$.


Since a generic perturbation is a superposition of all the eigenmodes multiplied by $e^{i\sigma t}=e^{i\sigma_\Re t}e^{ -\sigma_\Im t}$, we classify the stability of generic perturbations in the $(R_1,T_0)$ parameter space based on the value of $\sigma_\Re+ i\sigma_\Im$ for the smallest $\sigma_\Im$ at a given $(R_1,T_0)$:
\begin{enumerate}
  \setlength\itemsep{-.2em}
    \item $\sigma_\Im> 0$: stable,
    \item $\sigma_\Im=0$: stability boundary location, 
    \item $\sigma_\Im<0$ and $\sigma_\Re=0$: divergence (``static'' instability),
    \item $\sigma_\Im<0$ and $\sigma_\Re\neq 0$: flutter and divergence.
\end{enumerate}

\section{Fixed-fixed membranes}\label{sec:linearizedfifi}

\begin{figure}[H]
    \centering

    \includegraphics[width=.49\textwidth]{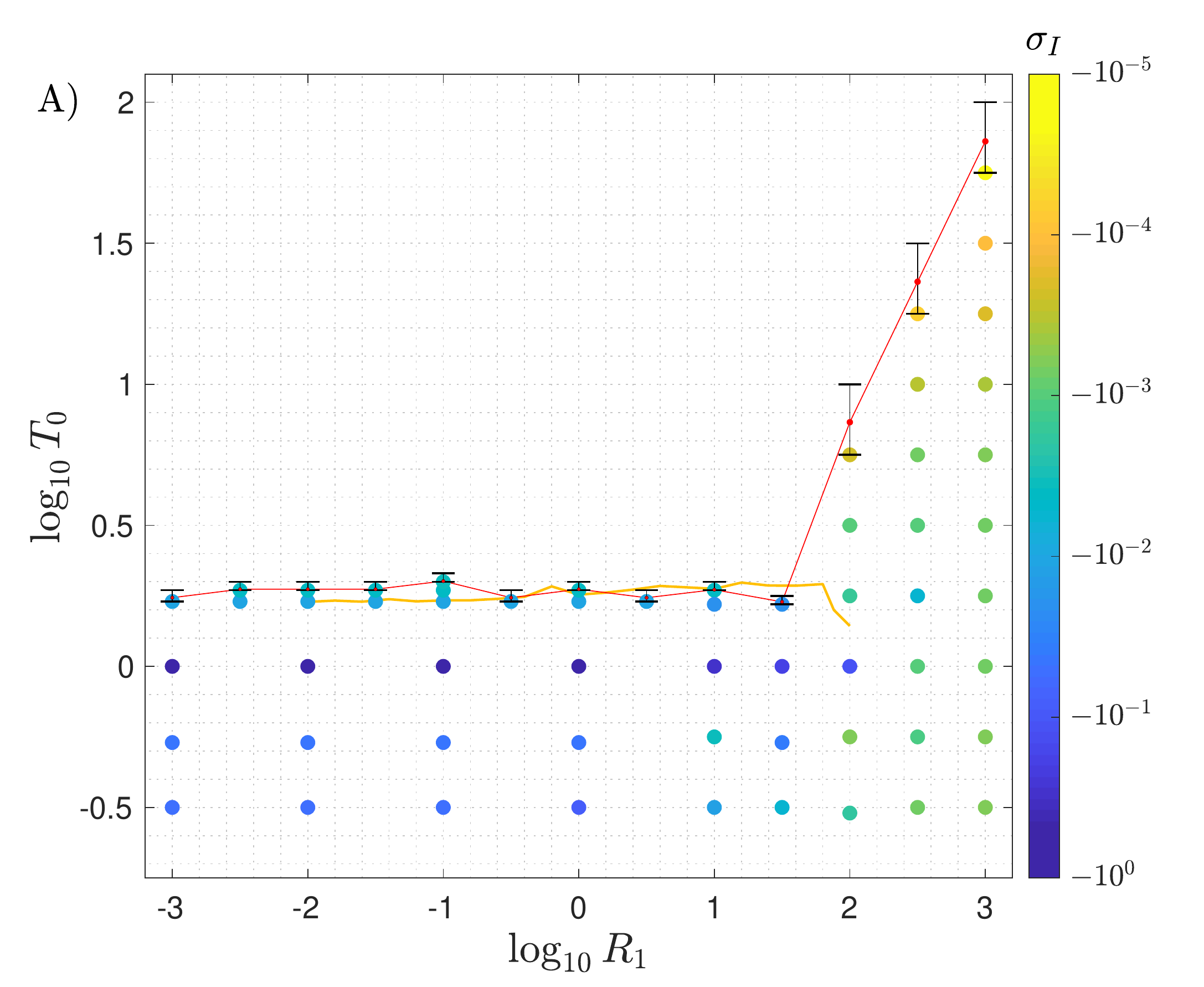}
    \includegraphics[width=.49\textwidth]{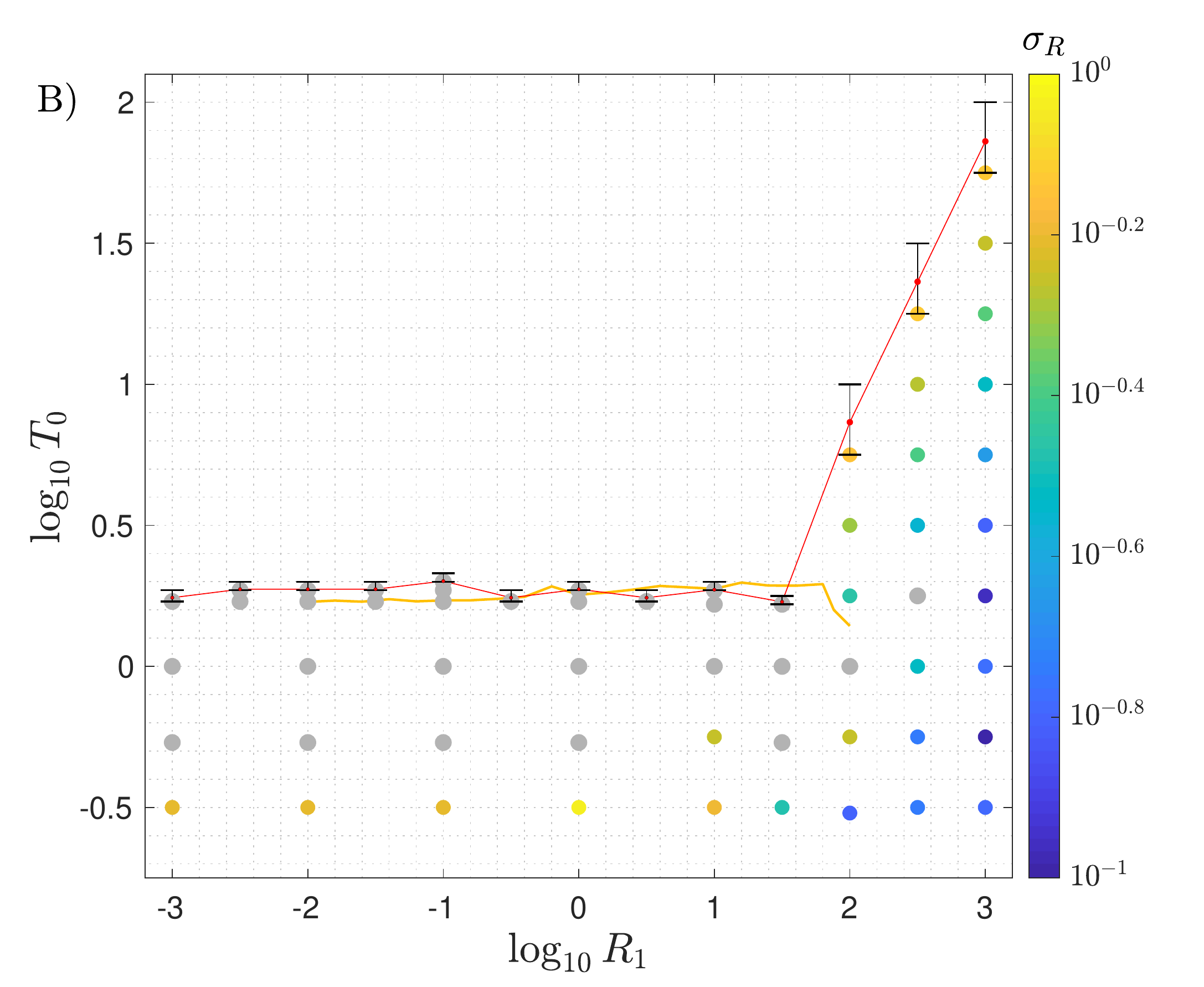}
    \caption{The region in $R_1$--$T_0$ space in which the fixed-fixed membrane is unstable. The red line and red dots indicate the position of the stability boundary computed by linear interpolation between $\sigma_\Im$ of the smallest $T_0$ that gives a stable membrane and the $\sigma_\Im$ of the largest $T_0$ that gives an unstable membrane (shown in the error bars). The colors of the dots below the stability boundary label: A) The imaginary parts of the eigenvalues ($\sigma_\Im$) corresponding to the most unstable modes. They represent the temporal growth rate. B) The real parts of the eigenvalues ($\sigma_\Re$) for the most unstable modes, representing the angular frequencies. The gray dots correspond to modes that lose stability by divergence and have $\sigma_\Re\leq 10^{-9}$. The orange line that spans $\log_{10}R_1\in[-2,2]$ represents the stability boundary computed numerically in~\cite{mavroyiakoumou2020large}. }
    \label{fig:ScatterPlotsFixedFixed}
\end{figure}

We start with membranes that have both edges fixed at zero deflection (satisfying~\eqref{eq:bcfifi}). We plot the stability boundary as the red dots connected by red lines in figure~\ref{fig:ScatterPlotsFixedFixed}A and B. 
Below and to the right is the unstable region. The red dots are computed by linear interpolation of $\sigma_\Im$ between neighboring $T_0$ values (shown by the horizontal black bars) that bracket the boundary: all $\sigma_\Im$
are positive at the larger of the $T_0$ values and above, but one $\sigma_\Im$ is negative at the smaller of the $T_0$ values. 

The stability boundary (red line) agrees well with that of our nonlinear time-stepping simulations (orange line, from~\cite{mavroyiakoumou2020large}) and with the results of~\cite{tiomkin2017stability}.
For each $R_1$, an eigenmode first becomes unstable when the pretension~$T_0$ drops below a critical value $T_{0C}(R_1)$. 
For $R_1< 10^2$, $T_{0C}(R_1) \in[1.7,2]$, 
almost independent of $R_1$.
In our nonlinear, unsteady simulations \cite{mavroyiakoumou2020large}, we found a similar range of $T_{0C}(R_1)$,  $[1.7,1.92]$, for $R_1< 10^{1.5}$. The small discrepancy could arise from the $\delta$-smoothing on the free vortex sheet (that is not used in the eigenvalue solution but is used in the time-stepping simulation). Another possible explanation (as stated in~\cite{alben2008ffi}) is that in the time-stepping simulation~\cite{mavroyiakoumou2020large} the wake grows from zero length but in the current eigenvalue problem the wake has fixed length~$\ell_w$. In our simulations we use $\ell_w=39$, and the modes are essentially unchanged at larger~$\ell_w$. In \cite{mavroyiakoumou2020large} we were not able to compute the upward sloping portion of the stability boundary for $R_1 > 10^{1.5}$ using the unsteady simulations, due to the slow growth/decay of small perturbations with large $R_1$.  

\begin{figure}[H]
    \centering
      \includegraphics[width=\textwidth]{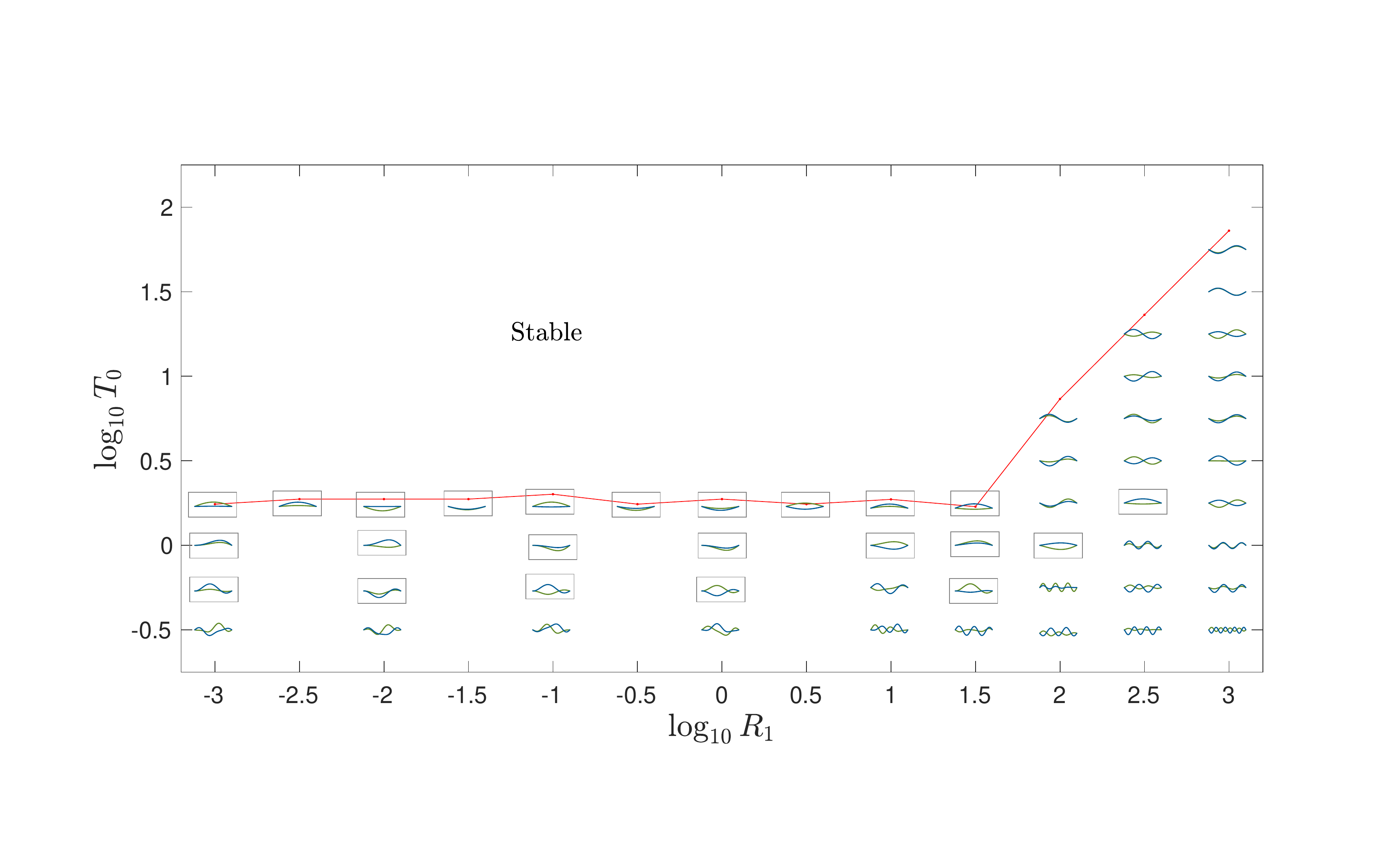}
    \caption{The shapes $Y(x)$ of the most unstable eigenmodes as a function of $R_1$ and $T_0$ in the fixed-fixed case. 
    The real and imaginary parts of $Y(x)$ are shown in green and blue, respectively. Each shape is scaled both vertically and horizontally
    to fit within the plot. 
    The shapes are superposed on the same stability boundary (red line)  as in figure~\ref{fig:ScatterPlotsFixedFixed}.
    Modes exhibiting a divergence instability have a gray rectangle outline.}
    \label{fig:stabFixedFixed}
\end{figure}

In figure~\ref{fig:ScatterPlotsFixedFixed} the colored dots give the imaginary (panel A) and real parts (panel B) of the most unstable eigenvalues (with corresponding eigenmodes shown later, in figure~\ref{fig:stabFixedFixed}). The gray dots in panel B indicate negative $\sigma_{\Im}$ and nearly zero~$\sigma_{\Re}$ ($\sigma_\Re\leq 10^{-9}$) for the most
unstable eigenmode, which corresponds to divergence.
The colored dots in panel B indicate a nonzero real part
(value in colorbar at right) for the most unstable eigenmode, corresponding to flutter and divergence. 
Within the instability region (region below the red line) we find that for a fixed~$T_0$, the fastest growing mode has a growth rate ($\sigma_{\Im}$) that generally decreases in magnitude as $R_1$ increases.
We also find that membranes with $R_1\lesssim 10^{1.5}$, in general, lose stability by divergence 
for $T_0\in[10^{-0.5}, T_{0C}(R_1)]$ but then for a smaller $T_0$ ($\lesssim 10^{-0.5}$), by flutter and divergence.  
Heavier membranes generally lose stability by flutter and divergence for $T_0\in( 10^{0.25},T_{0C}(R_1)]$. 

In the limit $R_1,T_0\gg 1$, the fluid pressure is negligible and the linearized membrane equation~\eqref{eq:linearmembrane} reduces to the homogeneous wave equation
\begin{equation}\label{eq:wave}
    R_1\partial_{tt}y-T_0\partial_{xx}y=0,
\end{equation}
which after substituting~\eqref{eq:yexp} becomes 
\begin{equation}\label{eq:vacuum}
-\sigma^2 R_1 Y-T_0\partial_{xx} Y=0.
\end{equation}
The eigenmodes are linear combinations of $\cos(kx)$ and $\sin(kx)$, with $k=\pm\sigma \sqrt{R_1/T_0}$, satisfying the boundary conditions~\eqref{eq:bcfifi}.
Nontrivial linear combinations exist for $k$ values for which the determinant of the matrix
\begin{equation}\label{eq:matrix}
\begin{pmatrix}
\sin(-k)&\cos(-k)\\
\sin(k)&\cos(k)
\end{pmatrix}
\end{equation}
vanishes, which occurs at $k=n\pi/2$ for $n\in\mathbb{Z}_{>0}$. Each $k$ gives a pair of eigenvalues:
\begin{equation}\label{eq:sigmafifi}
\sigma=\pm k \sqrt{\frac{T_0}{R_1}},
\end{equation}
and eigenmodes of the form
\begin{equation}\label{eq:eigenmodesfifi}
    Y(x)=\sin\left(\frac{n\pi}{2}(x+1)\right),
\end{equation}
for $n\in \mathbb{Z}_{>0}$ and $-1\leq x\leq 1$, where the amplitude is arbitrary.

Similar to~\cite{alben2008ffi}, in the limit of $R_1,T_0\gg 1$ equation~\eqref{eq:sigmafifi} shows that the frequency scales as $\sqrt{T_0/R_1}$. We have observed this in our simulations: $\sigma_\Re$ is approximately constant along lines of constant $T_0/R_1$ in the upper right portion of figure~\ref{fig:ScatterPlotsFixedFixed}B (toward the vacuum limit). At smaller~$R_1$, the angular frequency is less sensitive to the membrane pretension.

The numerical results for the eigenvalues (red dots) shown in figure~\ref{fig:pcolorfifi3and15}C show excellent agreement with the analytical form~\eqref{eq:sigmafifi} of $\sigma$ (black plusses), with very small imaginary parts (vertical axis).  In panel D there are two unstable modes ($n = 2$ and 4 in equation~\eqref{eq:eigenmodesfifi}), which are also the unstable modes that were found in~\cite{tiomkin2017stability} for large values of $R_1$ and~$T_0$.

In figure~\ref{fig:stabFixedFixed} we plot again the instability region in the $R_1$--$T_0$ parameter space for fixed-fixed membranes, but with the eigenmode shapes corresponding to the most unstable eigenvalues  in figure~\ref{fig:ScatterPlotsFixedFixed}. 
The real and imaginary parts of the eigenmode $Y(x)$ are shown in green and blue, respectively. We place gray rectangles around the modes that lose stability by divergence.
For $R_1< 10^{2}$ and $T_0$ just below $T_{0C}$, the unstable eigenmode is a single-hump shape that is nearly fore-aft symmetric. 
As the pretension is decreased further below $T_{0C}$ the divergent eigenmode becomes asymmetric, its maximum deflection point shifting towards the trailing edge. This agrees with~\cite[Fig.~10]{tiomkin2017stability}. 
In the divergence region of figure~\ref{fig:stabFixedFixed} 
when $T_0=10^0$ and $R_1$ decreases from $10^2$ 
to $10^{-1}$, the maximum deflection point also shifts from the midchord towards the trailing edge, in agreement with~\cite[Fig.\ 5]{tiomkin2017stability}. 
For heavier membranes ($R_1\geq 10^2$), the membrane loses stability with an even-numbered mode shape through flutter and divergence. In particular the second mode ($n=2$) is the most unstable mode for $R_1\geq 10^2$ and $T_0\in[10^{0.5},T_{0C}(R_1)]$, as well as $(R_1,T_0)=(10^2,10^{0.25})$ and $(10^3,10^{0.25})$. 
Decreasing the pretension value below $10^{0.25}$, the fourth mode $(n=4)$ becomes the most unstable for $R_1 > 10^2$, followed by the sixth mode at $(R_1,T_0)=(10^2,10^{-0.25})$, $(10^{2.5},10^{-0.5})$.
For heavy membranes with decreasing $T_0$, the most unstable mode apparently moves to progressively higher even-numbered modes.

We now study in more detail how the eigenvalues change in $R_1$--$T_0$ space by examining what happens when~$T_0$ passes through the stability boundary.
We track the stable and unstable modes using a grid of initial eigenvalue guesses in the complex plane covering $\sigma_\Re\in(0,8]$ and $\sigma_\Im\in[-3,3]$, with 160 values in the real direction and~13 values in the imaginary direction.
As can be observed in figure~\ref{fig:stabFixedFixed}, in general, as we move to smaller $T_0$ values higher wavenumber modes become the most unstable ones. We now consider the instability of higher wavenumber modes as we cross the stability boundary, by fixing two values of~$R_1$ and decreasing $T_0$, while tracking the real and imaginary parts of the computed eigenvalues. 

\begin{figure}[H]
    \centering
    \includegraphics[width=.49\textwidth]{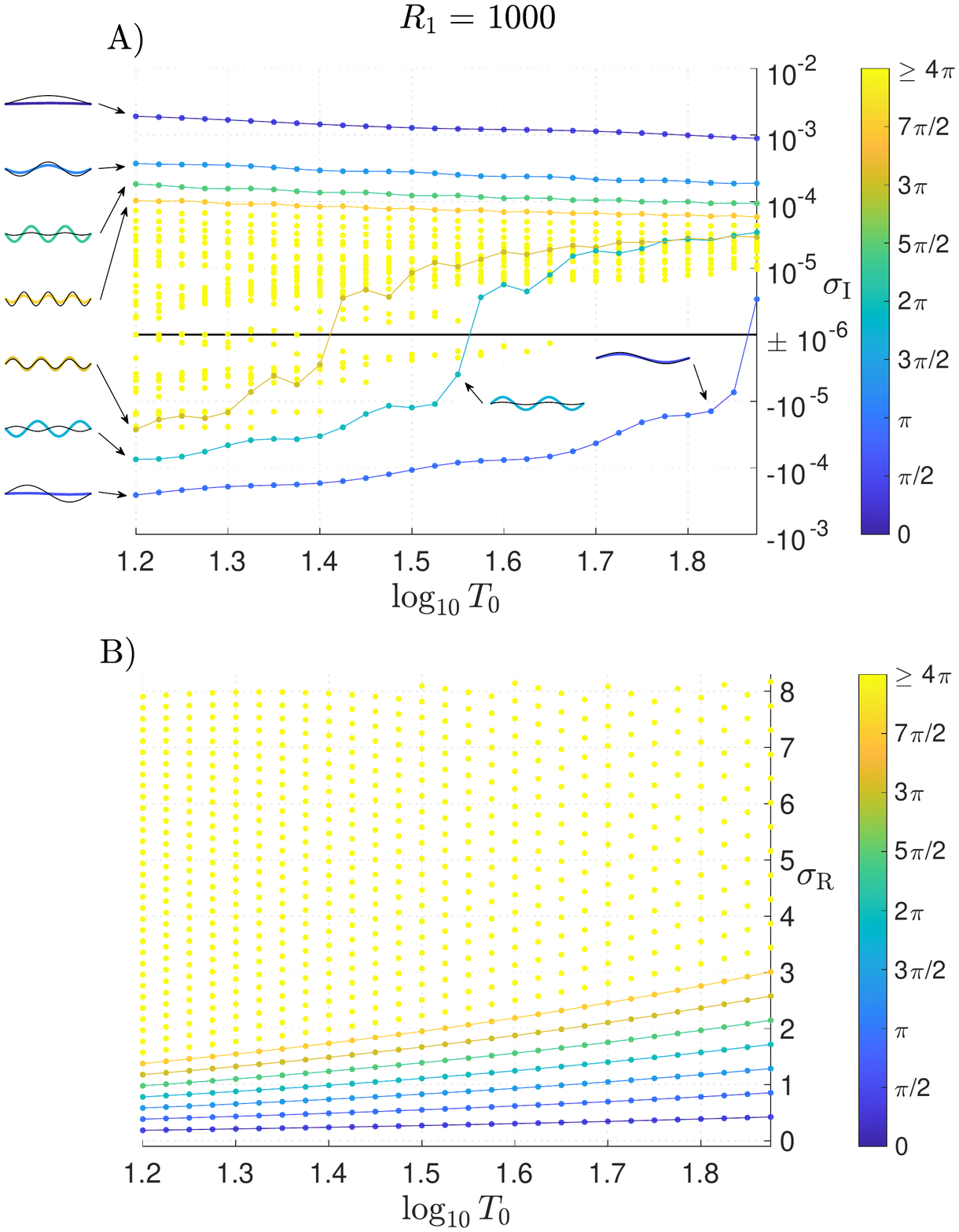}
    \includegraphics[width=.49\textwidth]{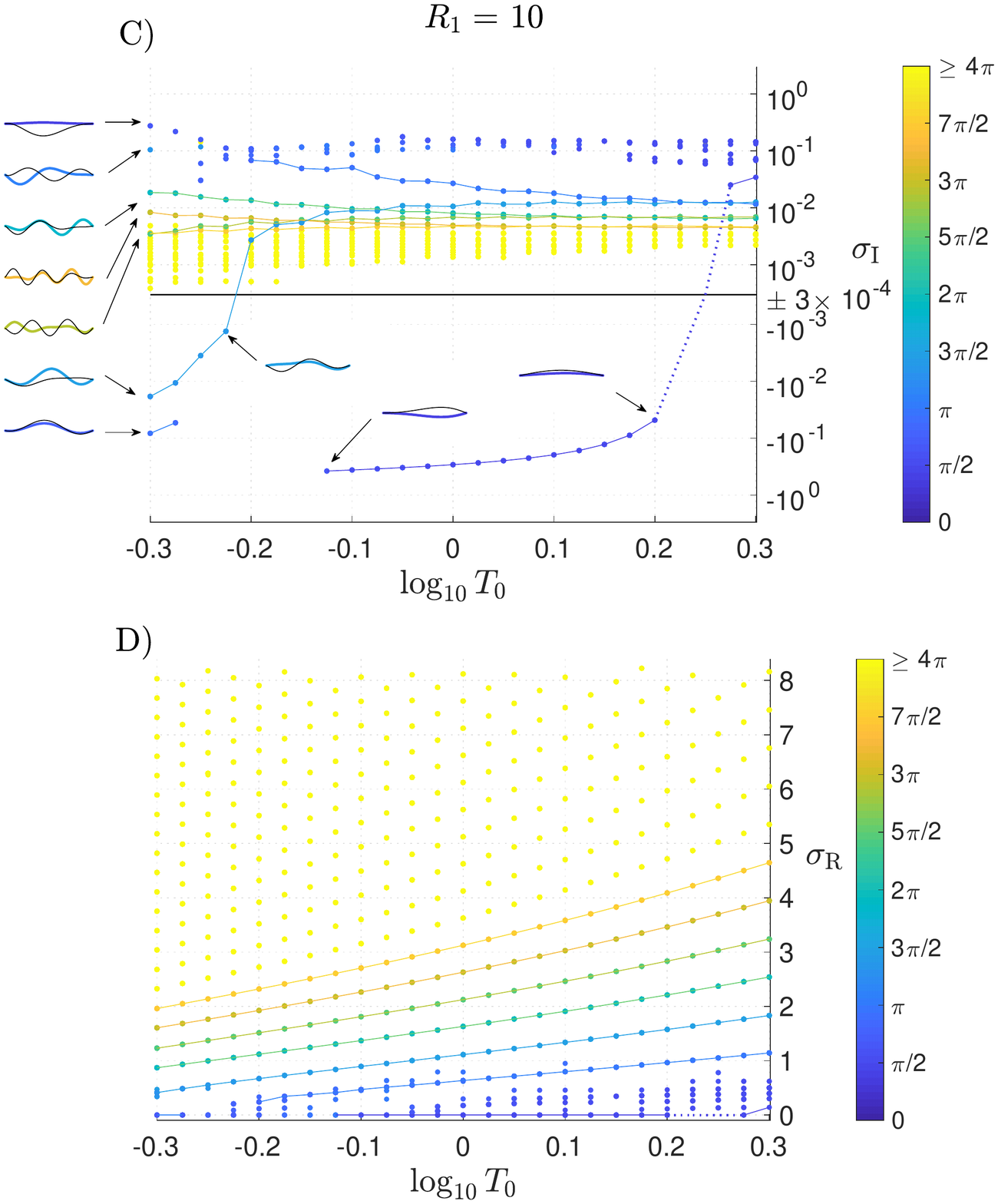}
    \caption{For two values of membrane mass ($R_1$), $10^3$ (left column) and $10^1$ (right column), the imaginary (A, C) and real parts (B, D) of the eigenvalues versus pretension ($T_0$) for fixed-fixed membranes. The coloring represents the RMS of the membrane's slope, $Y'_{\mathrm{RMS}}$~\eqref{eq:colorRMS} for each ($R_1,T_0$) pair. The horizontal black lines in the top panels located at A) $\sigma_\Im=\pm 10^{-6}$ and B) $\sigma_\Im=\pm 3\times 10^{-4}$ distinguish stable modes (above) and unstable modes (below). To the left of and within panels A and~C, we show typical modes for branches with $Y'_\mathrm{RMS}<4\pi$.}
    \label{fig:bif3fifi}
\end{figure}

In figure~\ref{fig:bif3fifi} we show the real (bottom row) and imaginary parts (top row) of the eigenvalues for $R_1=10^3$ (left column) and $10^1$ (right column), while decreasing $T_0$.  The colors show the normalized root mean square (RMS) slope of each membrane eigenmode on the Chebyshev mesh, defined by
\begin{equation}\label{eq:colorRMS}
Y'_{\mathrm{RMS}}:=\sqrt{\int_{-1}^1\left|\frac{\d Y}{\d x}\right|^2\,\d x\; \bigg/ \; \int_{-1}^1\left|Y\right|^2\,\d x},
\end{equation}
which is a measure of the ``waviness" of each mode. Each branch that possesses approximately the same color (lying in a particular, small range of $Y'_\mathrm{RMS}$) indicates a distinct mode. 
At the highest mass ($R_1=1000$), panels~A and B, we connect the eigenvalues by polygonal lines for the modes that are sufficiently distinct from the others---the seven lowest wavenumber modes. The branches in panel A are somewhat jagged when
$|\sigma_\Im|$ drops below $10^{-5}$. The corresponding $\sigma_\Re$ (panel B) vary much more smoothly, probably because their magnitudes are larger relative to numerical errors.
The blue branch with the most negative $\sigma_\Im$ values first becomes unstable ($\sigma_\Im$ changes from positive to negative) at $T_0\approx 10^{1.87}$, which coincides with the loss of stability in figure~\ref{fig:stabFixedFixed}. The mode associated with this blue branch is the second mode ($n=2$ in equation~\eqref{eq:eigenmodesfifi}).
The next branch to become unstable corresponds to the fourth mode ($n=4$) at $T_0\approx 10^{1.56}$ and then the sixth mode ($n=6$) at $T_0\approx 10^{1.41}$. Representative mode shapes at the smallest $T_0=10^{1.2}$ are shown to the left of panel A, for the three unstable branches ($n = 2$, 4, and 6) and four stable branches ($n = 1$, 3, 5, and 7). The $Y'_\mathrm{RMS}$ values that correspond to these seven lowest wavenumber modes are approximately those of the analytical eigenmodes in~\eqref{eq:eigenmodesfifi}, $n\pi/2$ for $n=1,2,\dots,7$. We also illustrate examples at larger~$T_0$ values for the $n=2$ and~4 branches, and find that the mode shapes are almost unchanged. In particular, we note that the seven modes shown to the left of $T_0=10^{1.2}$ all remain approximately the same across the corresponding colored branch for $T_0\in[10^{1.2},10^{1.875}]$. In figure~\ref{fig:bif3fifi} we focus on the lowest wavenumber shapes, as the higher wavenumber shapes (yellow dots) are not numerically resolved.
The odd-numbered modes remain stable for all values of $T_0$ shown. As we decrease the pretension~$T_0$ the number of distinct modes found---with the range of initial guesses that we are using---increases.
This is indicated by the higher density of dots at smaller $T_0$ in figure~\ref{fig:bif3fifi}B. 
Panels C and D show the corresponding data for a smaller membrane mass, $R_1 = 10$. The modes deviate more from the analytical expression of equation~\eqref{eq:eigenmodesfifi}, and change more significantly across $T_0$, compared to panel~A. Representative modes at the smallest $T_0=10^{-0.3}$ are shown at the left side of panel C.
The shape of the curves that connect the real part of the eigenvalues associated with a particular mode shape (lower panels) seems to be similar for the two mass densities. 
However for the smaller mass ($R_1=10$) there is a ``disordered" band of dark blue dots (with $Y'_\mathrm{RMS}<\pi/4$) that are stable ($\sigma_\Im \approx 10^{-1}$ is panel C) and have low frequency ($\sigma_\Re \lesssim 1$ in panel D).


\section{Fixed-free membranes}\label{sec:linearizedfifr}

We now investigate the stability of membranes with the leading edge fixed and the trailing edge free to move vertically, i.e.\ satisfying the boundary conditions~\eqref{eq:bcfifr}. In~\cite{mavroyiakoumou2020large} we found that with one end free, the membrane has a wider range of unsteady dynamics. In particular, in the steady-state large-amplitude regime we showed in~\cite{mavroyiakoumou2020large} that this set of boundary conditions has a mixture of periodic and chaotic dynamics as opposed to the steady single-hump solutions observed in fixed-fixed membranes. In the small-amplitude (growth) regime we will now show that the eigenmodes can also be somewhat more complicated.

\begin{figure}[H]
    \centering
     \includegraphics[width=.49\textwidth]{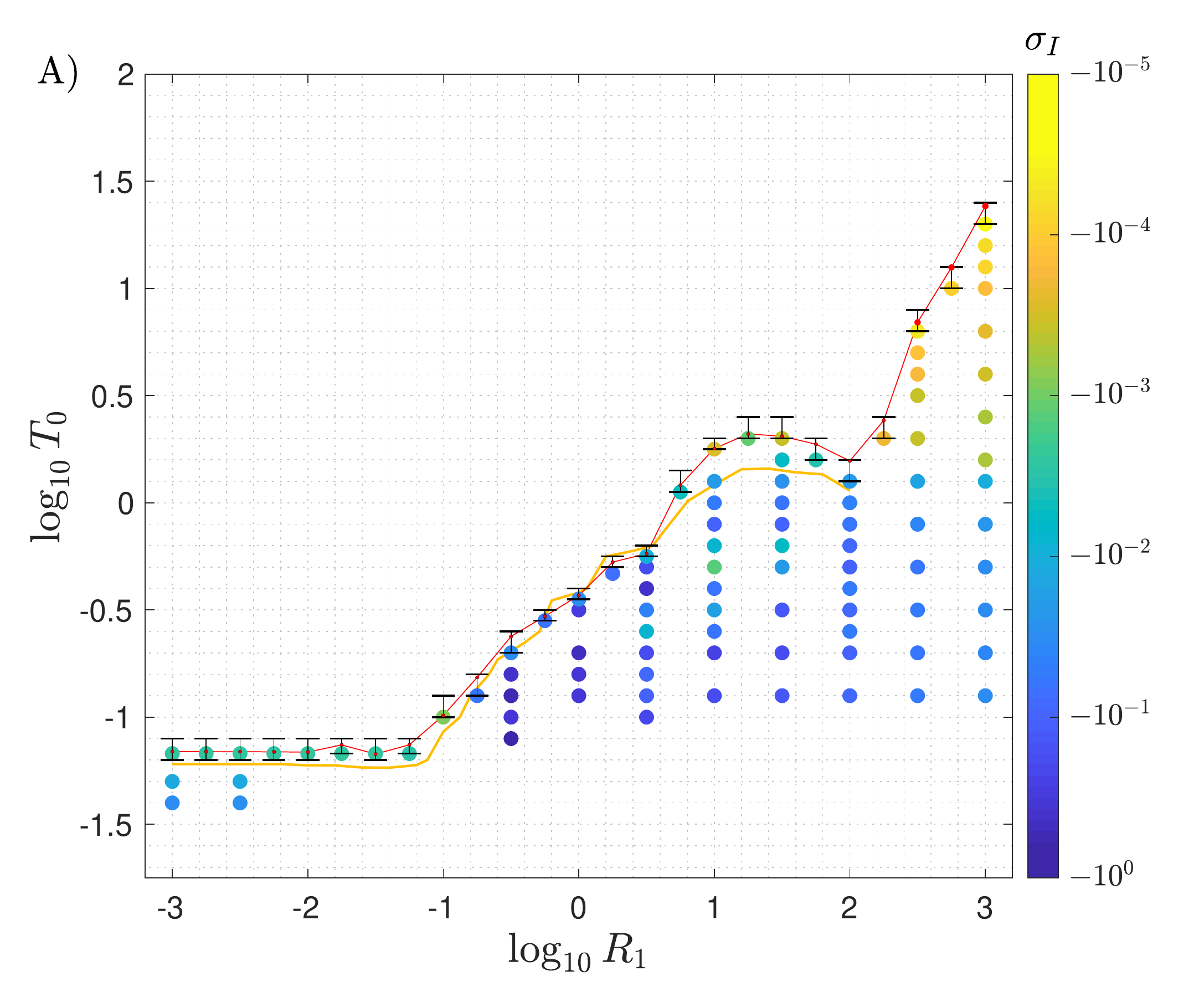}
    \includegraphics[width=.49\textwidth]{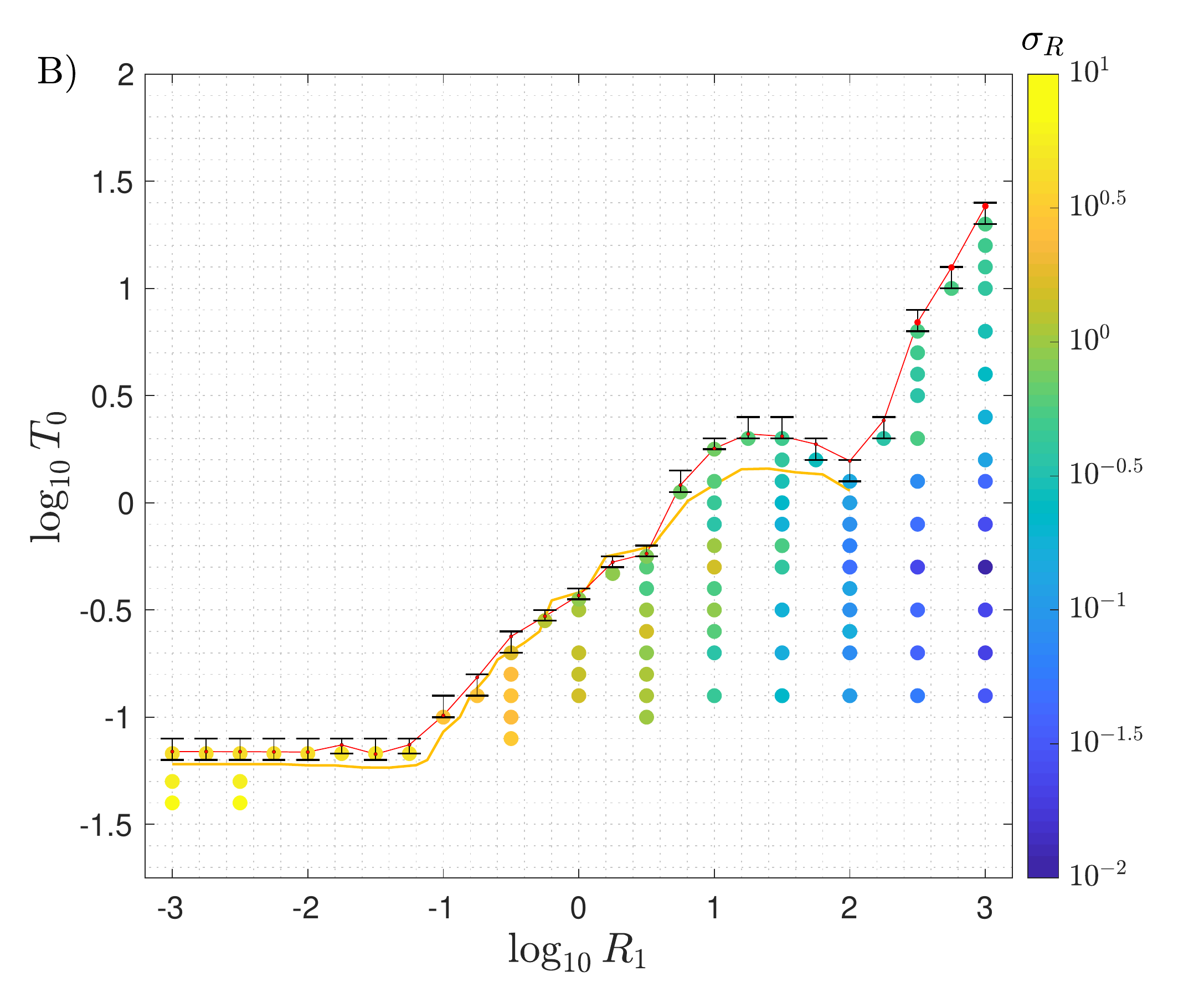}
    \caption{The region in $R_1$--$T_0$ space in which the fixed-free membrane is unstable. The red line and red dots indicate the position of the stability boundary computed using linear interpolation between $\sigma_\Im$ of the smallest $T_0$ that gives only stable eigenmodes and the $\sigma_\Im$ of the largest $T_0$ that gives an unstable eigenmode (shown in the error bars). The color of the dots below the stability boundary labels: A) The imaginary part of the eigenvalue~($\sigma_\Im$) corresponding to the most unstable modes. It represents the temporal growth rate. B) The real part of the eigenvalues~($\sigma_\Re$) for the most unstable mode, representing the angular frequency. The orange line that spans $R_1\in[10^{-3},10^2]$ represents the stability boundary computed numerically in~\cite{mavroyiakoumou2020large}. }
    \label{fig:ScatterPlotsFixedFree}
\end{figure}

In figure~\ref{fig:ScatterPlotsFixedFree} we plot the imaginary (panel A) and real parts (panel B) of the most unstable eigenvalues in the region of instability for the fixed-free membranes in $R_1$--$T_0$ space. The red line marks the boundary where the eigenvalues change from all $\sigma_\Im>0$ (stable membranes) to at least one $\sigma_\Im<0$ (unstable membranes), analogous to figure~\ref{fig:stabFixedFixed}. As in the fixed-fixed case, the stability boundary moves to larger pretension~($T_0$) values with increasing membrane mass ($R_1$), but starting at much smaller $R_1$ now ($\geq 10^{-1}$). As $R_1$ decreases below $10^{-1}$, the critical pretension reaches a lower plateau. 

The stability boundary of the current study is compared against the boundary from the nonlinear study  in~\cite{mavroyiakoumou2020large} (orange line). Their shapes are very similar and there is good agreement especially for $R_1\in[10^{-0.75},10^{0.5}]$. As in figure~\ref{fig:stabFixedFixed}, the discrepancy may be due to $\delta$-smoothing used on the free vortex sheet of~\cite{mavroyiakoumou2020large}, the choice of the vortex wake $\ell_w$, or the number of Chebyshev nodes ($m+1$) on the membrane. In the unsteady simulations (orange line) we used $m=40$ because the simulations require more computing time, but in the current work (red line) we used $m=120$. The eigenvalue solver shows that the boundary slopes upward over $R_1\in[10^{2},10^{3}]$, where it was difficult to obtain accurate results with
the unsteady simulations.  





The trends of the most unstable eigenvalues (colored dots) are similar to the fixed-fixed case (figure \ref{fig:ScatterPlotsFixedFixed}) in some ways: the growth rates $\sigma_\Im$ generally become larger in magnitude at smaller~$T_0$ and smaller~$R_1$ (panel~A), and the growth rates vary nonmonotonically with $T_0$ at intermediate $R_1$ ($[10^{0.5},10^{1.5}]$ for fixed-free, and smaller~$R_1$ for fixed-fixed). A difference is the slight decrease in growth rates as $R_1$ decreases below $10^{-1}$ for the fixed-free case, which does not occur in the fixed-fixed case. For $R_1\in[10^{2},10^{3}]$, the fixed-free growth rates are qualitatively similar to those in the fixed-fixed case above $T_0 = 10^{0.1}$. Below this value, however, the fixed-free growth rates jump by more than an order of magnitude. In both cases, the real parts of the eigenvalues (the angular frequencies $\sigma_\Re$, panel~B) generally decrease with decreasing $T_0$ and with increasing $R_1$, particularly at the largest $R_1$. Below $R_1 = 10^{1.5}$, the frequencies are very different: divergence ($\sigma_\Re \approx 0$) does not occur in the fixed-free case, but is common in the fixed-fixed case. 




To consider the eigenmodes in the fixed-free case we again start with $R_1$ and $T_0 \gg 1$, so the fluid forcing is negligible and the 
eigenmodes are again solutions of~\eqref{eq:wave}, 
i.e.\ nontrivial linear combinations of $\cos(kx)$ and $\sin(kx)$, with $k=\pm\sigma\sqrt{R_1/T_0}$, but satisfying the boundary conditions~\eqref{eq:bcfifr} now. The $k$ are now those for which the determinant of
\begin{equation}
\begin{pmatrix}
\sin(-k)&\cos(-k)\\
\cos(k)&-\sin(k)
\end{pmatrix}
\end{equation}
is zero, which leads to $k=(n-1/2)\pi/2$ for $n\in\mathbb{Z}_{> 0}$, corresponding eigenvalues $\sigma=\pm k \sqrt{T_0/R_1}$, and eigenmodes now of the form
\begin{equation}\label{eq:eigenmodesfifr}
    Y(x)=\sin\left(\left(n-\frac{1}{2}\right)\frac{\pi}{2}(x+1)\right),
\end{equation}
for $n\in \mathbb{Z}_{>0}$ and  $-1\leq x\leq 1$. Each mode has one quarter wavelength less than that of the corresponding fixed-fixed mode, so that the trailing edge has zero slope.

\begin{figure}[H]
    \centering
    \includegraphics[width=\textwidth]{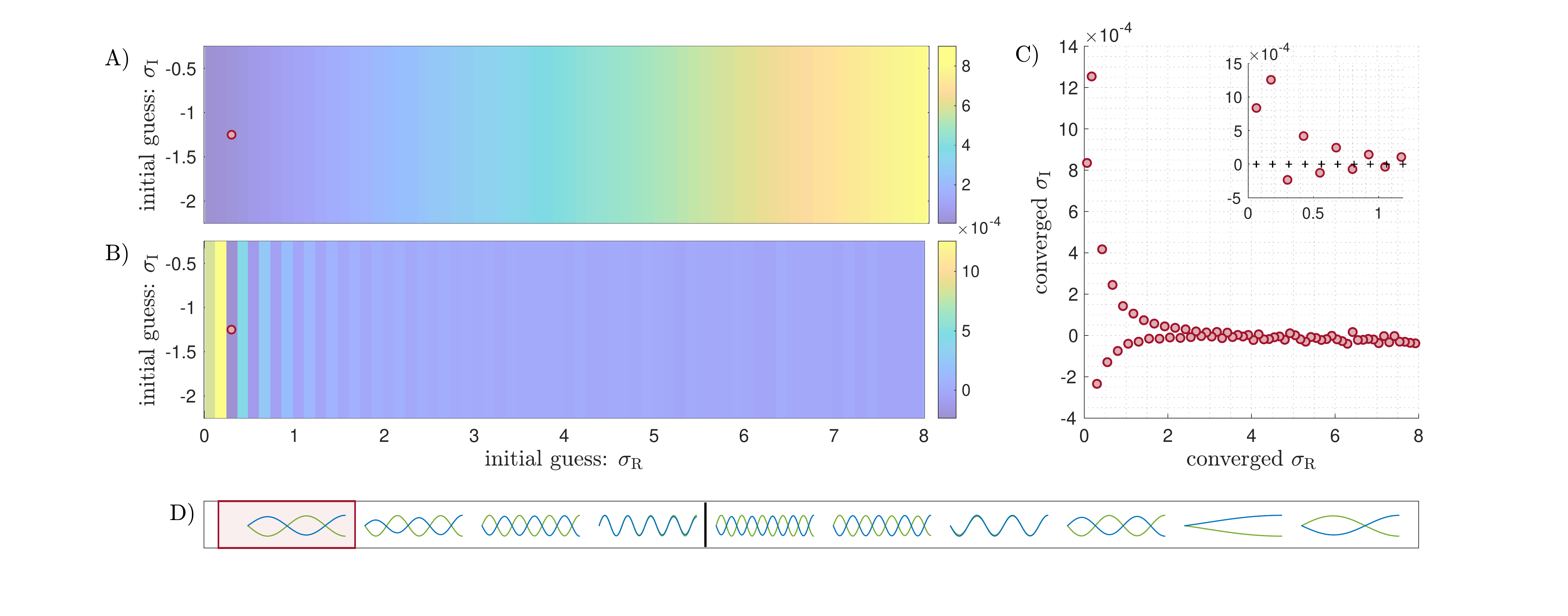}
    \caption{Fixed-free eigenvalues and eigenmodes with $R_1=10^{3}$ and $T_0=10^{0.8}$. 
    Computed~$\sigma_{\Re}$ (panel A, values in colorbars at right) and computed~$\sigma_\Im$ (panel~B, values in colorbars at right), both in the initial guess complex plane. C) The computed eigenvalues replotted as red dots in the $(\sigma_\Re,\sigma_\Im)$ plane. 
    The inset in panel C shows the ten computed eigenvalues (red $\circ$) that correspond to the eigenmodes shown in panel D. The analytical form of the eigenvalues is $\sigma=((n-1/2)\pi/2)\sqrt{T_0/R_1}=((n-1/2)\pi/2)\sqrt{10^{0.8}/10^{3}}$ for $n=1,\dots,10$ (black plusses). 
    }
    \label{fig:pcolorfifr3and08}
\end{figure}

Figure~\ref{fig:pcolorfifr3and08} shows an example of how the computed eigenvalues (real parts in panel A and imaginary parts in panel~B) vary over a grid of initial guesses in the complex plane, for a fixed-free membrane with $(R_1,T_0)=(10^3,10^{0.8})$, in the large $R_1$ region near the stability boundary. The quantities plotted are analogous to those in figure~\ref{fig:pcolorfifi3and15}.
The grid of initial eigenvalue guesses
in the complex plane covers $\sigma_\Re\in(0,8]$ and $\sigma_\Im\in[-2,-0.5]$, spanning 640 values in the real direction and~4 values in the imaginary direction. As in figure~\ref{fig:pcolorfifi3and15}, we see that for large $R_1$ ($10^3$) and moderately large $T_0$ ($10^{0.8}$) the eigenvalues obtained by the numerical method depend mainly on the real part of the initial eigenvalue guess. However, here we see that there is more variation in the computed eigenvalues with respect to the choice of initial~$\sigma_\Re$ compared to figure~\ref{fig:pcolorfifi3and15}, where the vertical bands of constant real (panel~A) and imaginary parts (panel~B) of~$\sigma$ are wider. This may be due to the smaller value of~$T_0$ considered in figure~\ref{fig:pcolorfifr3and08} ($10^{0.8}$ as opposed to $10^{1.5}$ in figure~\ref{fig:pcolorfifi3and15}). As we decrease the membrane pretension~($T_0$) the number of distinct modes found (with our range of initial guesses) typically increases (e.g.\ figure~\ref{fig:bif3fifi}A and~B).
The numerically computed eigenvalues from figure~\ref{fig:pcolorfifr3and08}A and B are replotted as red dots in the $(\sigma_\Re, \sigma_\Im)$ plane in panel C, and those at the smallest $\sigma_\Re$, shown in the inset, agree closely with the analytical form~\eqref{eq:sigmafifi} with $k=(n-1/2)\pi/2$ for $n\in\mathbb{Z}_{> 0}$ (black plusses in inset; note there is close agreement in the imaginary part due to the small axis scale).  Many eigenmodes are found with wavelengths decreasing down to the mesh scale, but in figure~\ref{fig:pcolorfifr3and08}D we show the ten modes with largest wavelengths (i.e.\ $n=1,\dots, 10$ in~\eqref{eq:eigenmodesfifr}), those that are best resolved numerically. Starting from the left, the most unstable modes have $n$ = 3, 5, 7, and 9, while $n$ = 10, 8, 6, 4, 1 and 2 are stable. Except for $n = 1$, the modes with even and odd $n$ have the opposite stability behavior. Here we omit the computed modes with highest wavenumbers because they (and the corresponding eigenvalues) are not numerically converged. For the large-$R_1$, large-$T_0$ limit solved analytically in equation~\eqref{eq:eigenmodesfifr}, we have a quadratic eigenvalue problem. When discretized by the numerical method in \S~\ref{sec:numerical}, we have $2m+2$ eigenmodes $Y(x)$ varying from low wavenumber modes to very high wavenumber modes that oscillate on the mesh scale (due to the discretized second $x$-derivative). For more general $R_1$ and $T_0$, we have a nonlinear eigenvalue problem, but still have eigenmodes that oscillate on the mesh scale, and are thus not resolved (i.e.\ not close to a continuum solution). Therefore, we focus on the lower wavenumber eigenmodes---those with $Y'_{\mathrm{RMS}}$ (defined in \eqref{eq:colorRMS}) below a threshold near $4\pi$, or about four wavelengths for a sinusoidal $Y(x)$---which we can resolve well with $m = 120$ grid points. 

\begin{figure}[H]
    \centering
\includegraphics[width=\textwidth]{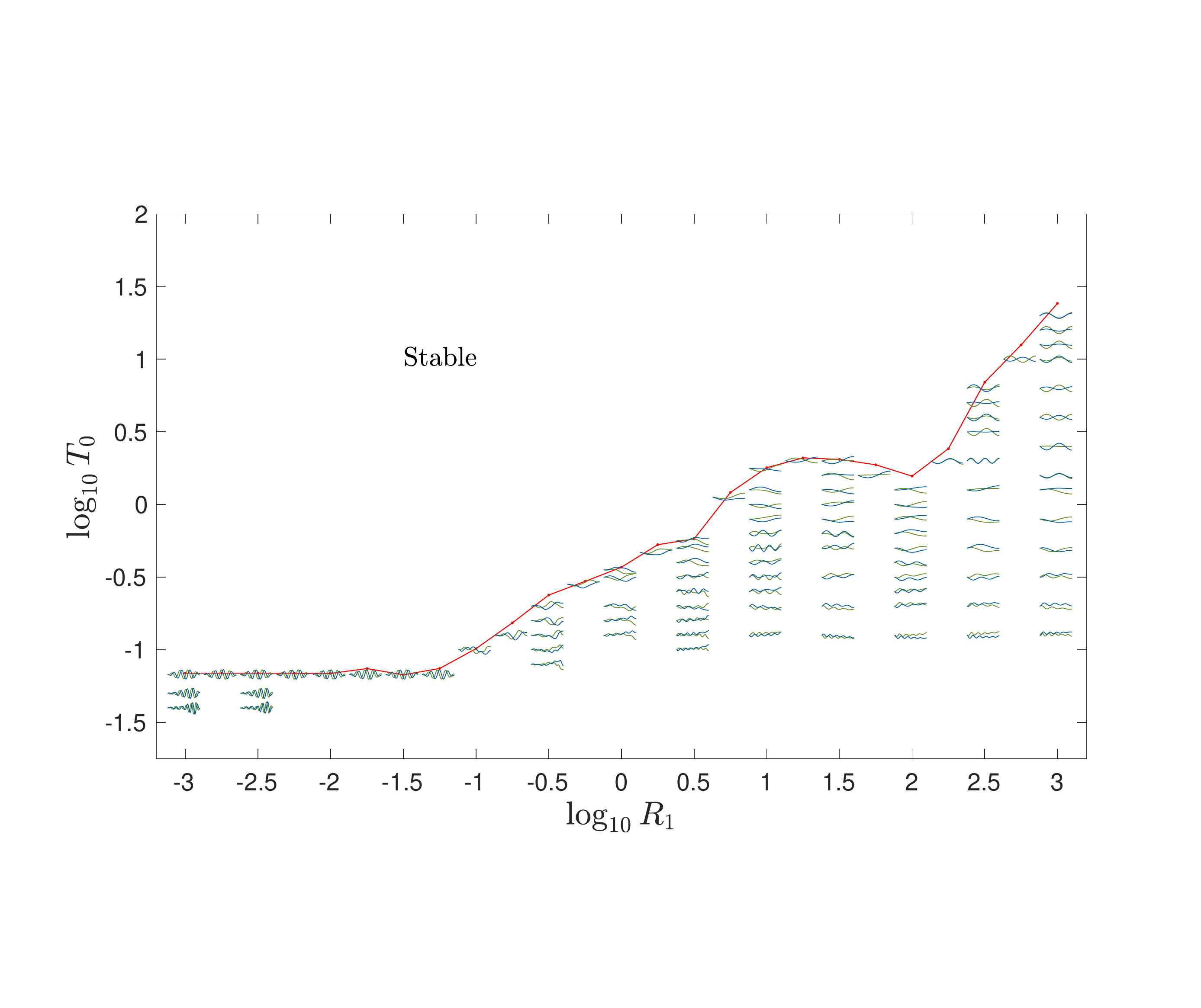}
    \caption{
    The shapes $Y(x)$ of the most unstable eigenmode as a function of $R_1$ and $T_0$ in the fixed-free case. The real part of $Y(x)$ is shown in green and the imaginary part of $Y(x)$ is shown in blue. Each shape is scaled, both vertically and horizontally, to fit within the plot.
    The shapes are superposed on the same stability boundary (red line) as in figure~\ref{fig:ScatterPlotsFixedFree}.
    }
    \label{fig:stabFixedFree}
\end{figure}

In figure~\ref{fig:stabFixedFree} we examine the variations in the most unstable eigenmodes in the same $(R_1,T_0)$ space as figure~\ref{fig:ScatterPlotsFixedFree}, corresponding to the eigenvalues shown there. 
There is a singular behavior observed at $(R_1,T_0)=(10^0,10^{-0.6})$, where our initial guesses all produced $\sigma_{\Im}>0$ and so the mode there is omitted. The shapes do not change noticeably for the more irregular motions at $R_1\in[10^{-3},10^{-1.25}]$ (the eigenvalues in figure~\ref{fig:ScatterPlotsFixedFree} were also nearly constant in this region). At these smallest $R_1$ values the deflection at the free end is nearly zero. As we decrease $T_0$ for $R_1\leq 10^{-2.5}$, the ripples move toward the trailing edge of the membrane while maintaining nearly zero deflection at that end.  Close to the stability boundary, all the shapes for $R_1\in[10^{0.75},10^{2}]$ are also nearly alike.  
At moderate values of $R_1$ ($[10^{-1},10^2]$) the maximum deflection occurs in most cases at the trailing edge of the membrane. At these and larger values of~$R_1$, the mean slope of the membrane is nonzero.
In a similar region of~$R_1$ (i.e.\ $[10^{-1},10^{1.75}]$) fixed-fixed membranes become unstable with a single hump, losing stability via divergence. Fixed-free membranes, however, become unstable by flutter and divergence.
When $T_0$ is below $10^{-0.2}$ the most unstable mode changes to a ``wavier" profile---the mode wavenumber increases with decreasing~$T_0$. Similar to the fixed-fixed case where even-numbered modes become unstable for large~$R_1$, we see in figure~\ref{fig:stabFixedFree} that heavy fixed-free membranes ($R_1>10^2$) with $T_0\in[10^{0.2},T_{0C}(R_1)]$,
become unstable with an odd-numbered mode---the third mode (the first mode is stable). At $T_0<10^{0.2}$ we are no longer in the vacuum limit ($R_1\gg 1$ but $T_0$ is not). Thus, the mode shape is not a simple sinusoidal function of the form~\eqref{eq:eigenmodesfifr}, but the waviness still increases with decreasing $T_0$ for heavy membranes.





\begin{figure}[H]
    \centering
    \includegraphics[width=.49\textwidth]{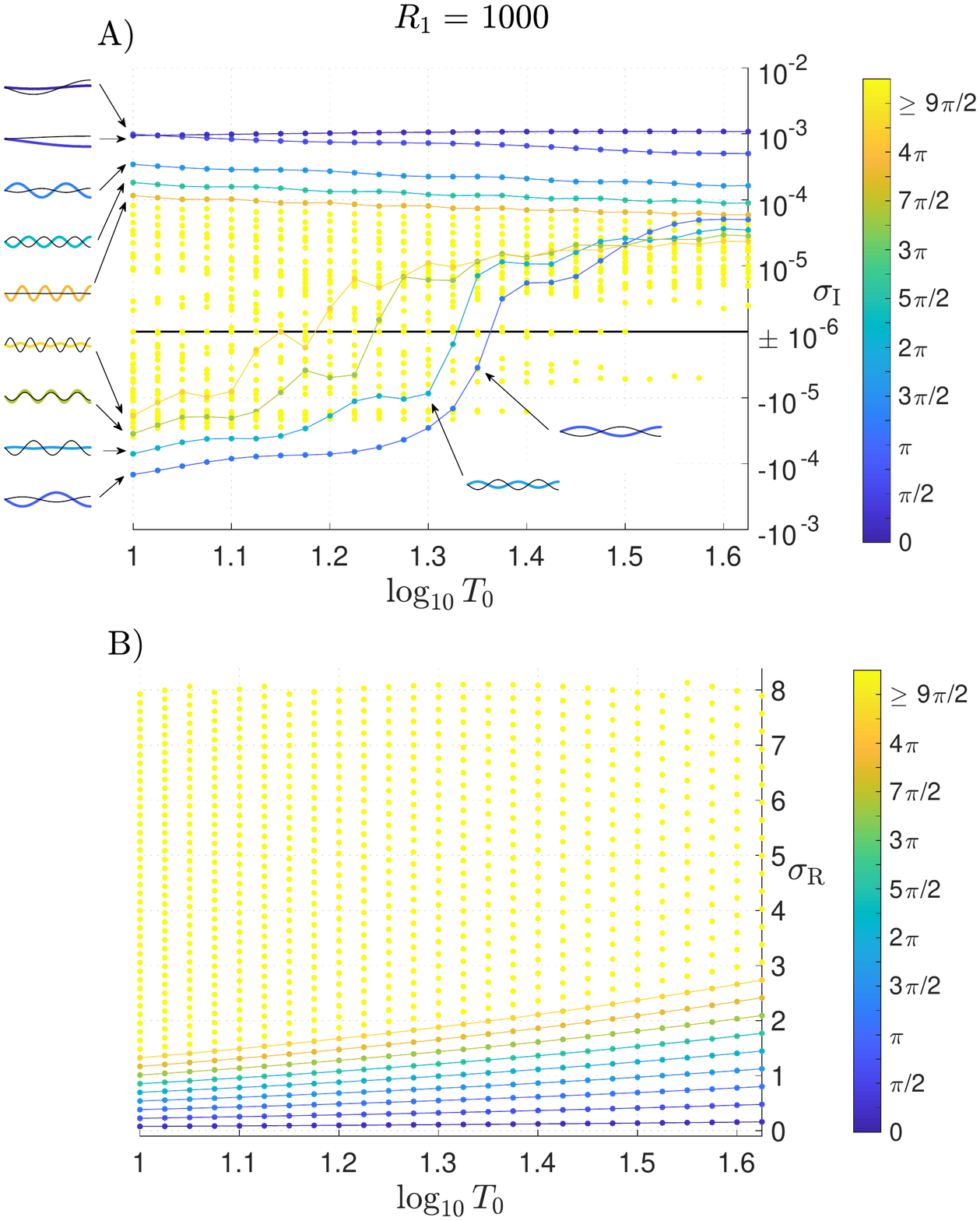}
    \includegraphics[width=.49\textwidth]{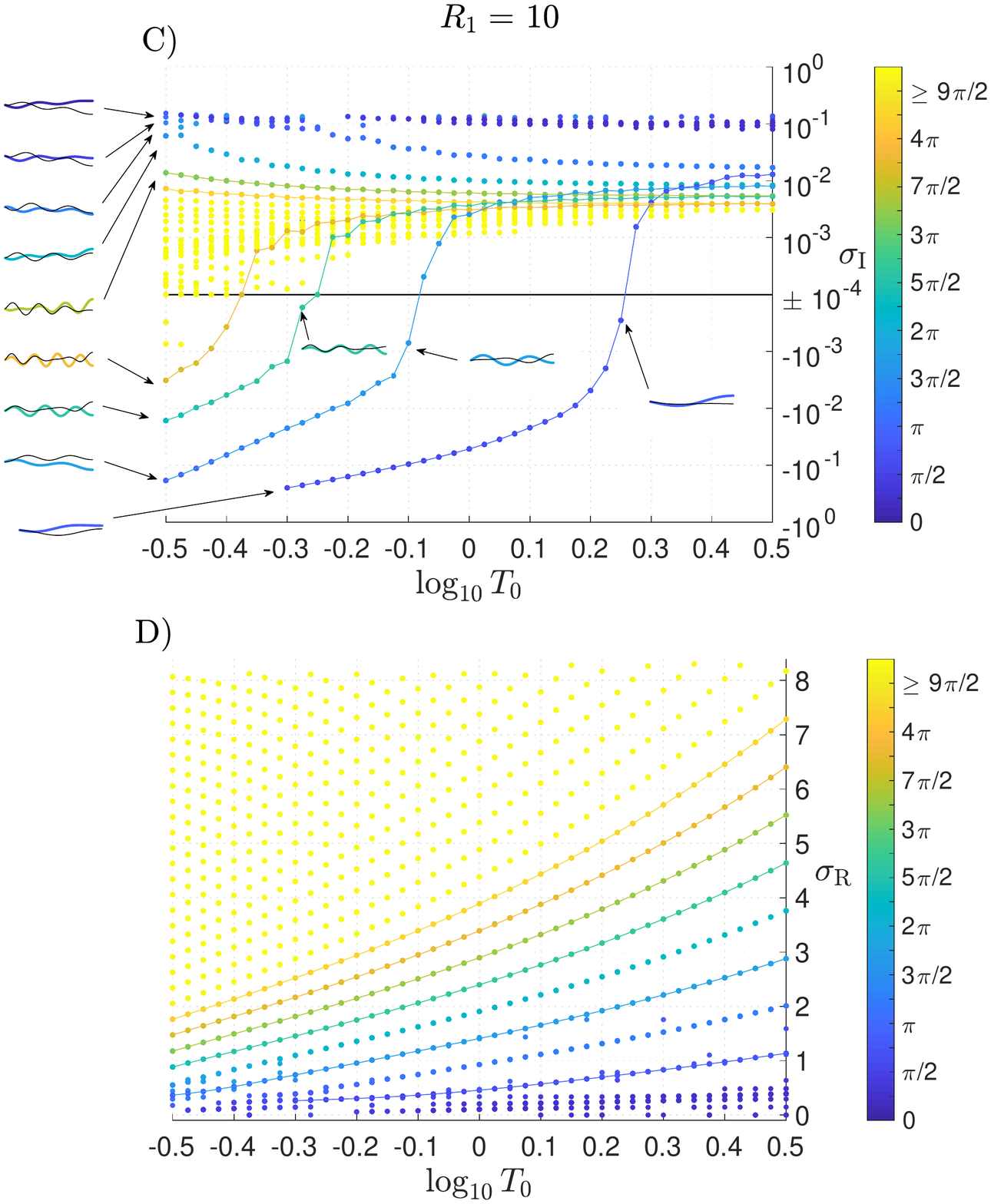}
    \caption{For two values of membrane mass ($R_1$), $10^3$ (left column) and $10^1$ (right column), the imaginary (A, C) and real parts (B, D) of the eigenvalues versus pretension ($T_0$) for fixed-free membranes. The coloring represents the RMS of the membrane's slope, $Y'_\mathrm{RMS}$, for each ($R_1,T_0$) pair given by~\eqref{eq:colorRMS}. The horizontal black line in the top panels located at A) $\sigma_\Im=\pm 10^{-6}$, B) $\sigma_\Im=\pm 10^{-4}$ distinguishes stable modes (above) and unstable modes (below). We also show typical modes that correspond to each branch with $Y'_\mathrm{RMS}< 9\pi/2$.}
    \label{fig:bifurcationfifr}
\end{figure}

We now consider the changes in the eigenvalues and associated eigenmode shapes as we pass through the stability boundary for a fixed mass density, the fixed-free analog of figure~\ref{fig:bif3fifi}. In figure~\ref{fig:bifurcationfifr} the colors label~$Y'_\mathrm{RMS}$, given by~\eqref{eq:colorRMS}. 
For the larger $R_1$, 1000 (panel A) the unstable modes are odd-numbered and they become unstable in order of increasing $n$. The third mode becomes unstable first, at $T_0\approx 10^{1.36}$---consistent with figure~\ref{fig:stabFixedFree}. Then the fifth mode ($n=5$) becomes unstable at $T_0\approx 10^{1.33}$, the seventh mode at $T_0\approx 10^{1.24}$, and the ninth mode at $T_0\approx 10^{1.18}$. The even-numbered modes and the first mode remain stable for all values of~$T_0$. Contrary to the fixed-fixed case with $R_1=1000$ (figure~\ref{fig:bif3fifi}A) where the four branches with the largest positive~$\sigma_\Im$ correspond to modes $n=1,3,5,7$, the branches with largest $\sigma_\Im$ in the fixed-free case correspond to modes $n=1,2,4,6,8$.   This additional branch with opposite parity ($n=1$) in panel A has a slightly smaller $\sigma_\Im$ than the second mode at the smallest $T_0=10^1$ shown. Above a certain $T_0$ value the $n=1$ branch acquires the largest $\sigma_\Im>0$. 

We show the membrane shapes of the nine lowest wavenumber modes to the left of panel~A at the lowest $T_0=10^1$, but also examples of membrane shapes at a couple of larger~$T_0$ values for the first two unstable branches and observe that the mode shapes are almost unchanged.  The $Y'_\mathrm{RMS}$ values that correspond to these nine lowest wavenumber modes are approximately those of the analytical eigenmodes in~\eqref{eq:eigenmodesfifr},  $(n-1/2)\pi/2$ for $n=1,2,\dots,9$.
Even though higher wavenumber shapes (yellow dots) appear to become unstable at a larger $T_0$ value, such cases are not numerically resolved and are thus not used in determining $T_{0C}$ here.
At $R_1=1000$, the branches with the largest $\sigma_\Im>0$ are all continuous but at $R_1=10$, the same branches (blue dots at the top of panel C and bottom of panel D) are more scattered. There, the numerical method gives individual eigenvalues that do not seem to follow a particular branch, as was also found for fixed-fixed membranes at $R_1 = 10$. This could potentially be due to our choice for the range and density of the mesh of initial eigenvalue guesses. 
The loss of stability in figure~\ref{fig:bifurcationfifr}C occurs at $T_0\approx 10^{0.26}$. The imaginary parts of the eigenvalues (panel C) are about two orders of magnitude higher
than in panel~A. 
At $R_1=10$ we see four branches that fall below $\sigma_{\Im}=0$, each having approximately its own distinct value of $Y'_\mathrm{RMS}$. If we consider smaller values of $T_0$ we would expect to observe more branches becoming unstable. As opposed to panel~A, we see in panel~C that the yellow dots (higher wavenumber modes) are mostly stable. Similar to the fixed-fixed case in figure \ref{fig:bif3fifi}
we see that the curves connecting the $\sigma_\Re$ associated with a particular mode shape appear to be steeper in panel~D than in panel~B. 



\section{Free-free membranes}\label{sec:linearizedfrfr}

\begin{figure}[H]
    \centering
    \includegraphics[width=.49\textwidth]{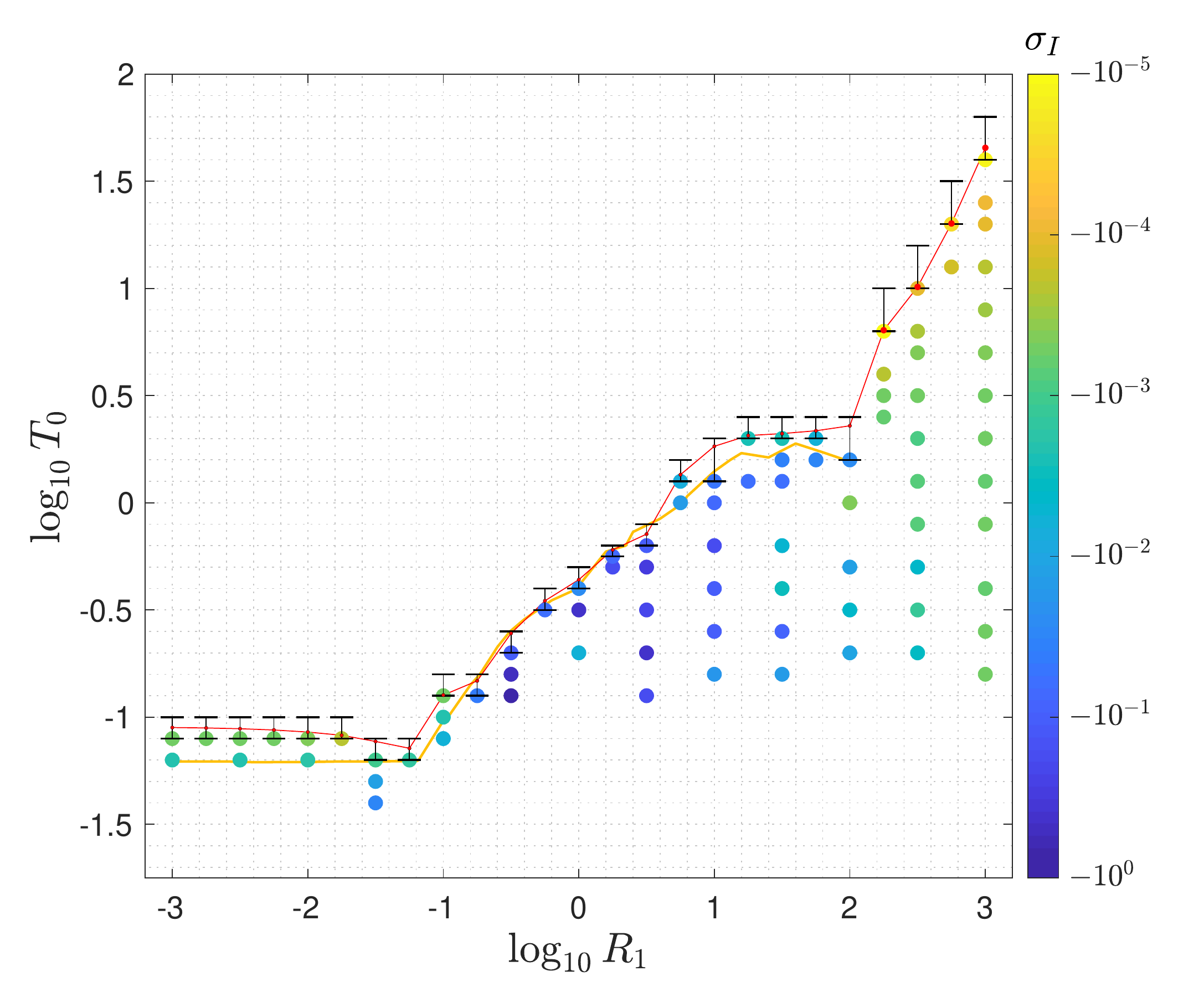}
    \includegraphics[width=.49\textwidth]{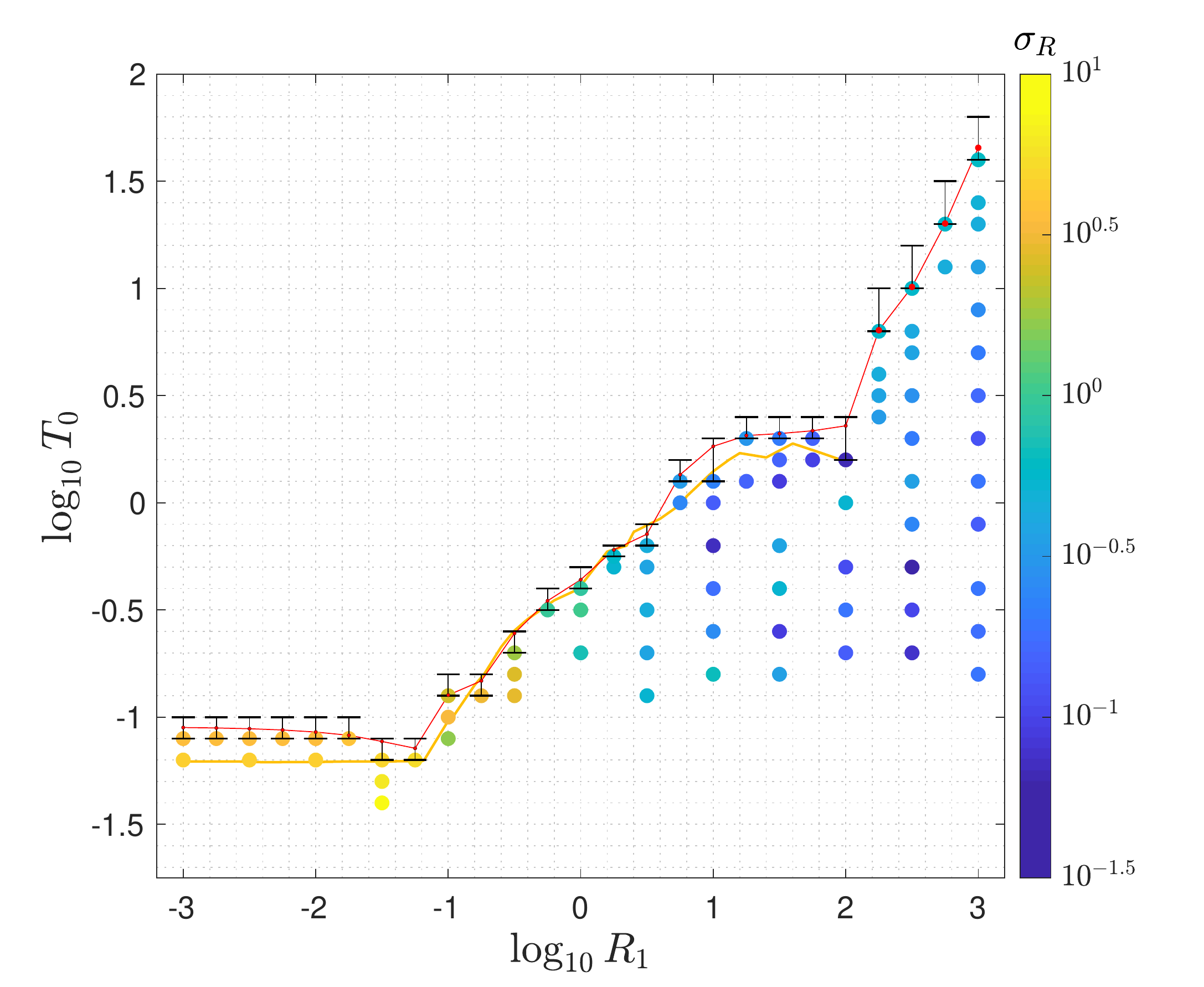}
    \caption{The region in $R_1$--$T_0$ space in which the free-free membrane is unstable. The red line and red dots indicate the position of the stability boundary computed using linear interpolation between $\sigma_\Im$ of the smallest $T_0$ that gives a stable membrane and the $\sigma_\Im$ of the largest $T_0$ that gives an unstable membrane (shown in the error bars). The color of the dots below the stability boundary labels: A) The imaginary part of the eigenvalue~($\sigma_\Im$) corresponding to the most unstable modes. It represents the temporal growth rate. B) The real part of the eigenvalues~($\sigma_\Re$) for the most unstable mode, representing the angular frequency. The orange line that spans $R_1\in[10^{-3},10^2]$ represents the stability boundary computed numerically in~\cite{mavroyiakoumou2020large}.}
    \label{fig:ScatterPlotsFreeFree}
\end{figure}

We have found that allowing the trailing edge to deflect freely in the vertical direction dramatically changes the instability region and the membrane dynamics. As a natural next step, we now study the effect of making both ends free, satisfying the boundary conditions~\eqref{eq:bcfrfr}. The stability boundary (red line) and most unstable eigenvalues are shown in figure~\ref{fig:ScatterPlotsFreeFree}. The stability boundary is similar to the fixed-free case (figure~\ref{fig:ScatterPlotsFixedFree}): the critical pretension increases with mass when $R_1> 10^{2}$, it decreases as we decrease $R_1$, and it plateaus when $R_1\ll 1$.
In figure~\ref{fig:ScatterPlotsFreeFree} we show that there is close agreement for $R_1\in[10^{-0.75},10^{0.5}]$ between the stability boundary computed here and in~\cite{mavroyiakoumou2020large} using unsteady simulations (orange line). For smaller~$R_1$ $([10^{-3},10^{-1}])$ and larger~$R_1$ $([10^{0.75},10^2])$, the red line has slightly higher $T_0$. As noted in \S~\ref{sec:linearizedfifr} the difference in $m$ (40 in~\cite{mavroyiakoumou2020large} versus 120 here) may be the main cause. As for the fixed-free case, we will show that the most unstable eigenmodes have higher wavenumbers at the smallest $R_1$,
so numerical resolution is an issue there:
in~\cite{mavroyiakoumou2020large} we found that the small- and large-amplitude motions were not converged with $m$ = 40 for $R_1 < 10^{-1}$.

We can again use the imaginary (panel A) and real parts (panel B) of the eigenvalues to characterize the instability in ($R_1, T_0$) space. 
Within the region of instability (below the red line) a comparison with fixed-fixed (figure~\ref{fig:ScatterPlotsFixedFixed}) and fixed-free membranes (figure~\ref{fig:ScatterPlotsFixedFree}) reveals that the colored dots (most unstable eigenvalues) have the same general behavior: the temporal growth rates (panel A) increase in magnitude with decreasing $R_1$ and $T_0$, but vary nonmonotonically with $T_0$ at moderate values of $R_1$ ($[10^0,10^2]$). 
The growth rates of free-free heavy membranes ($R_1\in[10^{2},10^{3}]$) are qualitatively similar to those in the fixed-free case in the same region. The angular frequencies ($\sigma_\Re$, panel B) are also larger for smaller~$R_1$, but vary nonmonotonically with~$T_0$. Similar to the fixed-free case, we observe that membranes exhibit the flutter and divergence instability but do not lose stability solely by divergence (i.e.\ with $\sigma_\Re \approx 0$) for any $(R_1,T_0)$ pair. In the region $R_1\leq 10^{-1.25}$ the eigenvalues just below the stability boundary are nearly constant; observed also in the fixed-free case (figure~\ref{fig:ScatterPlotsFixedFree}).

For $R_1,T_0\gg 1$ the eigenvalues are the same as for the fixed-fixed case~\eqref{eq:sigmafifi}, with the addition of zero. The free-free eigenmodes are given by
\begin{equation}\label{eq:eigenmodesfrfr}
    Y(x)=\cos\left(\frac{(n-1)\pi}{2}(x+1)\right),
\end{equation}
for $n\in\mathbb{Z}_{> 0}$ and $-1\leq x\leq 1$, where the amplitude is arbitrary.

\begin{figure}[H]
    \centering
    \includegraphics[width=\textwidth]{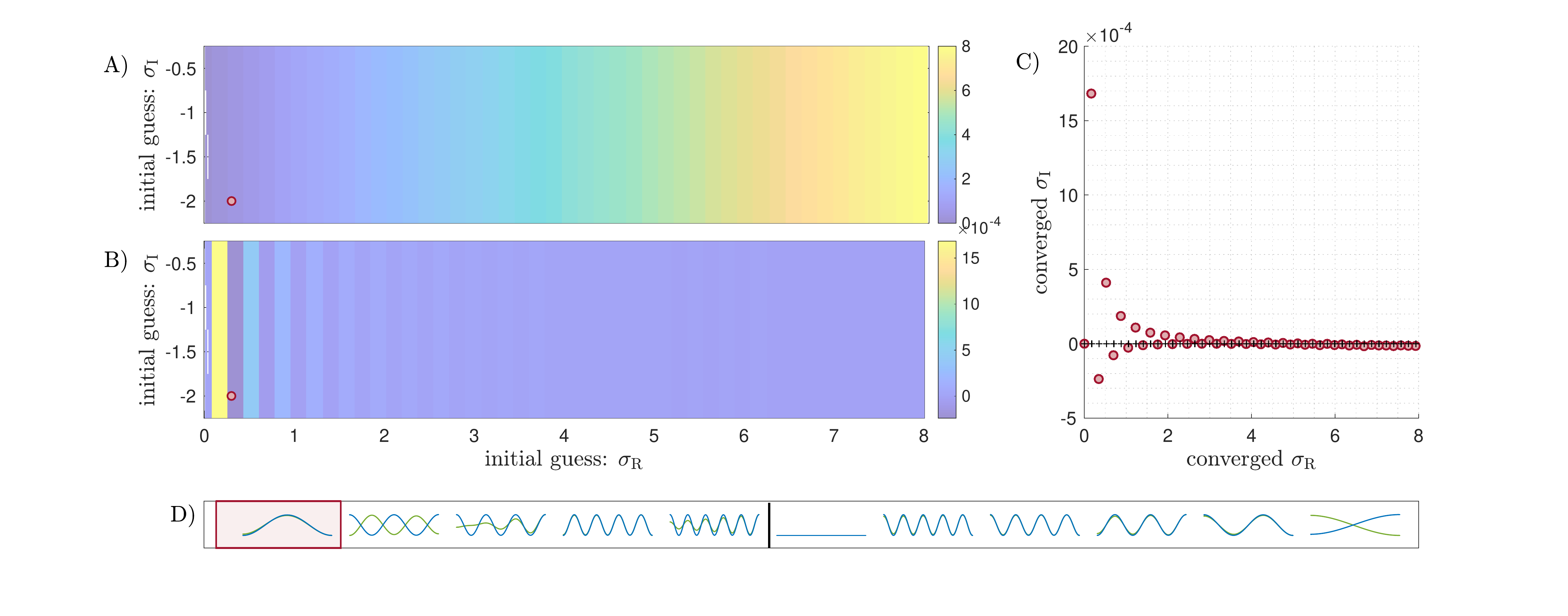}
    \caption{Free-free eigenvalues and eigenmodes with $R_1=10^{3}$ and $T_0=10^{1.1}$.
    Computed~$\sigma_{\Re}$ (panel~A, values in colorbars at right) and computed~$\sigma_\Im$ (panel~B, values in colorbars at right), both plotted in the initial guess complex plane.  C) The distinct eigenvalues generated by the numerical method plotted as red dots in the $(\sigma_\Re,\sigma_\Im)$ plane. The analytical form of the eigenvalues is $\sigma=((n-1)\pi/2)\sqrt{T_0/R_1}$ for $n=1,\dots,46$ (black plusses). 
    D) The eleven lowest wavenumber eigenmodes ($\mathrm{Re}(Y(x))$ in green, $\mathrm{Im}(Y(x))$ in blue), from the most unstable (most negative~$\sigma_{\Im}$) on the left to the most stable (largest positive $\sigma_{\Im}$) on the right. The vertical black line separates unstable modes (on its left) and stable modes (on its right).}
    \label{fig:pcolorfrfr3and11}
\end{figure}

Figure \ref{fig:pcolorfrfr3and11} shows an example of how the computed eigenvalues (real parts in panel~A and imaginary parts in panel~B) vary over a grid of initial guesses in the complex plane for a free-free membrane with $R_1=10^{3}$ and $T_0=10^{1.1}$, with the same mesh as in the fixed-free case of figure~\ref{fig:pcolorfifr3and08}.
We take $R_1$ and $T_0\gg 1$ (vacuum limit) to compare with the analytical values~\eqref{eq:sigmafifi} with $k=(n-1)\pi/2$ for $n\in\mathbb{Z}_{>0}$ (panel C). 
In panel~D we show the eleven lowest wavenumber modes. Starting from the left, the most unstable modes are $n=3$, 5, 7, 9, and 11 whereas $n=1$, 10, 8, 6, 4, and 2 are stable. 
The sixth shape from the left that is displayed is flat ($n=1$), with corresponding  $\sigma_\Re$ and  $\sigma_\Im\approx 10^{-8}$.

\begin{figure}[H]
    \centering
 \includegraphics[width=\textwidth]{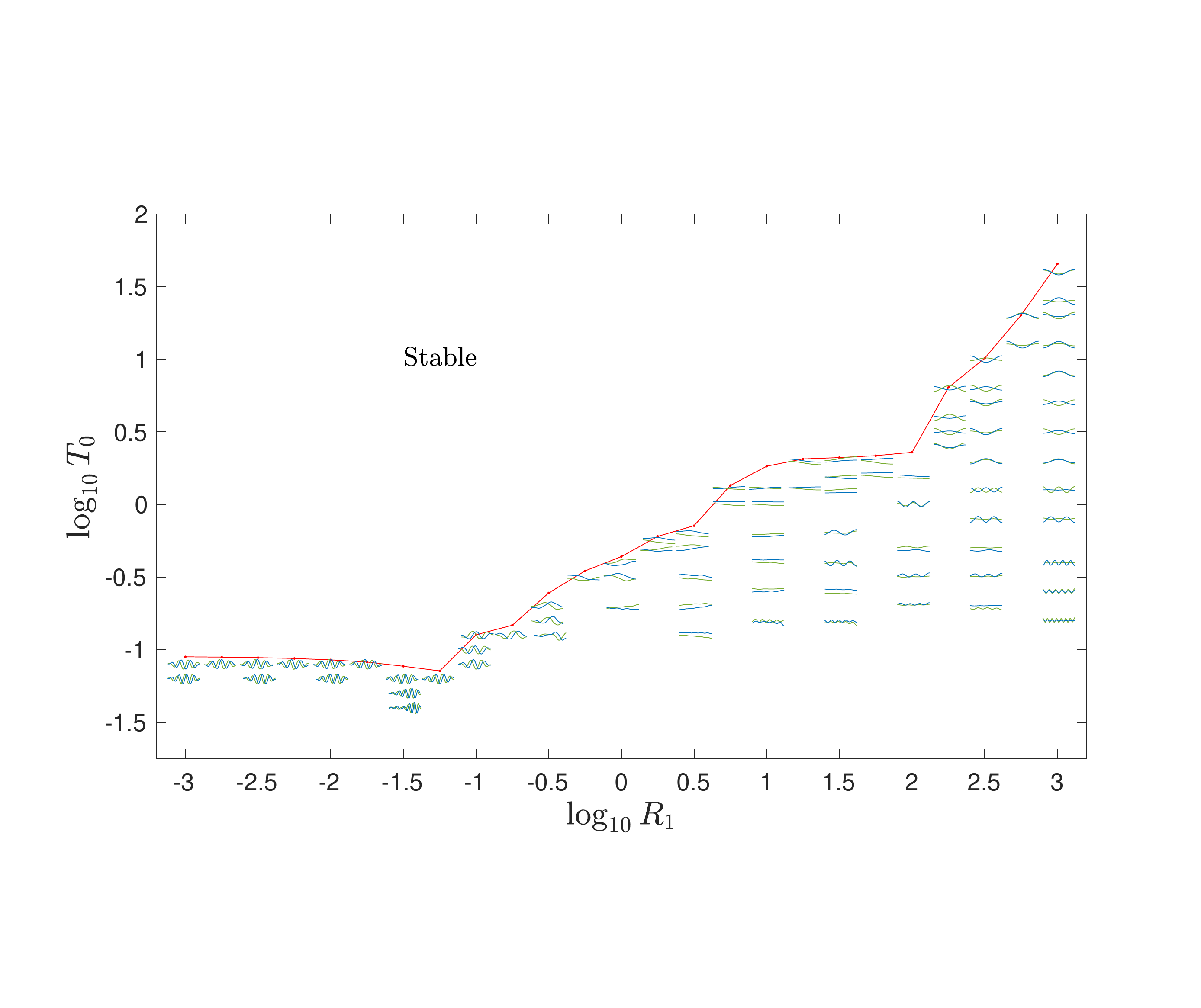}
    \caption{The shapes $Y(x)$ of the most unstable eigenmode as a function of $R_1$ and $T_0$ in the free-free case. The real part of $Y(x)$ is shown in green and the imaginary part of $Y(x)$ is shown in blue. Each shape is scaled, both vertically and horizontally, to fit within the plot. The shapes are superposed on the same stability boundary (red line)  as in figure~\ref{fig:ScatterPlotsFreeFree}.
    }
    \label{fig:stabFreeFree}
\end{figure}

In figure \ref{fig:stabFreeFree} we show the most unstable eigenmodes across $(R_1,T_0)$ space. The mode shapes of light membranes $(R_1\leq 10^{-1.75})$ just below the stability boundary seem very similar to fixed-free membranes with the same mass but have one less peak and one less trough. Decreasing the pretension values for membranes with $R_1\leq 10^{-1.5}$, not only makes the membrane profile  more wavy but also causes the ripples in the membrane shape to move rearward to the trailing edge. Mode shapes with nearly zero deflection at the free ends exist up to $R_1 = 10^{-0.75}$, slightly higher than in the fixed-free case (figure~\ref{fig:stabFixedFree}). 
When the mass density is between $10^{0.75}$ and $10^2$ and the pretension is between~$10^0$ and~$T_{0C}(R_1)$, the membranes are somewhat straighter than in the fixed-free case. Finally, heavy membranes ($R_1> 10^2$) with~$T_0$ between ~$10^0$ and~$T_{0C}(R_1)$ (the stability boundary) all lose stability with the third mode, $n = 3$ in equation~\eqref{eq:eigenmodesfrfr} (the highlighted mode in figure~\ref{fig:pcolorfrfr3and11}D). 

\begin{figure}[H]
    \centering
    \includegraphics[width=.49\textwidth]{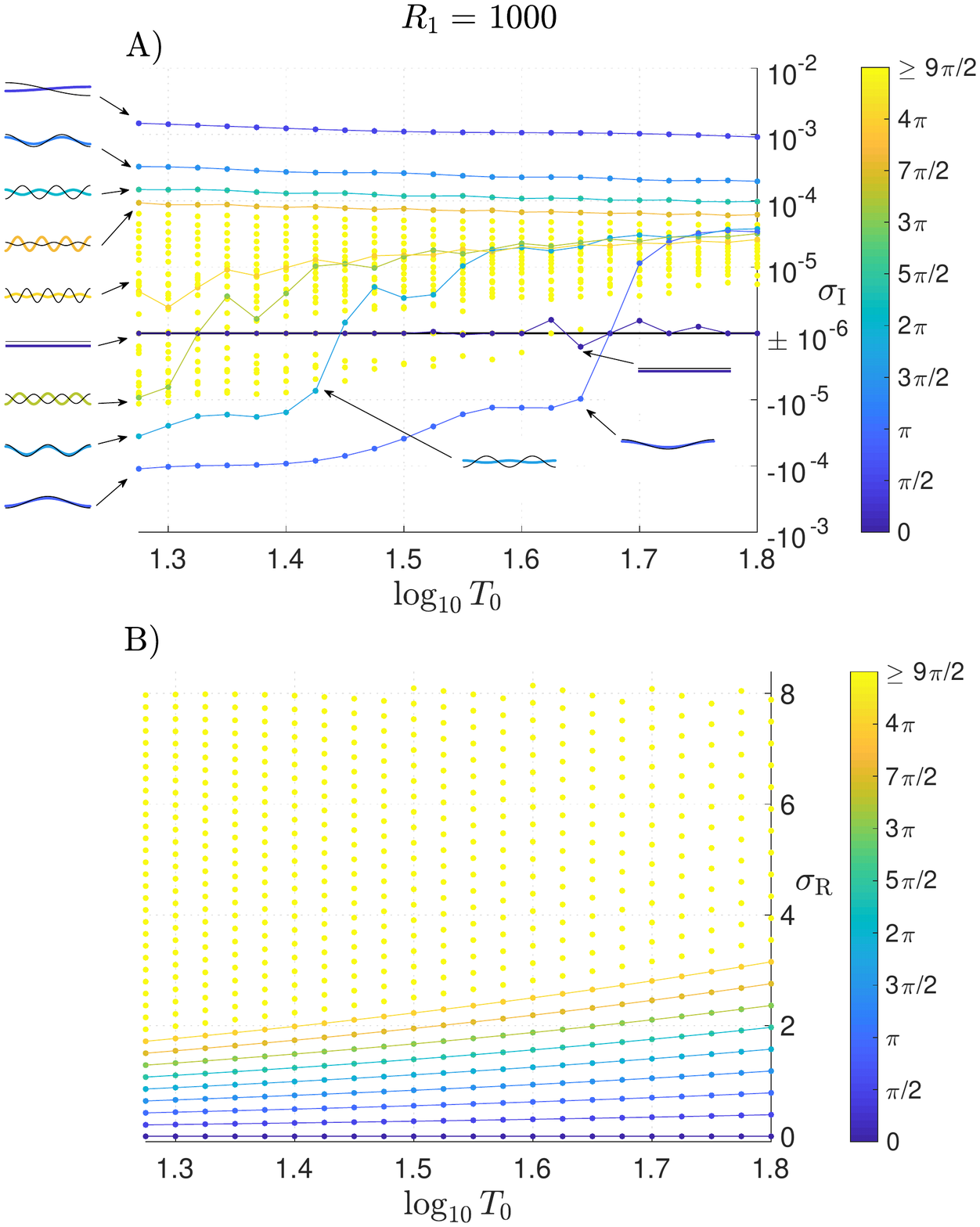}
    \includegraphics[width=.49\textwidth]{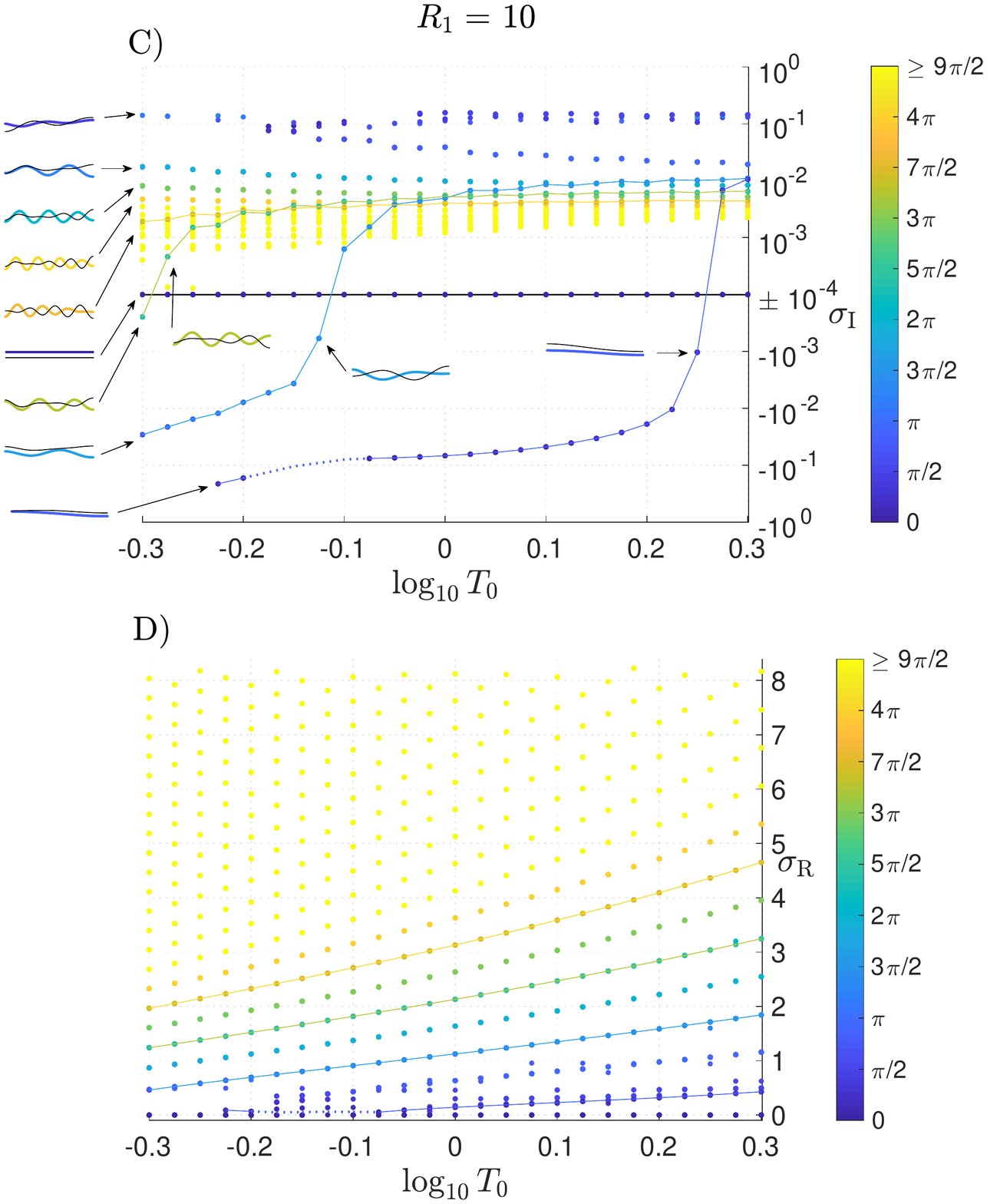}
    \caption{For two values of membrane mass ($R_1$), $10^3$ (left column) and $10^1$ (right column), the imaginary (A, C) and real parts (B, D) of the eigenvalues versus pretension ($T_0$) for free-free membranes. Numerical results are shown as points with color coded according to the value of the RMS of the membrane's slope for each ($R_1,T_0$) pair given by~\eqref{eq:colorRMS}, as given in the colorbar. The horizontal black line in the top panels located at A) $\sigma_\Im=\pm 10^{-6}$, B) $\sigma_\Im=\pm 10^{-4}$ distinguishes stable modes (above) and unstable modes (below). We also show typical modes that correspond to each branch with $Y'_\mathrm{RMS}< 9\pi/2$.}
    \label{fig:bif3frfr}
\end{figure}

In figure~\ref{fig:ScatterPlotsFreeFree} we studied how the most unstable eigenvalues change in the $R_1$--$T_0$ parameter space, and in figure~\ref{fig:stabFreeFree} we investigated the trends in the corresponding most unstable eigenmodes. Now in figure~\ref{fig:bif3frfr} we show the changes in the eigenvalues and associated eigenmode shapes as we pass through the stability boundary for two fixed values of mass density, as for the fixed-fixed and fixed-free cases (figures~\ref{fig:bif3fifi} and~\ref{fig:bifurcationfifr}, respectively). Each dot's color is used to label~$Y'_\mathrm{RMS}$ (equation~\eqref{eq:colorRMS}). 
For the largest $R_1=1000$ (panel A) the unstable modes are odd-numbered. The first branch to become unstable is the third mode ($n=3$ in equation~\eqref{eq:eigenmodesfrfr}) at $T_0\approx 10^{1.68}$---consistent with figure~\ref{fig:stabFreeFree} (for the same $R_1$). Then the fifth mode ($n=5$) becomes unstable at $T_0\approx 10^{1.45}$, and the seventh mode at $T_0\approx 10^{1.33}$. The even-numbered modes are all stable for the entire range of~$T_0$ values considered here. 
We show the membrane mode shapes that correspond to the nine lowest wavenumber modes to the left of panel~A at the lowest $T_0=10^{1.275}$. 
 The $Y'_\mathrm{RMS}$ values that correspond to these nine lowest wavenumber modes are approximately those of the analytical eigenmodes in~\eqref{eq:eigenmodesfrfr}, $(n-1)\pi/2$ for $n=1,2,\dots,9$.
We also show instances of membrane shapes at a couple of larger~$T_0$ values for the first two unstable branches and the flat mode. We see that in all cases, these mode shapes have the same features as at the smallest $T_0$. The branch corresponding to the flat mode ($n=1$) in figure~\ref{fig:bif3frfr}A oscillates about $\sigma_\Im= \pm 10^{-6}$ at $T_0\geq 10^{1.6}$ (while $\sigma_\Re$ lies on $\pm 10^{-4}$)---it is essentially zero. As in the fixed-fixed and fixed-free cases at $R_1=1000$, the branches with the largest $\sigma_\Im>0$ are all continuous but at the smaller  $R_1$ (i.e.\ $10$), the same branches (blue dots at the top of panel C and bottom of panel D) appear more disordered. 
The loss of stability in figure~\ref{fig:bifurcationfifr}C occurs at $T_0\approx 10^{0.275}$. The values of $\sigma_\Im$ in panel C are about two orders of magnitude higher than those in panel~A (as for fixed-free membranes at the same membrane masses). The downward tendency of the darker orange branch when $\sigma_\Im$ drops below $10^{-2}$ (panel C) suggests that the mode may be the next to become unstable as $T_0$ decreases.
Contrary to panel~A, we see in panel~C that the yellow dots (higher wavenumber modes) are mostly stable.  The free-free angular frequency ($\sigma_\Re$) behaves similarly to fixed-fixed and fixed-free membranes: the curves connecting $\sigma_\Re$ associated with particular modes are steeper for $R_1=10$ (panel D) compared to $R_1=1000$ (panel B).
The dotted part of the most unstable branch shown in figures~\ref{fig:bif3frfr}C and D is used to bridge a gap in $T_0$ in which we did not find eigenvalues and eigenmodes for the lowest branch.

\section{Comparison with unsteady and large-amplitude simulations}\label{sec:comparisonIVP}

\begin{figure}[H]
    \centering
    \includegraphics[width=\textwidth]{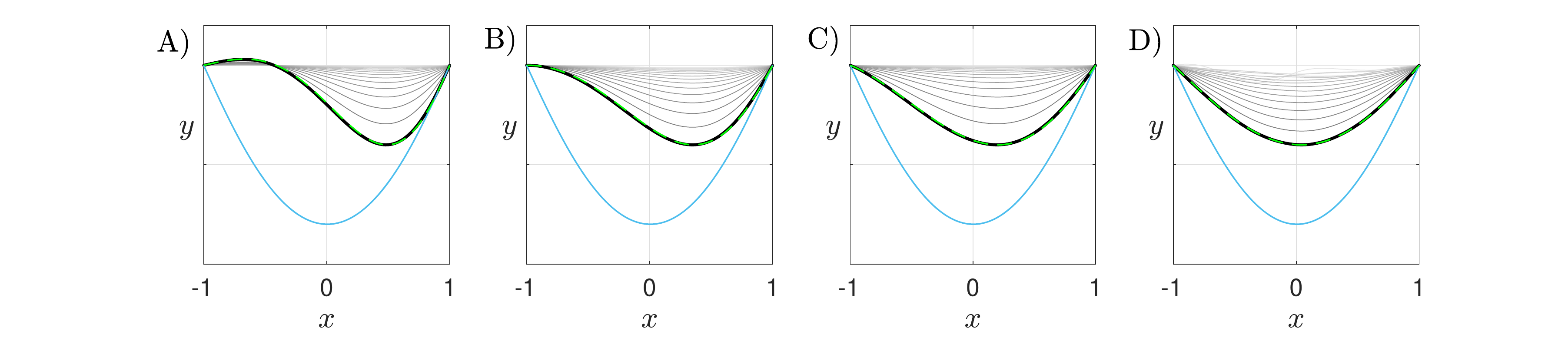}
\caption{Fixed-fixed membranes at $R_1=10^{-1}$ and A) $T_0=10^{-0.1}$, B) $T_0=10^0$, C) $T_0 = 10^{0.1}$, and D) $T_0=10^{0.2}$. These membranes lose stability by divergence. We compare the most unstable modes obtained from the eigenvalue analysis (dashed green lines) to the membrane shapes of the time-stepping simulations in the small-amplitude (growth) regime---in each panel, 15 equally spaced snapshots are shown in the growth regime, gray and then black at the last time. The arbitrary amplitudes of the green lines are set to match those of the black lines. The light blue curves indicate shapes in the large-amplitude steady state regime. }
    \label{fig:comparisonfifi}
\end{figure}

We now compare the most unstable eigenmodes, in a few cases, with the corresponding small-amplitude motions as well as the eventual large-amplitude steady-state motions in the unsteady time-stepping simulations of~\cite{mavroyiakoumou2020large}. The main differences are that in the eigenvalue problem the free vortex wake has a finite length~$\ell_w$ whereas in the unsteady simulations it grows from zero length, and has $\delta$-smoothing to avoid chaotic dynamics. 
For fixed-fixed membranes, figure \ref{fig:comparisonfifi} compares eigenmodes (dashed green lines) with snapshots of time-stepping simulations in the small-amplitude growth regime (sequence of gray lines ending with black lines) and the time-stepping simulations' eventual large-amplitude steady states (blue lines).
The comparison is made at $R_1=10^{-1}$ with $T_0$ increasing: (A) $10^{-0.1}$, (B) $10^0$, (C) $10^{0.1}$, and (D) $10^{0.2}$, the last value close to the stability boundary. Here we have a divergence instability, so the imaginary parts of the eigenmodes are zero; the green lines show the real parts.  As $T_0$ increases, the small-amplitude membrane shapes change gradually, from ones with both downward and upward curvature (A) to a nearly fore-aft symmetric hump with upward curvature only (D). The close agreement between the green and black lines shows that the linearized model captures the small-amplitude unsteady dynamics well. Here the initial deflection is $y(x,0) = 10^{-12}\sin(\pi x)$, but we find essentially the same agreement with a different form of the initial perturbation, in which the leading edge is moved slightly upward and then back to $y = 0$. In this case the membrane initially forms a small bump near the trailing edge as it evolves under the nonlinear membrane equation \eqref{eq:membrane}.
Both types of initial deflections are much smaller than the gray shapes in figure \ref{fig:comparisonfifi}, and eventually converge to them as the fastest growing mode outgrows the other modal components of the initial deflections. At large amplitudes, all the unsteady shapes converge to steady humps (blue lines), nearly fore-aft symmetric, despite the early-time differences. The magnitudes of the humps' deflections are set by the nonlinear stretching resistance in (\ref{eq:membrane}), the term proportional to the stretching modulus~$R_3$. Here~$R_3$ is set to 10 but only the magnitudes of the humps, and not their shapes, change much over the range $R_3 \geq 10$~\cite{mavroyiakoumou2020large}.


We now investigate membranes with the leading edges fixed and the trailing edges free. Now the membranes lose stability through divergence and flutter, so the eigenmodes are complex. They are determined only up to a complex constant, with both a magnitude and a phase that need to be matched to a given time-stepping simulation of~\cite{mavroyiakoumou2020large}.
In appendix~\ref{app:comparison} we give details about how the matching is done.

\begin{figure}[H]
    \centering
    \includegraphics[width=.7\textwidth]{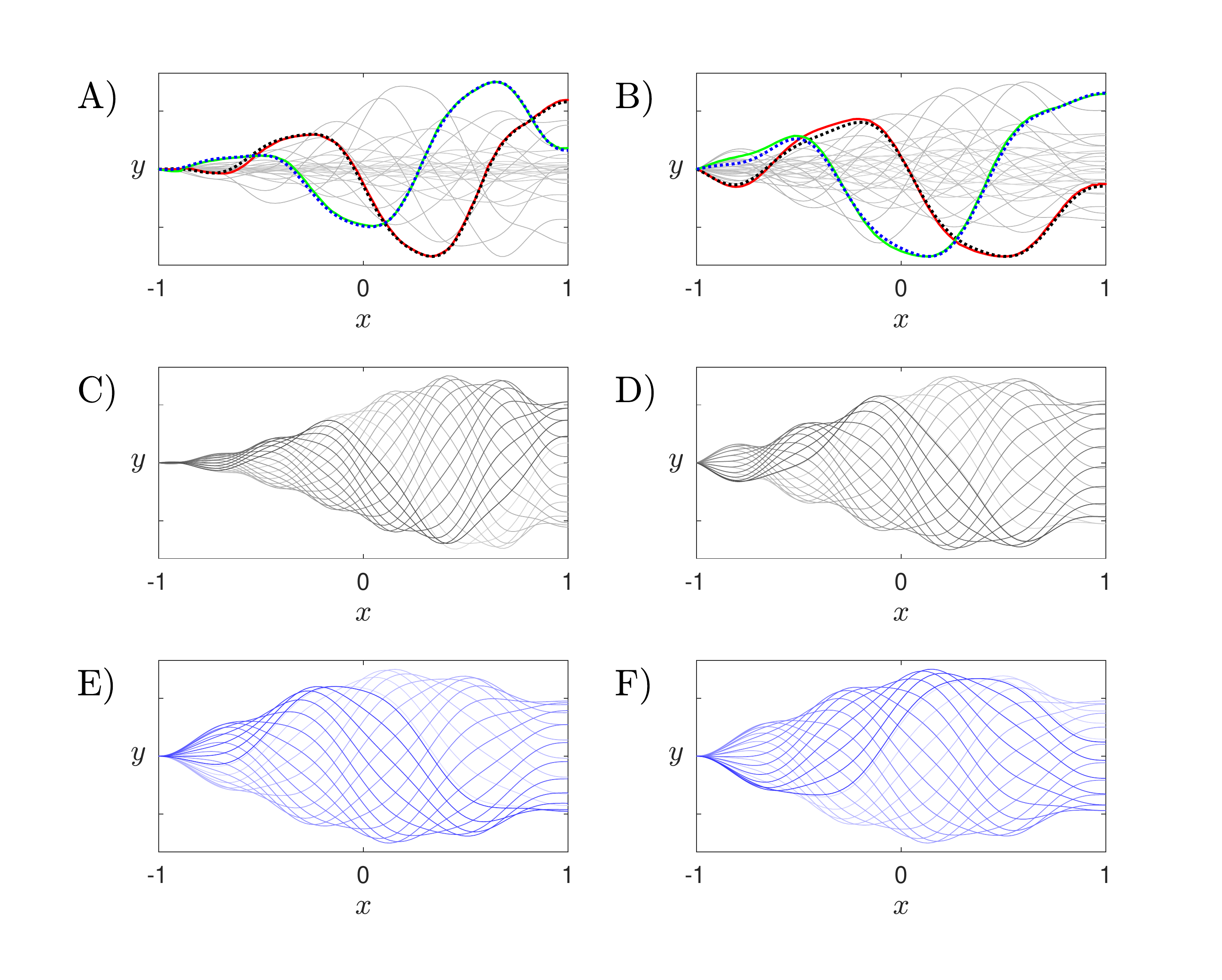}
\caption{Fixed-free membranes at $R_1=10^{-0.5}$ and $R_3 = 10^{1.5}$, with $T_0=10^{-0.8}$ for A, C, E and $T_0=10^{-0.7}$ for B, D, and F. In panels A and B the solid red lines are $\real(y_{\nonlin}(\alpha))$ estimated from the time-stepping simulation, which are close to  $\real(Y(x))$ from the eigenvalue problem (dotted black lines). The solid green lines are $\imag(y_{\nonlin}(\alpha))$, close to $\imag(Y(x))$ from the eigenvalue problem (dotted blue lines). The gray lines are a subset of snapshots in the linear growth regime. In panels C and D we show snapshots during the small-amplitude (growth) regime, but with the exponential growth removed. Panels E and F show snapshots during the steady-state large-amplitude motions. We show 20 equally spaced snapshots of membranes over a period, ranging from light blue at earlier times to dark blue at the last time.}
    \label{fig:comparisonfifrMinus05}
\end{figure}

In figure~\ref{fig:comparisonfifrMinus05},
we compare two cases slightly below the stability boundary at $R_1=10^{-0.5}$: 
$T_0=10^{-0.8}$ (panels A, C, and E) and
$T_0=10^{-0.7}$ (panels B, D, and F).
In panels A and B, the gray lines again show sequences of snapshots from the time-stepping simulations. 
We fit the values $y(\alpha, t)$ for such a sequence to a function of the form 
$\real\left([\real(y_{\nonlin}(\alpha))+i\imag(y_{\nonlin}(\alpha))]e^{i\sigma  t}\right)$. First
$\sigma_\Im$ and $\sigma_\Re$ are estimated. Then
for each $\alpha$, the real and imaginary parts of $y(\alpha, t)e^{-i\sigma  t}$ are estimated (in amplitude-phase form; see appendix~\ref{app:comparison}),
giving $\real(y_{\nonlin}(\alpha))$ (red solid lines in panels A and B) and $\imag(y_{\nonlin}(\alpha))$ (green solid lines). The most unstable eigenmode $Y(x)$ is arbitrary up to a complex constant. The function $y_{\nonlin}(\alpha)$ contains a complex factor (magnitude and phase) that depends on the initial conditions of the time-stepping simulation. To account for this, we scale $Y(x)$ by the complex factor that gives the best $L^1$-fit with $y_{\nonlin}(\alpha)\approx y_{\nonlin}(x)$ (see appendix~\ref{app:comparison})
and plot the resulting $\real(Y)$ and $\imag(Y)$ as dotted black and blue lines respectively, in panels A and B. The fit between $Y(x)$ and $y_{\nonlin}(x)$ is nearly as good as in the steady fixed-fixed cases (figure \ref{fig:comparisonfifi}). The slight increase of error in the fit may be due to the extra steps involved in fitting the fixed-free eigenmodes because they are complex.

In panels C and D, we show 20 snapshots from the
time-stepping simulations, but multiplied by our estimate of $e^{\sigma_\Im t}$, which should remove the exponential growth. This shows the mode shapes much more clearly than in panels A and B. The rescaled shapes are equally spaced over our estimate of one time period. They appear to follow an up-down symmetric, periodic (as expected) oscillation with (A) seven and (B) five ``necks" in their envelopes, respectively. 
Panels E and F show snapshots in the eventual large-amplitude periodic steady-state. The shapes are qualitatively similar to those in C and D, but the numbers of necks are reduced to four in both E and F. The shapes are nearly the same in both panels; as in the fixed-fixed case (figure \ref{fig:comparisonfifi}) the differences in the small-amplitude shapes disappear at large amplitude. This may be because the $T_0$ term in (\ref{eq:membrane}) is subdominant to the $R_3$ term at large amplitudes, even at $T_{0C}(R_1)$, the largest $T_0$ where the membranes are unstable.

\begin{figure}[H]
    \centering
    \includegraphics[width=.7\textwidth]{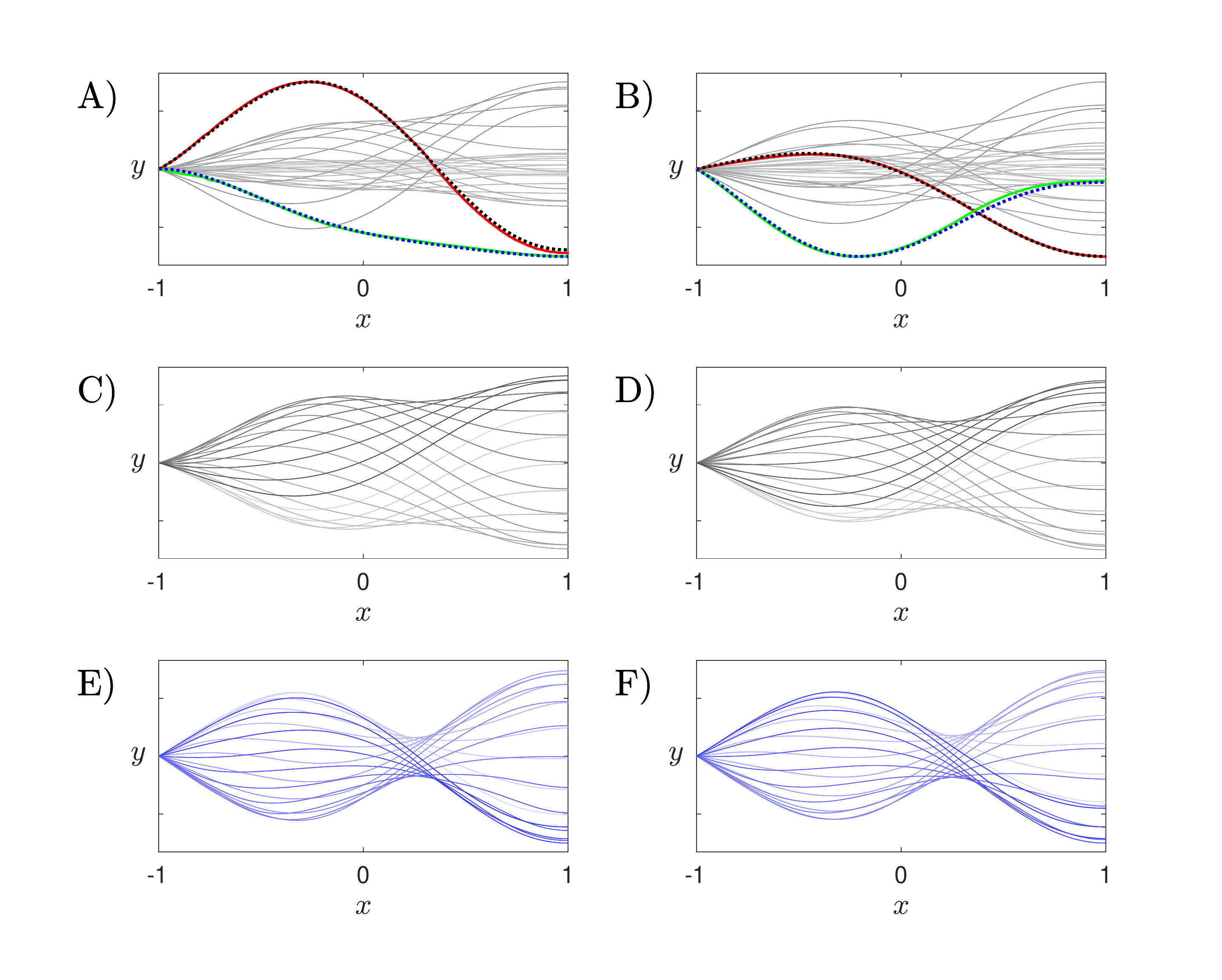}
\caption{Same quantities as described in figure~\ref{fig:comparisonfifrMinus05} but with $R_1=10^{1}$ and $R_3 = 10^{1.5}$, and $T_0=10^{-0.1}$ (for A, C, and~E); $T_0=10^0$ (for B, D, and F).
}
    \label{fig:comparisonfifr10}
\end{figure}

We show the same comparisons at larger $R_1$ (10) in figure~\ref{fig:comparisonfifr10}, at two $T_0$ values near the stability boundary. The wave numbers of the shapes are much reduced---only one neck appears in each envelope now---but otherwise many of the same features carry over from the previous figure. There is again good agreement between the eigenmodes and the versions estimated from the time-stepping simulations (panels A and B). The periodic parts of the small-amplitude motions have small but noticeable differences in panels C and D---in particular, the widths of the necks relative to the maximum widths of the envelopes. The large-amplitude motions (E and~F) are again nearly indistinguishable, however.

We obtain similar levels of agreement in the free-free case; two examples are shown in appendix \ref{frfrcomp}.

\section{Conclusions}\label{sec:conclusions}

To summarize, we have used a linearized model and a nonlinear eigenvalue solver to study small amplitude membrane motions, including the onset of membrane instability, in inviscid fluid flows. We characterized the different types of possible motions with respect to the two key dimensionless parameters---membrane mass and pretension---and for three sets of boundary conditions: ``fixed-fixed," ``fixed-free," and ``free-free" leading and trailing edges.
Previous work by other groups was limited to the fixed-fixed case, and a smaller range of membrane mass densities, and our own previous time-stepping simulations~\cite{mavroyiakoumou2020large} were
unable to resolve the small-amplitude motions at small mass densities due to limited spatial resolution, and at large mass densities due to the very slow growth of instabilities.

For each of the three sets of boundary conditions, when membrane inertia and pretension dominate fluid pressure forces, the eigenmodes tend towards neutrally stable sinusoidal functions with half-integer or quarter-integer numbers of wavelengths. When the fluid forces are small but nonnegligible, the mode shapes are similar, but the even- (for fixed-fixed) or odd-numbered modes (for fixed-free and free-free) become unstable, starting with the second and third modes, respectively. For the fixed-fixed case, there are roughly two regimes: small membrane density, where divergence occurs and the most unstable mode becomes more fore-aft asymmetric as one moves further into the instability region; and large membrane density, where flutter and divergence occur with approximately sinusoidal modes. In both regimes, the most unstable modes become wavier at smaller $T_0$, akin to the most unstable beam modes at smaller bending rigidity in \cite{alben2008ffi}. These results agree with those of \cite{tiomkin2017stability} in the same parameter regimes.

The stability boundaries for the fixed-free and free-free cases resemble the fixed-fixed case at large membrane densities, showing an upward slope for $R_1 \geq 10^2$ (which we were not able to compute using time-stepping simulations).
The fixed-free and free-free stability boundaries
differ strongly from the fixed-fixed case at moderate and small membrane densities.  There the membranes remain stable down to smaller pretension values, and eventually become unstable by flutter and divergence. For $10^{-3} \leq R_1 \leq 10^{-1}$, the most unstable mode is very wavy, and we were unable to resolve it with the time-stepping simulations in \cite{mavroyiakoumou2020large}. Here we find that the most unstable eigenmodes have small deflection at the leading and trailing edges, despite the free boundary conditions. For $10^{-1} \leq R_1 \leq 10^{2}$, the modes are wavy shapes (wavier at smaller $T_0$) superposed on background shapes with nonzero slopes (fixed-free) and/or deflections (free-free). By tracking the eigenmodes across the stability boundaries, we found that at moderate membrane densities, the modes resemble the sinusoidal shapes at large densities, but with more disorder, and the appearance of irregular bands of stable low-wavenumber modes that are difficult to associate with a particular branch. 

Finally, we compared the eigenmodes with the membrane motions in the time-stepping simulations, and found very good agreement with the small-amplitude portion of the time-stepping simulations in examples with the three different boundary conditions. In all the examples, the large-amplitude motions qualitatively resembled those in the small amplitude regime in terms of the number of necks in the deflection envelopes, but had clear differences in the envelopes' shapes and the relative sizes of maxima and minima.

\subsection*{Acknowledgments}
We acknowledge support from the NSF Mathematical Biology program under award number DMS-1811889, and from a Catalyst grant from the Michigan Institute for Computational Discovery and Engineering (MICDE).

\appendix

\section{Convergence with respect to number of Chebyshev nodes}\label{sec:convergencem}

In this work we have computed the membrane eigenmodes and eigenvalues using $m$ = 120 (121 Chebyshev points on the membrane) and a free vortex wake of length $\ell_w=39$. We consider here the effect of varying the former. The effect of varying the  vortex wake length was explored in~\cite[Sec.\ V]{alben2008ffi}, and here we find that the results in the unstable regime are basically unchanged when $\ell_w$ is as large as 39, given the exponential decay of circulation in the wake (except right on the stability boundary, but there is still algebraic decay of the induced velocity by an alternating-sign wake).

\begin{figure}[H]
    \centering
    \includegraphics[width=.8\textwidth]{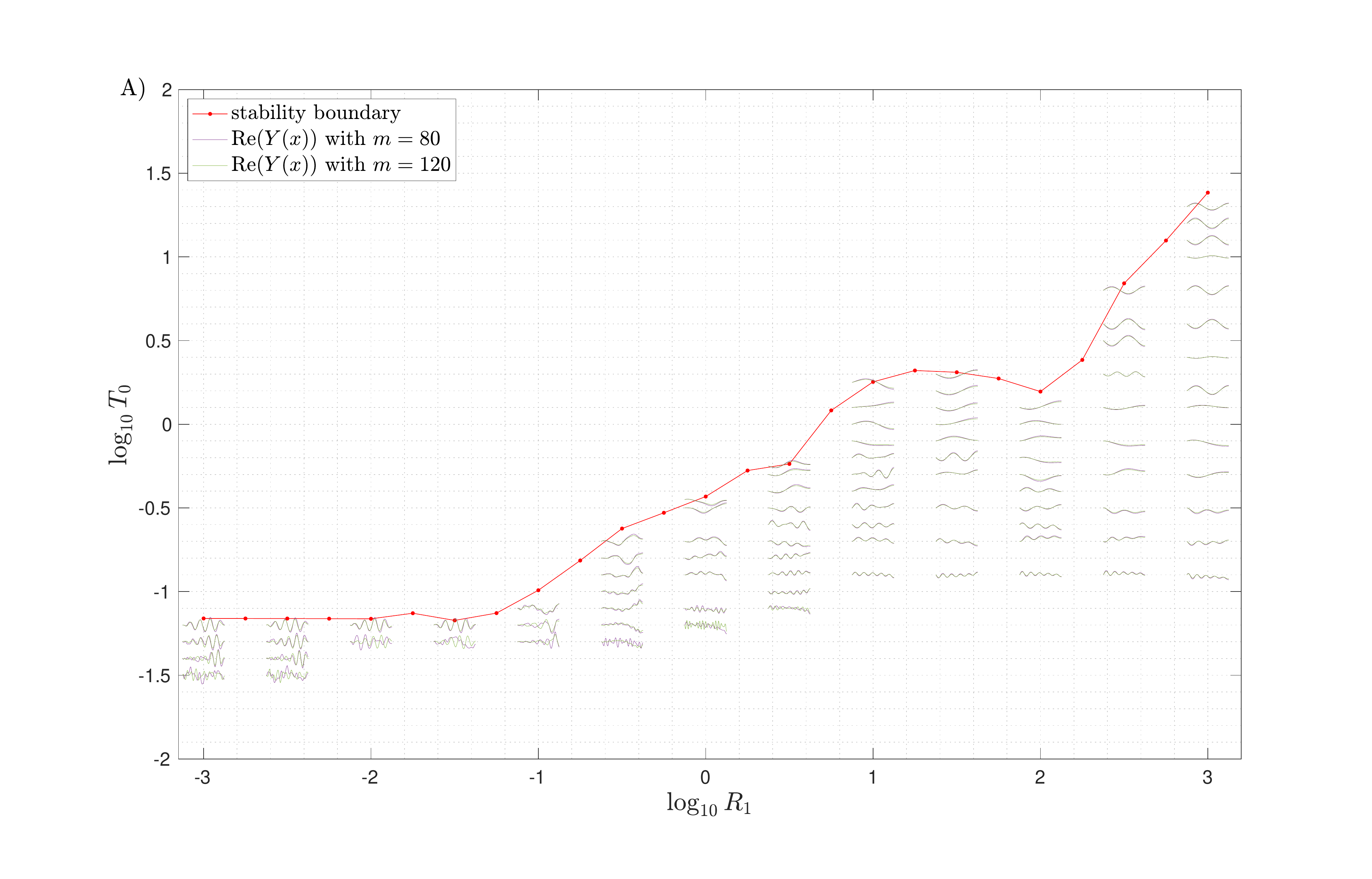}
    \includegraphics[width=.8\textwidth]{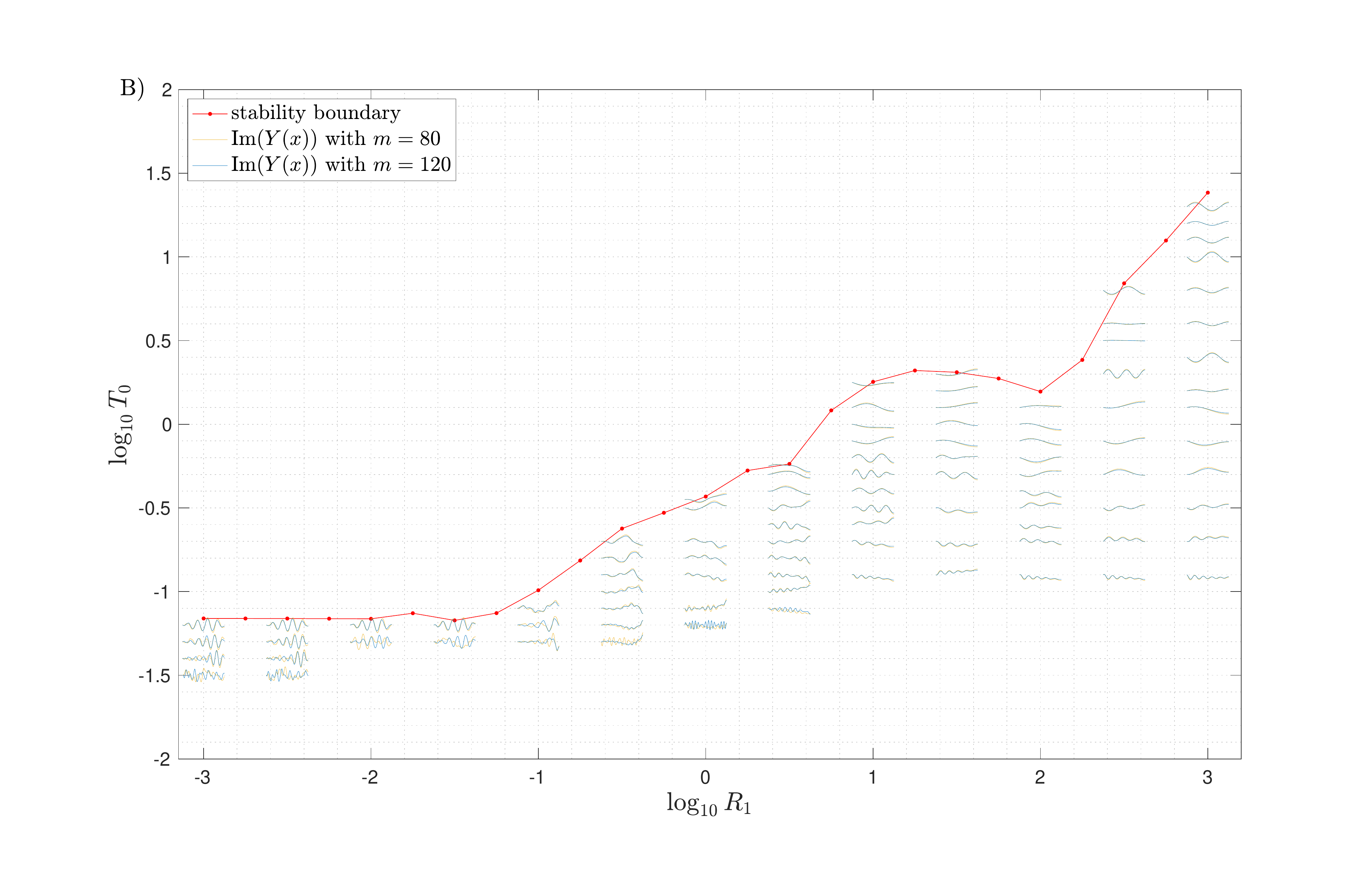}
    \caption{Comparison between the A) real and B) imaginary parts of the eigenmodes with fixed-free boundary conditions, using grids with $m=80$ and $m=120$. Each shape is scaled in both vertical and horizontal directions to fit within the plot. The red dots indicate the position of the stability boundary (same as in figure~\ref{fig:stabFixedFree}).}
    \label{fig:comparison80and120}
\end{figure}

\begin{figure}[H]
    \centering
    \includegraphics[width=.75\textwidth]{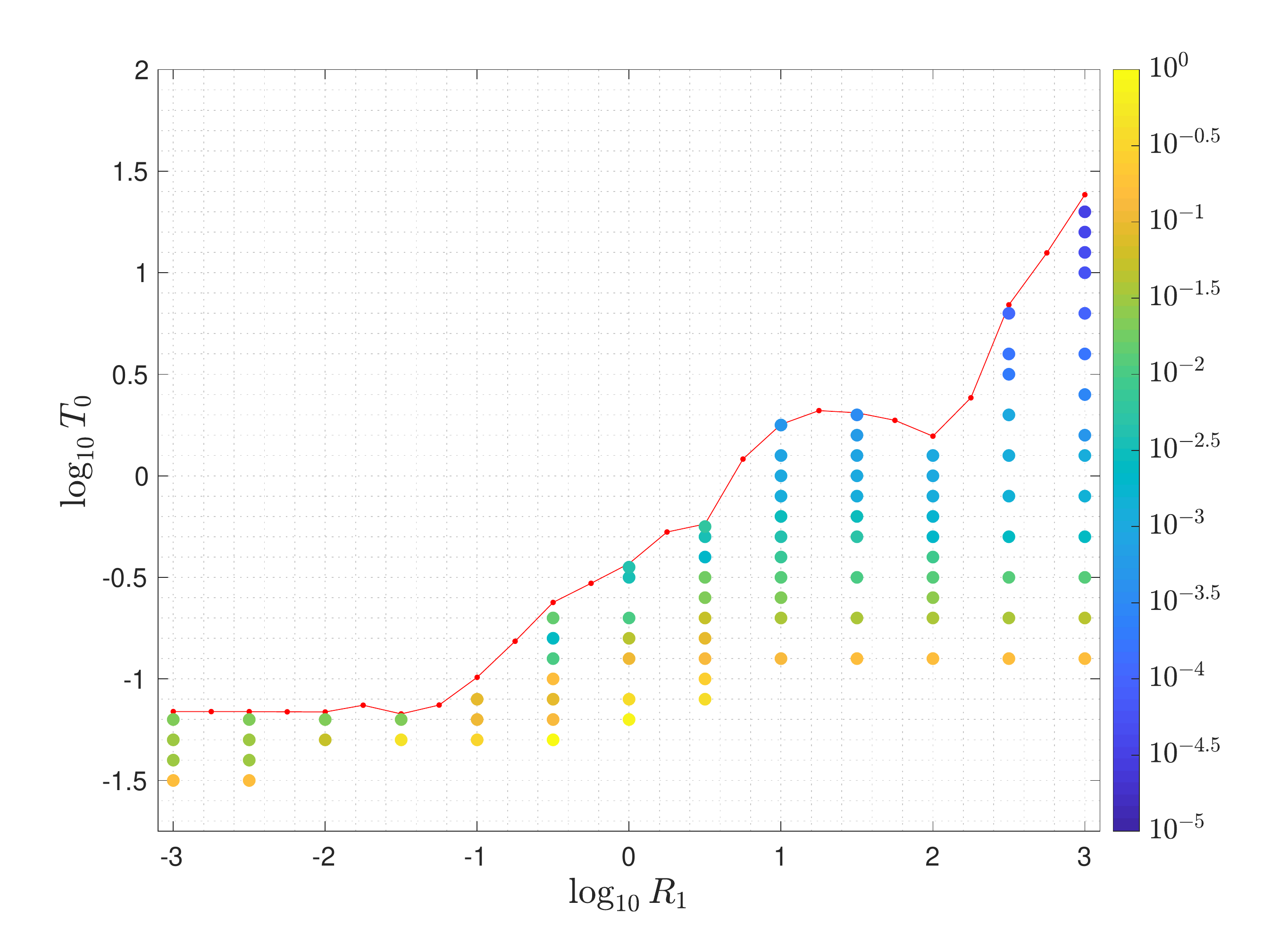}
    \caption{At each $(R_1,T_0)$ in the instability region (below red line), the relative error~\eqref{eq:relativeErr} in the eigenvalues when using $m=80$ and $m=120$ Chebyshev points on the fixed-free membrane is plotted as a colored dot. }
    \label{fig:evalue80and120RelativeError}
\end{figure}

\begin{figure}[H]
    \centering
    \includegraphics[width=.3\textwidth]{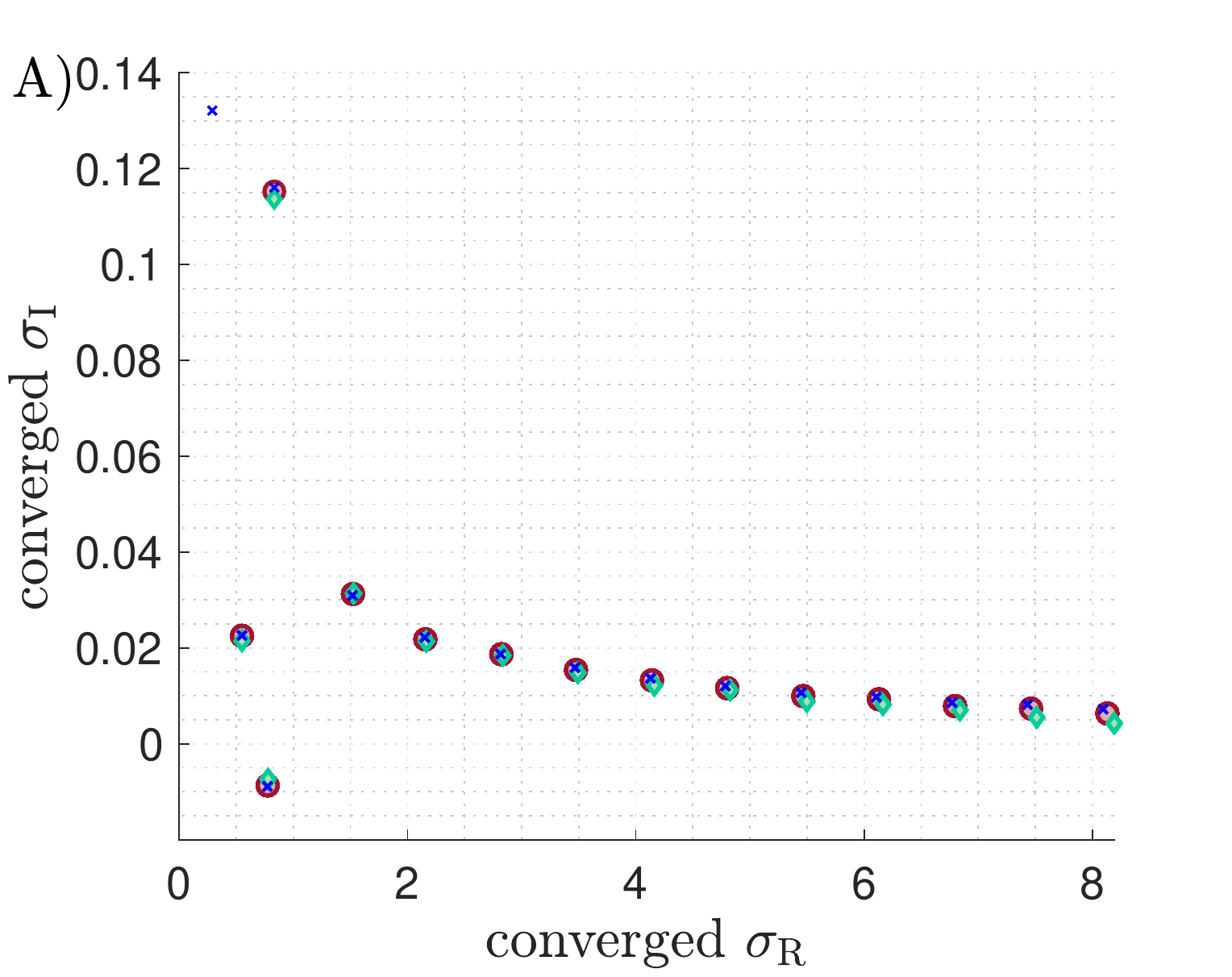}
     \includegraphics[width=.36\textwidth]{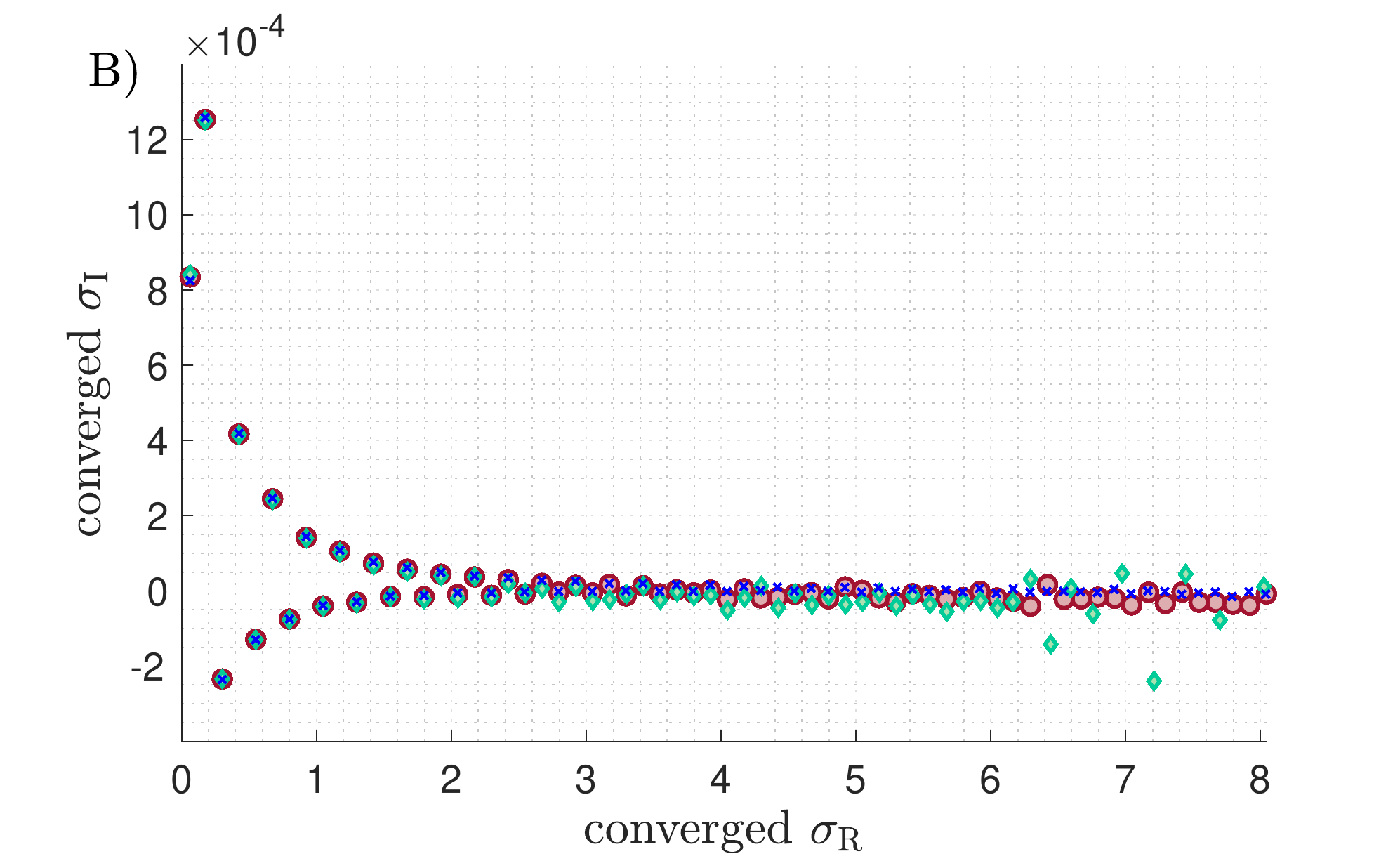}
     \includegraphics[width=.3\textwidth]{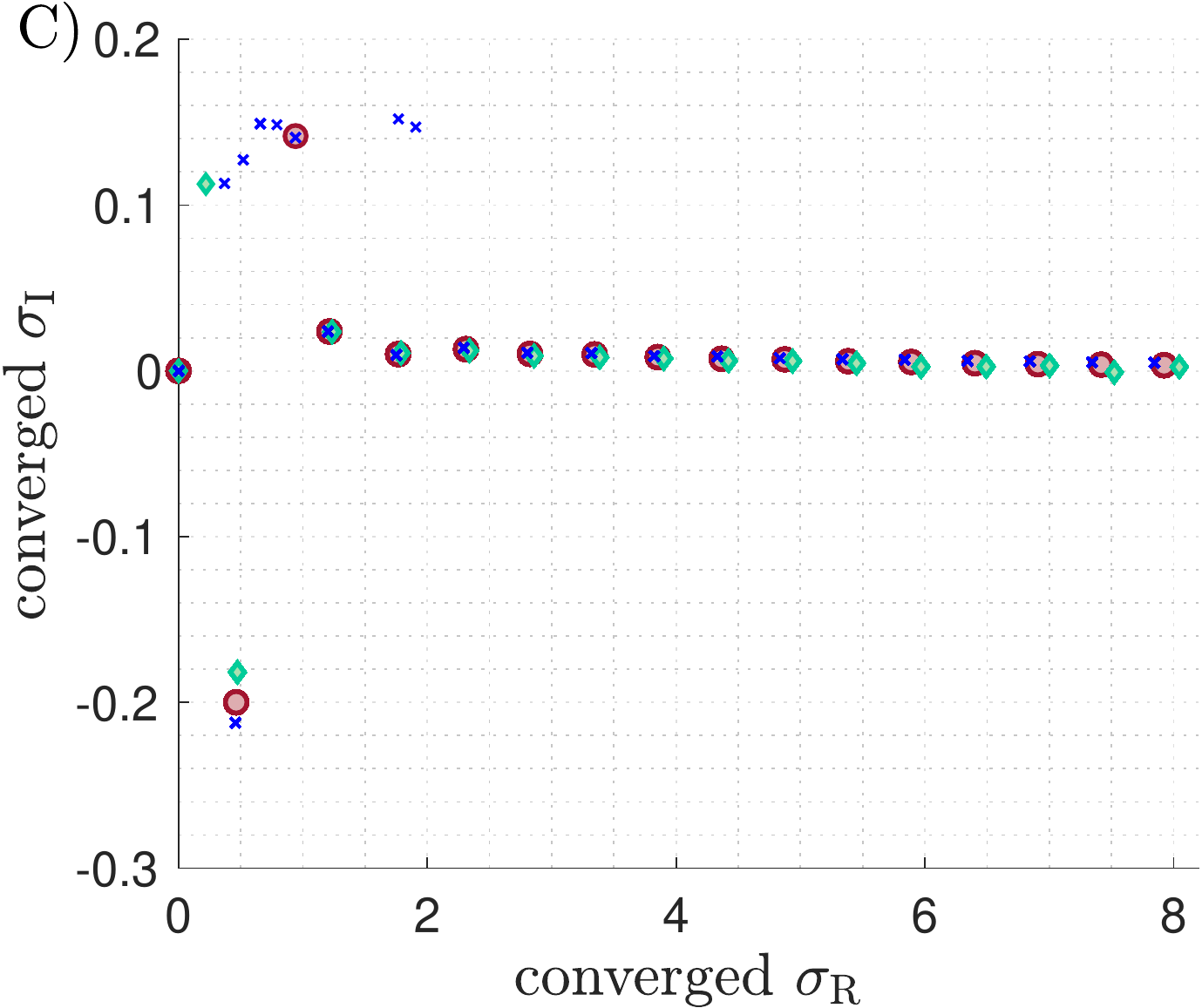}
    \caption{Spectrum of eigenvalues for $m$ = 80 (green diamonds), 120 (red circles), and 240 (blue crosses) for a fixed-free membrane at A) $(R_1,T_0)=(10^{0.5},10^{-0.25})$, B) $(R_1,T_0)=(10^3,10^{0.8})$, and a free-free membrane at C) $(R_1,T_0)=(10^{0.5},10^{-0.5})$.}
    \label{fig:evalue80and120spectrum}
\end{figure}

To compare the eigenmodes obtained when using $m=80$ versus 120, we remove the arbitrary phase shift from the eigenmode solver by finding $\phi\in[0,2\pi]$ that solves 
\begin{equation}\label{eq:shift80and120}
    \min\limits_\phi \int_{-1}^1 |Y_{80}(x)-Y_{120}(x)e^{i\phi}|\,\d x.
\end{equation} 
To perform the subtraction in~\eqref{eq:shift80and120} we interpolate $Y_{80}$ using shape-preserving piecewise cubic interpolation onto the 120-point grid.
 In figure~\ref{fig:comparison80and120} we compare the real (panel A) and imaginary parts (panel B) of the fixed-free eigenmodes when using $m=80$ and 120 across an array of $(R_1,T_0)$ pairs. The eigenmodes agree well except in some cases at the smallest values of $T_0$ for each $R_1$, where the modes are also more wavy and difficult to resolve numerically.
In figure~\ref{fig:evalue80and120RelativeError} we present the relative error in the eigenvalues when $m$ = 80 and 120. This quantity is computed as
\begin{equation}\label{eq:relativeErr}
    \text{relative error}=\left|\frac{\sigma_{80}-\sigma_{120}}{\sigma_{120}}\right|.
\end{equation}
The errors are typically $10^{-2}$--$10^{-5}$ near the stability boundary, and gradually increase to $10^{-1}$--$10^0$ as we decrease $T_0$, eventually reaching a point where the solutions are underresolved (as in figure~\ref{fig:comparison80and120}).
Figure~\ref{fig:evalue80and120spectrum} shows three examples of the computed eigenvalues using grids of initial guesses with $m$ = 80 (green diamonds), 120 (red circles), and 240 (blue crosses). Panel A corresponds to a fixed-free membrane at a moderate value of $R_1$ ($10^{0.5}$) and $T_0=10^{-0.25}$. The eigenvalues agree well except for an additional stable eigenmode found when $m=240$ (small blue cross located at $(\sigma_\Re,\sigma_\Im) \approx(0.29,0.13)$). Panel B corresponds to a fixed-free membrane with a larger value of $R_1$ ($10^3$), and $T_0 = 10^{0.8}$. The eigenvalues agree well at the three values of $m$ when $\sigma_\Re \lesssim 2.5$, approximately the 15 lowest modes. As $\sigma_\Re$ increases, the modes are eventually underresolved and the eigenvalues deviate significantly, beginning with $m = 80$ (green triangles). Panel C corresponds to a free-free membrane at $R_1 = 10^{0.5}$ as in panel~A, but with $T_0$ slightly smaller, $10^{-0.5}$. As in panel A, there are extra stable eigenvalues (with $\sigma_\Re<2$ and $\sigma_\Im>0.1$) most with $m=240$ (blue crosses) and one with $m = 80$ (green diamond). 
These eigenvalues are similar to those in the irregular bands of stable eigenvalues in figures \ref{fig:bif3fifi}, \ref{fig:bifurcationfifr}, and
\ref{fig:bif3frfr} when $R_1 = 10$.
We have good agreement among the eigenvalues that are unstable or close to neutrally stable.
In each case, the most unstable modes (i.e.\ the modes associated with smallest---or most negative---$\sigma_{\Im}$) change little when~$m$ increases from 120 to 240, and they are the focus of this paper. 

\section{Method for comparing the eigenvalue analysis results to time-stepping simulations}\label{app:comparison}

Here we outline the method for comparing mode shapes obtained using the eigenvalue analysis developed in the current paper and mode shapes obtained from the time-stepping simulations in~\cite{mavroyiakoumou2020large}.

We first determine the regime of exponential growth in the nonlinear simulations by plotting the trailing edge deflection as a function of time, i.e.\ $y(1,t)$. The time it takes to reach the large-amplitude steady state regime depends on the magnitude of the initial perturbation. To extend the time spent in the small-amplitude regime, we start with a very small perturbation, $O(10^{-12})$. This is important particularly when the growth rate is large, i.e.\ for unstable membranes that are far from the stability boundary in parameter space.

\begin{figure}[H]
    \centering
    \includegraphics[width=\textwidth]{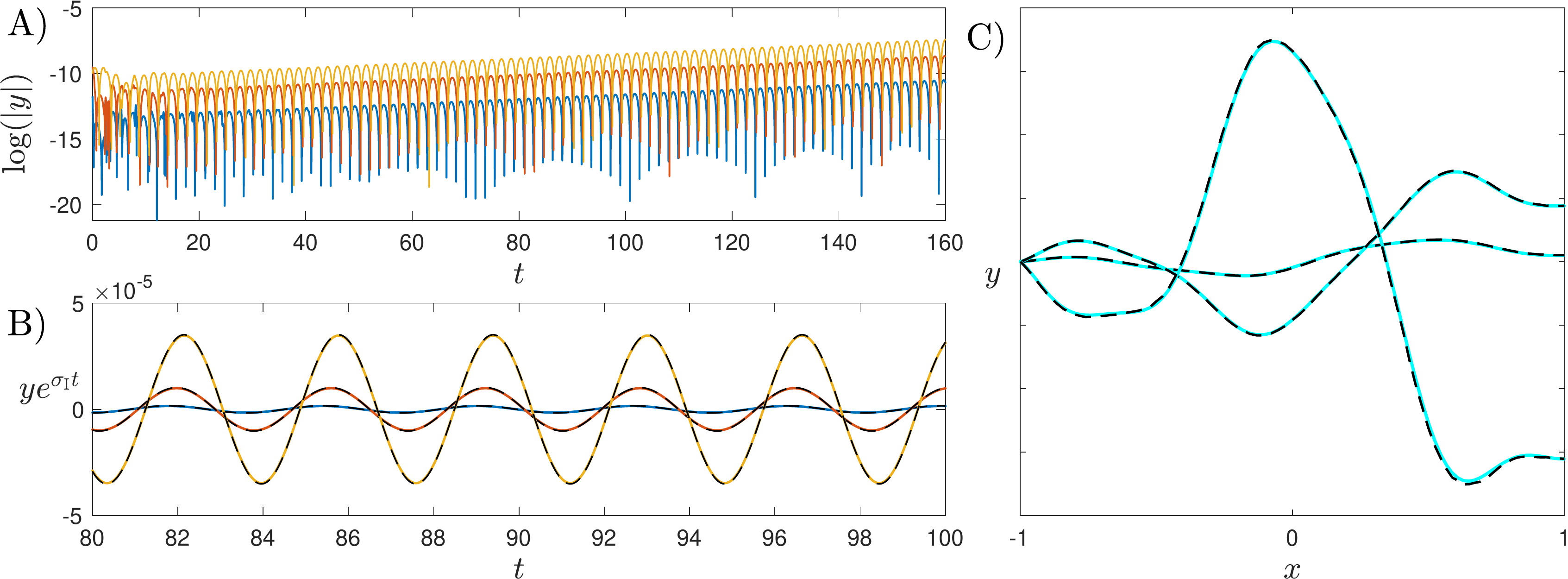}
    \caption{Example of the comparison method using data from fixed-free membranes with $(R_1,T_0)=(10^{-0.5},10^{-0.7})$. A) $\log(|y|)$ versus time for the 10th (blue), 30th (red), and 100th (yellow) grid points on the membrane. B) A portion of the time series of $ye^{\sigma_{\Im}t}$ at the 10th (blue), 30th (red), and 100th (yellow) grid points. This corresponds to part of the small-amplitude regime but with the growth removed. The black dashed lines represent the constructed $R(\alpha)\cos(\sigma_{\Re}t+\phi(\alpha))$ at the same grid points. C) The reconstructed data $\real\left([\real(y_{\nonlin}(\alpha))+i\imag(y_{\nonlin}(\alpha))]e^{i\sigma  t}\right)$ (black dashed lines) compared against the data $y(\alpha,t)$ (cyan solid lines) at the times $t=20,100,160$. The initial perturbation here is $\zeta(\alpha,0)=\eta\sin(\pi \alpha)$ where $\eta$ is chosen as 0.0001. Note that the axes are not to scale.}
    \label{fig:comparisonmethod}
\end{figure}

During this exponential growth (with flutter) regime, we approximate the computed $y(\alpha,t)$ as 
\begin{equation}
    y(\alpha,t)\approx \real\left([\real(y_{\nonlin}(\alpha))+i\imag(y_{\nonlin}(\alpha))]e^{i\sigma t}\right).
\end{equation}
 To obtain $\sigma$ and $y_{\nonlin}(\alpha)$, we first obtain $\sigma_{\Im}$ as the negative of the slope of $\log(|y|)$ versus time and subsequently compute $y(\alpha,t)e^{\sigma_{\Im}t}$. For each grid point $1,\dots,m+1$ in $\alpha\in[-1,1]$, this is a function that oscillates sinusoidally in time but does not grow (figure~\ref{fig:comparisonmethod}B). 
 We estimate the frequency $f$ of these functions as the reciprocal of the time between the peaks of the sinusoidal function. The frequency should be the same for all $\alpha\in[-1,1]$ according to our Ansatz, and the computed values vary only slightly due to numerical errors. We use the average over $\alpha$ as our estimate of the single, global frequency. 
 We then define $\sigma_{\Re} := 2\pi f$ and denote the amplitudes of these sinusoidal functions $R(\alpha)$. We denote by $t_{\mathrm{peak}}(\alpha)$ the times at which they reach their peaks and define the phase as $\phi(\alpha) := -\sigma_{\Re}\cdot t_{\mathrm{peak}}(\alpha)$. Thus, we have:
 \begin{align}
     \real(y_{\nonlin}(\alpha)) &= R(\alpha)\cos(\phi(\alpha)),\label{eq:realynonlin}\\
     \imag(y_{\nonlin}(\alpha)) &= R(\alpha)\sin(\phi(\alpha)).\label{eq:imagynonlin}
 \end{align} 
We show in figure~\ref{fig:comparisonmethod}C the reconstructed data $\real\left([\real(y_{\nonlin}(\alpha))+i\imag(y_{\nonlin}(\alpha))]e^{i\sigma  t}\right)$ (black dashed line) compared to $y(\alpha,t)$ (cyan solid lines) at three times.

Finally, we choose a phase $\phi$ that gives the best match between~$Y(x)$ from the eigenvalue analysis described in \S~\ref{sec:linearized} and $y_{\nonlin}(\alpha)$ from the nonlinear simulations. We do this by solving the following optimization problem:
\begin{equation}
    \min\limits_{\phi}\int_{-1}^1\left|\frac{[\real(y_{\nonlin}(x))+i\imag(y_{\nonlin}(x))]}{\max(\left|[\real(y_{\nonlin}(x))+i\imag(y_{\nonlin}(x))]\right|)}-\frac{Y(x)e^{i\phi}}{\max(|Y(x)e^{i\phi}|)}\right|\d x
\end{equation}
for $\phi\in[0,2\pi]$.

\section{Comparison of eigenmodes and
time-stepping simulations with free-free
boundary conditions \label{frfrcomp}}

\begin{figure}[H]
    \centering
    \includegraphics[width=.7\textwidth]{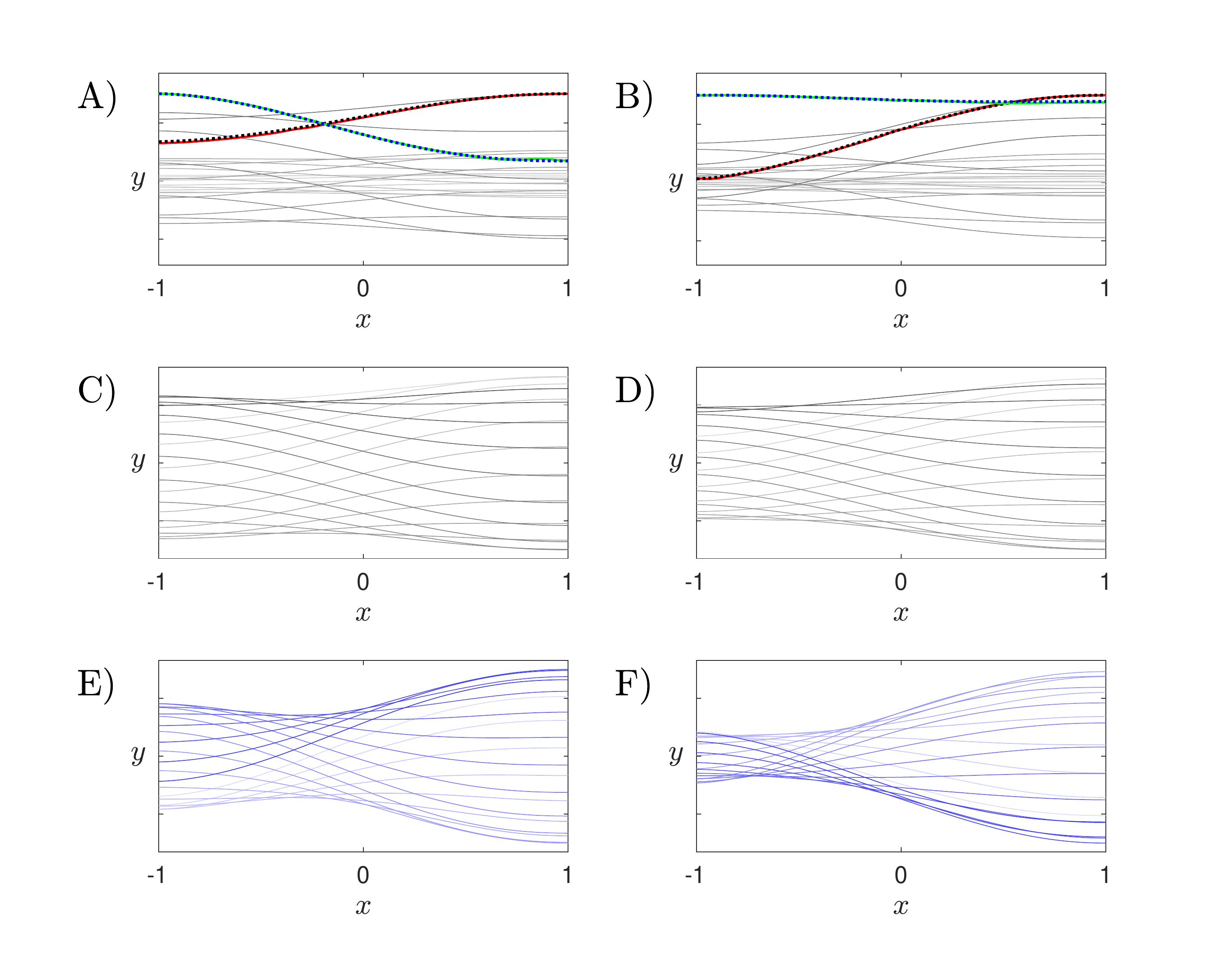}
\caption{Free-free membranes at $(R_1,T_0)=(10^{1},10^{0.1})$ for A, C, E and at $(R_1,T_0)=(10^{1.5},10^{0.2})$ for B, D, and F, with $R_3 = 10^{1.5}$ in both cases.
These membranes lose stability by flutter and divergence. In panels A and B the solid red lines are $\real(y_{\mathrm{ynonlin}}(\alpha))$ estimated from the time-stepping simulation, which are close to $\real(Y(x))$ from the eigenvalue problem (dotted black lines). The solid green lines are $\imag(y_{\mathrm{ynonlin}}(\alpha))$, close to $\imag(Y(x))$ from the eigenvalue problem (dotted blue lines). The gray lines are a subset of snapshots in the linear growth regime. In panels C and D we show the snapshots during the small-amplitude (growth) regime, but with the exponential growth removed. Panels E and F show snapshots during the steady-state large-amplitude motions. Shades of gray (and blue) increase from light to dark as 20 membrane positions cycle through a period.}
    \label{fig:comparisonfrfr}
\end{figure}

In figure~\ref{fig:comparisonfrfr}, we compare the eigenmodes to the time-stepping simulations for two cases of free-free membranes: $(R_1,T_0)=(10^{1},10^{0.1})$ (panels A, C, and E) and $(10^{1.5},10^{0.2})$ (panels B, D, and F) at $R_3 = 10^{1.5}$ in both cases. The comparison methods for fixed-free membranes (see appendix~\ref{app:comparison}) are used again here. In A and B we see close agreement between the real and imaginary parts of the eigenmodes obtained from the two methods. In panels C and D a point of inflection occurs close to the midpoint of the membrane, and migrates closer to the leading edge at large amplitude (panels E and F). The small- and large-amplitude shapes are similar in terms of the number of local maxima and minima of deflection (typically one of each).

\bibliographystyle{plain}  
\bibliography{biblio.bib} 

\end{document}